# From phase analysis to phase tracking: a robust approach to atomic-scale strain mapping

Ying Han[1] and Yuan Tian[2]*

[1] School of Materials Science and Engineering, Tianjin University, Tianjin, China.

[2] Department of Materials Science and Engineering, Southern University of Science and Technology, Shenzhen, China.

*Corresponding author: Yuan Tian (tiany3@sustech.edu.cn).

ORCID: Y.H. 0000-0002-6621-2351; Y.T. 0000-0002-7074-5468.

## Abstract

Atomic-scale strain governs the electronic, ferroic and mechanical properties of materials, and atomic-resolution microscopy measures it most directly. Geometric phase analysis (GPA) is the standard tool, but it fails when a Bragg reflection moves outside its fixed Fourier aperture under large strain or rotation, or in misoriented grains. Here we introduce geometric phase tracking analysis (GPTA), whose Fourier aperture tracks the Bragg reflection at each pixel in real space. A narrow aperture can therefore reject noise without losing a displaced reflection. On 13,795 simulated and experimental atomic-resolution images under various imaging conditions, GPTA reduces GPA's failure rate from 27% to 11.5% with up to 335 times better accuracy. It also gives strain and dislocation maps across grains, resistance to dynamical-scattering extinction, and a confidence map, capabilities that GPA lacks. Tracking the aperture combines reciprocal-space dose efficiency with real-space spatial resolution and extends atomic-scale strain metrology to beam-sensitive materials and challenging imaging conditions.

## Introduction

Band structure, ferroic order and mechanical response are governed by strain on the atomic scale[1–5]. Strain mapping from atomic-resolution transmission electron microscopy (TEM) images[6] gives the most direct measurement and resolves strain at the unit-cell scale. Diffraction-based mapping sacrifices this resolution for field of view[7–9]. Geometric phase analysis (GPA)[10–12], the established method for this analysis, uses two reflections from the Fourier transform of an atomic-scale image to obtain the distribution of the complete strain tensor. This simple, model-free construction is sufficiently precise for picometer metrology[13–15], and GPA has remained the principal tool for strain mapping for more than two decades[16–19].

GPA operates in reciprocal space, measuring strain from deviations of the lattice from perfect periodicity. It can therefore remain effective at doses and under contrast conditions where individual atomic columns are not resolved. Complementary methods operate in real space, including peak-pair analysis[20–26], which localizes individual atomic columns and derives strain from their positions. The two approaches have distinct limitations. Real-space methods degrade when columns are obscured by low dose, thickness or defocus, since each column must be individually resolved. Bragg filtering before peak detection and machine-learning-based denoising can recover some columns, but introduce assumptions from either the filter or the training set[27–30]. GPA and related reciprocal-space methods recover phase through demodulation using a fixed Fourier aperture to select a Bragg reflection and its reference (the uniform mean lattice of the field). The fringe phase is measured relative to this reference. The method fails when the reflection moves outside this aperture, whether because of large strain, lattice rotation, a misoriented grain, extinction by thickness or bending, or low dose[31–33]. In the low-dose regime, the aperture width also sets both the noise level and the spatial resolution, creating an intrinsic trade-off between them[16,34,35]. These limitations have been eased individually over decades of work. In windowed forms of GPA[36,37], transforms of a sliding window are used in place of the single Fourier transform of the whole image. The lattice frequency measured in each window is used as the local reference, allowing the analysis to follow a lattice that varies slowly across the field of view. Adaptive references[38–41] update the reference lattice from the displacement already measured, allowing the analysis to follow distortions well beyond the linear range of a single fixed reference. Aperture optimization at a given dose[35,42] accounts for the trade-off between the phase noise suppressed by a narrow aperture and the spatial resolution preserved by a wide aperture. It selects the width best suited to that dose. Scan-distortion correction and multi-frame registration[43–45] act on the image itself, removing the scan artifacts of the probe and aligning many low-dose frames into one image with a higher effective dose. The resulting cleaner image is then used for strain analysis. These advances constitute the present state of the art.

Two issues remain unresolved, the first being that each advance addresses one failure mode with a correction built for it, and none extends to images in which several coincide. More fundamentally, these corrections do not alter GPA's trade-off between spatial resolution and precision because they use the same information as GPA, i.e., one aperture around one reflection and one reference for the whole field. In an atomic-resolution image, the same lattice is represented in every unit cell and in every visible reflection. The precision available at a given spatial resolution depends on how much of that redundancy is used in the measurement.

Here we introduce geometric phase tracking analysis (GPTA), in which the Fourier aperture follows each Bragg reflection as its position changes with the local lattice. This tracking is possible because the reference is updated iteratively. In place of GPA's single fixed reference, GPTA uses a reference lattice tracked from the measurement in each iteration, which converges on the local lattice at every position in the field. The concept originates in radar interferometry, where a phase reference is repeatedly re-estimated from the measurement to recover terrain as phase[46,47]. When applied to a crystal lattice, the method uses both forms of redundancy in the image. Convergence at every position uses information from every unit cell, allowing a reflection displaced by strain, rotation or misorientation to be followed without being lost. Tracking the lattice variation also allows the aperture to be narrowed at low dose without losing the signal. In addition, confidence-weighted fusion of all visible reflections yields a single lattice basis with higher reliability. Suppression of one reflection by thickness or bending therefore no longer corrupts the map. On 13,795 simulated images spanning strain, defects, grain boundaries, foil geometry and dose, GPTA reduces the failure rate from 27% to 11.5% and halves the median noise level of successful maps. Experimental high-angle annular dark-field scanning TEM (HAADF-STEM) images confirm the same gains under low-dose conditions. By tracking the reference and fusing multiple reflections, GPTA combines the dose efficiency of reciprocal-space analysis with the spatial resolution of real-space methods.

# Results

## Geometric phase analysis (GPA) and the limits of a fixed reference

GPA measures strain as the phase of lattice fringes, each set of which corresponds to one Bragg reflection in the Fourier transform of an atomic-resolution image. GPA isolates a reflection $\boldsymbol{g}$ with a Gaussian aperture and transforms the masked diffractogram back to real space to obtain the Bragg-filtered image $H_{\boldsymbol{g}}$. The geometric phase is the phase of this image measured relative to the phase $2\pi\boldsymbol{g}\cdot\boldsymbol{r}$ of a uniform reference lattice (the mean lattice of the field)

$$\varphi_{\boldsymbol{g}}(\boldsymbol{r}) \;=\; \arg H_{\boldsymbol{g}}(\boldsymbol{r}) \;-\; 2\pi\,\boldsymbol{g}\cdot\boldsymbol{r} \;=\; -\,2\pi\,\boldsymbol{g}\cdot\boldsymbol{u}(\boldsymbol{r}) \tag{1}$$

with $\boldsymbol{u}$ denoting the displacement field[48,49] (formalism in Supplementary Note 1). The phases of two non-collinear reflections determine the full strain tensor (the complete workflow of GPA is shown in **Supplementary Fig. S1** and Supplementary Note 2). A Bragg reflection contains the summed signal of every atomic column in the field, allowing its phase to be measured with greater precision than is possible from any single column. GPA therefore remains precise at doses where individual columns are invisible. A perfectly periodic lattice concentrates each reflection into a sharp spot. Any departure from periodicity spreads it into a diffuse distribution whose extent grows with the distortion. This distribution is the strain signal in reciprocal space, and the aperture determines how much of it is included in the measurement (**Supplementary Fig. S2**). GPA is exact within four settings fixed before any measurement, i.e., the position of the aperture, its width, one reference lattice for the whole field, and exactly two reflections for the solution. **Figure 1** shows four common experimental conditions in which the assumptions behind these four settings are violated.

Under continuous lattice rotation (**Fig. 1a**), the lattice turns by 28° across a single crystal, as it does around a bend contour, a crack tip or an inclusion. The rotation smears each Fourier reflection into an arc. The fixed aperture passes the part of the arc that falls inside it, which accounts for 30% of the reflection power here. The rotation map is correct where the local lattice matches the mean orientation and breaks down elsewhere, with an error of 4.4° over a field whose true rotation span 28°. A lattice that turns by 37° reaches the same limit (**Supplementary Fig. S3**). Widening the aperture to $0.25|\boldsymbol{g}|$ for allowing the whole arc signal takes the cost of a coarser map and the noise passed by the aperture, i.e., a large aperture is limited by the coupling of one width to both the signal and the noise.

At low dose (**Fig. 1b**), the same rotating field is recorded at one tenth of the dose. The reflection is visible above the noise of the diffractogram (**Fig. 1b2**). The rotation map is obscured by noise, with an error of 6.3°, because the aperture now passes the stronger noise along with the arc. Narrowing the aperture to reject the noise would also reject part of the arc. This illustrates the resolution-precision trade-off created by the fixed aperture (aperture-width series in **Supplementary Fig. S4**).

At the 15° grain boundary of **Fig. 1c**, the distance between the two reflections from the two grains is 2.6 times the aperture width. When the aperture is centered on one grain, the other grain appears as noise. The grain boundary dislocations also distort the strain of the grain to which the reference belongs. Over the field, a 3% elastic strain is obscured by an error of 12% in the normal strain $E_{xx}$. The same mechanism occurs in all cases with misorientation, including a small-angle boundary (**Supplementary Fig. S5**) and a large-angle boundary such as a mirror twin (**Supplementary Fig. S6**).

The fourth condition is reflection extinction, in which variations in foil thickness, bending or tilt decrease the intensity of a reflection by dynamical diffraction while others remain visible[31,32,34] (**Supplementary Figs. S7 and S8**). The intensity is also decreased by the defocus-dependent contrast reversal of high-resolution TEM (HRTEM)[18,33]. In **Fig. 1d**, we impose this behavior analytically by modulating the amplitude of one fringe family with a periodic envelope whose contrast passes through zero and reverses sign. The region of near-zero amplitude is the reversal band (Methods). At positions where the fringe amplitude vanishes, GPA gives 16% root-mean-square (RMS) strain in a crystal whose true strain is zero. Two reflections give exactly as many constraints as the lattice basis has components, leaving no redundancy for error detection or repair if either contains an error. The error extends beyond noise because the phase is undefined where the fringe amplitude passes through zero. Noise at these positions creates pairs of phase singularities that are indistinguishable from dislocation cores[50] (**Supplementary Figs. S9 and S10**). Supplementary Note 3 shows how each of the four limits arises in the intermediate quantities. All four failures occur because GPA fixes its four settings before the measurement, and in each condition the image no longer matches one of them. The reflection has moved away from the aperture position, the noise requires a narrower aperture than the displaced reflection allows, the reflection belongs to a lattice other than the reference, or one of the two chosen reflections has disappeared. GPA is therefore limited by its fixed settings, not by its phase measurement. This observation motivates the method of this paper, geometric phase tracking analysis (GPTA). GPTA keeps the phase measurement of GPA and replaces each of the four fixed settings with a quantity that is estimated from the image itself and updated over a small number of iterations. The following sections build GPTA in three steps, first rewrite GPA as a demodulation against a reference lattice, then track that reference from the image, and finally combine all reflections above the noise level into one lattice basis.

## From a fixed to a tracked reference

To develop GPTA from GPA, we first rewrite GPA as an equivalent demodulation process. **Figure 2** shows the analysis of one reflection of the rotating lattice of **Fig. 1a** at a lower dose. In the textbook construction of GPA, masking $\boldsymbol{g}_1$ in the diffractogram (**Fig. 2c**) and transforming back gives the Bragg-filtered image $H_{\boldsymbol{g}}$, whose phase (**Fig. 2e**) contains the fringes of the reference lattice. Subtracting the reference phase $\Phi_{\boldsymbol{g}} = 2\pi\boldsymbol{g}\cdot\boldsymbol{r}$ of the mean lattice (**Fig. 2d**) gives the geometric phase $\varphi_g = \arg H_{\boldsymbol{g}} - \Phi_{\boldsymbol{g}}$ of Eq. (1) (**Fig. 2h**). Its gradient gives the local rotation $\theta = \arg(1 + \nabla\varphi_g/2\pi\boldsymbol{g}_1)$ (**Fig. 2k**). Two-dimensional vectors are written as complex numbers, and arg of a quotient then gives the angle between them. This process is equivalent to demodulation (**Fig. 2f–m**). The same map is obtained in one step by multiplying the image $I$ by $\exp(-i\Phi_{\boldsymbol{g}})$, which shifts the reflection to the origin of the diffractogram (**Fig. 2f**), and applying a low-pass Gaussian aperture of the same width around the origin (the operator LP of **Fig. 2**). **Supplementary Figs. S11 and S12** and Supplementary Note 4 show the detailed operations in one and two dimensions. The demodulated field $\psi_g = \mathrm{LP}[I\exp(-i\Phi_{\mathrm{ref}})]$, with $\Phi_{\mathrm{ref}}$ denoting the phase of any reference used, has the residual phase $\varphi_{\mathrm{res}} = \arg\psi_g$ (**Fig. 2g**), which is the phase of the lattice relative to the reference. The local wave vector is obtained as $\boldsymbol{k}_1 = \nabla(\Phi_{\mathrm{ref}} + \varphi_{\mathrm{res}})/2\pi$ (**Fig. 2i**) and the rotation as $\theta = \arg(\boldsymbol{k}_1/\boldsymbol{g}_1)$ (**Fig. 2j**). With the plane-wave reference $\Phi_{\mathrm{ref}} = \Phi_{\boldsymbol{g}}$, $\psi_g$ is $H_{\boldsymbol{g}}\exp(-i\Phi_{\boldsymbol{g}})$. The two routes coincide at each operation (Supplementary Note 1). This formulation shows that the reference is an input to the measurement. In GPA, the mean lattice is specified once and is not revised, and the aperture alone must therefore accommodate all lattice variations across the field.

GPTA revises it using the wave-vector field of the first iteration, which measures where the local lattice lies at each pixel. Smoothing and integration of this field give a phase correction $\varphi_{\mathrm{ref}}$ (**Fig. 2l**), and $\Phi_{\mathrm{ref}} = \Phi_{\boldsymbol{g}} + \varphi_{\mathrm{ref}}$ is a reference whose fringes turn with the lattice (**Fig. 2m**). For a step-by-step view, the same construction is demonstrated on a one-dimensional signal and on a dislocation in **Supplementary Figs. S13 and S14**. The next iteration demodulates the image using this reference. Because the rotation is now included in the reference, each pixel's reflection is brought to the origin by its own local reference wave ($\Phi_{\mathrm{ref}}$) rather than by the field average ($\Phi_{\boldsymbol{g}}$). The diffractogram of the demodulated field therefore contracts around zero frequency in successive iterations (**Figs. 2r, t, v**). This contraction is local, with the reference converging in each iteration on the part of the arc that belongs to the lattice at each position in the field. As a result, regions with different local orientations are each brought to the origin by their own part of the reference (**Supplementary Fig. S15** follows this convergence at three positions in the field). The reference tracks the reflection across one image and has converged after two iterations, and the aperture is narrowed in the third iteration (every intermediate array of the three iterations is shown in **Supplementary Fig. S16**).

Tracking allows the aperture to be narrowed because only the residual needs to pass through it. A fixed reference, however, requires an aperture wide enough to pass the whole arc. With the tracked reference, the aperture can be narrowed from $0.25|\boldsymbol{g}|$ to $0.125|\boldsymbol{g}|$ and $0.05|\boldsymbol{g}|$ over three iterations while the error in the rotation map falls from 4.2° to 2.2° and 1.6° (**Figs. 2s, u, w**). In comparison, the same $0.05|\boldsymbol{g}|$ aperture placed around the plane-wave reference removes the arc together with the noise and raises the GPA error to 5.1° (**Figs. 2p, q**). Tracking and narrowing are therefore complementary, since noise remains when the aperture is tracked at a fixed width, while narrowing the aperture removes the signal if the reference is fixed.

## The GPTA iteration step by step

**Figure 3** shows one iteration of GPTA on a simulated edge dislocation image, from the image to the reference for the next iteration. Before the iteration starts, the image is detrended, tapered with a Hann window and masked to the crystalline area (the pre-treatment of **Supplementary Fig. S17**). The basis reflections $\boldsymbol{g}_1$ and $\boldsymbol{g}_2$ are found from the diffractogram (**Figs. 3a, b**), and the noise level $N_0$ of every reflection is measured as the median spectral power in an annulus at $|\boldsymbol{g}|$ from which all reflections are excluded (Methods and **Supplementary Fig. S18**). The iteration consists of three steps, which are demodulation of each reflection, fusion of all reflections into one lattice basis, and integration of the fused basis into the next reference.

In the first step, each reflection $\boldsymbol{g}$ is demodulated using the current reference (**Figs. 3c, d**),

$$\psi_{\boldsymbol{g}}(\boldsymbol{r}) = \mathrm{LP}_b\left[I(\boldsymbol{r})\exp[-i(2\pi\boldsymbol{g}\cdot\boldsymbol{r} + \varphi_{\mathrm{ref}}(\boldsymbol{r}))]\right] \quad (2)$$

where $\mathrm{LP}_b$ is a Gaussian aperture of width $b|\boldsymbol{g}|$ around the origin of the demodulated diffractogram, and $\varphi_{\mathrm{ref}}$ is the phase correction added to the plane-wave phase $\Phi_{\boldsymbol{g}} = 2\pi\boldsymbol{g}\cdot\boldsymbol{r}$ of the mean lattice for this reflection (zero in the first iteration). The total reference is $\Phi_{\mathrm{ref}} = \Phi_{\boldsymbol{g}} + \varphi_{\mathrm{ref}}$. The demodulated field $\psi_g$ is complex. Both its phase arg $\psi_g$ and its modulus $|\psi_g|$ are used. Its phase is the residual phase $\varphi_{\mathrm{res}} = \arg \psi_g$ between the lattice and the reference. The gradient of this phase gives the local wave vector of the reflection, $\boldsymbol{k}_g = \boldsymbol{g} + \nabla(\varphi_{\mathrm{ref}} + \varphi_{\mathrm{res}})/2\pi$ (**Fig. 3e**). Its modulus $|\psi_g|$ is the local amplitude of the reflection, which GPA computes and discards. Divided by the noise level $N_0$, it gives the confidence $|\psi_g|^2/N_0$ (**Fig. 3f**), which is proportional to the inverse variance of the phase at that pixel. The confidence therefore maps the reliability of each reflection at each position in the field. It is high where the fringes of $\boldsymbol{g}$ are strong and clean, falls where they are weakened by thickness, bending or noise, and reaches zero at the dislocation core, where no single-valued phase exists.

In the second step, every reflection whose median confidence exceeds a threshold is demodulated in the same way, using the phase implied by the common reference for its own $\boldsymbol{g}$ (**Fig. 3g–i**, with the threshold decision shown in **Supplementary Fig. S18**). The wave-vector gradient of each reflection is averaged over a window of one to three lattice periods with the weight $|\psi_g|^2$. As a result, pixels where the reflection is strong dominate the average and pixels where it has vanished add no noise to their neighbors (Methods). Because all reflections encode the same displacement field, their wave vectors constrain one lattice basis ($\boldsymbol{v}_1$, $\boldsymbol{v}_2$), which is found by confidence-weighted least squares,

$$(\boldsymbol{v}_1, \boldsymbol{v}_2) = \mathrm{argmin} \sum_{mn} w_{mn} \|\boldsymbol{k}_{mn} - (m\boldsymbol{v}_1 + n\boldsymbol{v}_2)\|^2, \quad w_{mn} = |\psi_{mn}|^2/N_{0,mn} \quad (3)$$

where $\boldsymbol{k}_{mn}$ is the measured wave vector of the reflection $m\boldsymbol{g}_1 + n\boldsymbol{g}_2$ and $N_{0,\mathrm{mn}}$ is its noise level. Up to twelve reflections constrain the four components of the basis (**Figs. 3j, k**), making the inversion overdetermined, with three benefits that cannot be obtained with GPA. The noise of the fused basis is lower than that of any single reflection because independent measurements of the same basis are averaged with their proper weights. The solution is robust because a reflection that is weak or lost at some position receives a low weight there and the basis remains determined by the others. Thus, no single reflection can corrupt the map (the fusion with a suppressed reflection is shown in **Supplementary Fig. S19**). The residual $\chi^2$ of the fit indicates where the reflections disagree (**Fig. 3l**). Disagreement arises where the local image does not represent one lattice with one displacement field, for instance at a dislocation core, where two grains overlap at a boundary, or where the phase of a weak reflection is dominated by noise. The $\chi^2$ map therefore marks the pixels at which the strain value is not to be trusted (**Supplementary Fig. S20b** shows this map on a dissociated dislocation, whose partials cannot be described by a single lattice). It is included with every result together with the total confidence $\Sigma|\psi_i|^2/N_0$ (**Fig. 3m**), and the fused basis gives the strain tensor (**Figs. 3n, o**).

In the third step, the fused wave-vector field is smoothed by an edge-preserving filter and integrated into the phase correction $\varphi_{\mathrm{ref}}$ of the reference for the next iteration (**Fig. 3p**). Because the reference is integrated from a wave-vector field, a dislocation can be written into it. The integer winding of the core is detected in the fused field before integration and included in the reference as a phase vortex[51,52]. The residual around the core consequently stays single-valued and small, and the core is reported with its Burgers vector (the integral methods available for this step are compared in **Supplementary Fig. S21**). We call a reference that contains phase vortices in this way the topological reference and a reference integrated as a single-valued phase the curl-free reference. **Supplementary Fig. S20a** compares the two on a wall of three dislocations, and **Supplementary Fig. S22** shows the displacement fields, winding densities and Burgers vectors obtained with the topological reference.

The next iteration demodulates using the new reference with a narrower aperture. Three iterations are sufficient for the dislocation in **Fig. 3**, with the aperture narrowing from 0.25|***g***| to 0.125|***g***| and 0.09|***g***| and the strain error falling from 0.23% to 0.17% and 0.13%. The reference has converged after the second iteration, and the aperture is narrowed in the third. Iteration 1, before any revision and with the two basis reflections alone, is identical to GPA (**Figs. 2 and 4a, b** and Methods). The iteration process is given in more detail, with all its intermediate arrays, noise levels and weights, in **Supplementary Fig. S23** and Supplementary Note 5.

## Four components of the iteration

The four failures in **Fig. 1** were traced to four fixed settings of GPA. In each component of the GPTA iteration in **Fig. 3**, a quantity measured from the image is used in place of one fixed setting. The tracked reference is used in place of the fixed aperture position, and the aperture is narrowed from the fixed width. A separate reference is used for each lattice in place of the single reference lattice, and the joint inversion is used in place of the fixed pair of reflections. **Figure 4** shows the four images of **Fig. 1** again, with each setting examined individually to show how the corresponding component of GPTA overcomes the failure.

**The reference tracks the reflection.** In **Fig. 1a**, the reflection was smeared into an arc and only a fraction of it passed through a fixed aperture. This failure is overcome by tracking the reflection at each pixel in real space instead of widening the aperture. In iteration 1, every pixel of the field is demodulated using the center of the arc, which is identical to GPA at 0.25|***g***| (**Fig. 4a1**). The wave vector measured in that iteration already locates the part of the arc corresponding to each pixel. This information is included in the integrated reference used in iteration 2. From then on, each pixel is demodulated using its own local reference wave ($\Phi_{\mathrm{ref}} = \Phi_{\boldsymbol{g}} + \varphi_{\mathrm{ref}}$) updated in the previous iteration (**Fig. 4a2–a7**). Multiplying the image by $\exp(-\mathrm{i}\Phi_{\mathrm{ref}})$ shifts the reflection of every pixel to the origin of the demodulated diffractogram, so the aperture no longer has to span the arc. The rotation error falls from 0.89° in iteration 1 to 0.26° after three iterations, compared with 4.4° for the fixed 0.10|***g***| aperture of GPA in **Fig. 1a**. The same procedure is used for a 37° rotation (**Supplementary Fig. S3**) and a small-angle boundary whose second reflection lies at the edge of a fixed aperture (**Supplementary Fig. S5**).

**The aperture narrows at low dose.** In **Fig. 1b**, an aperture wide enough for the arc signal also passed the noise within it, which becomes more damaging at low dose. Tracking resolves this dilemma by incorporating the local wave vector into the reference of each pixel. The reflections that were spread along the arc are then all brought to the origin by the same multiplication by $\exp(-\mathrm{i}\Phi_{\mathrm{ref}})$ (**Fig. 4b1**). The aperture can therefore be narrowed around the reflections, from 0.25|***g***| to 0.125|***g***| and 0.05|***g***| over three iterations (**Figs. 4b2, b4, b6**). The narrow aperture rejects the noise and retains the rotation because the rotation is included in the reference. The rotation error falls from 4.0° to 0.67° (**Figs. 4b3, b5, b7**), compared with 6.3° for the fixed aperture of **Fig. 1b**. In the aperture-width series of **Supplementary Fig. S4**, 6% of the field has an error above 0.02|***g***| (the bad area, Methods) with the best fixed width. This fraction is 27% with the tracked reference and the same two reflections, and 0% with fusion of the four reflections above threshold (Supplementary Note 3).

**Each lattice receives its own reference.** In the bicrystal case of **Fig. 1c**, the reflections of the two grains lie 2.6 aperture widths apart, and one reference could not be used for both grains. For this case, GPTA uses a domain-aware reference initialization algorithm in which grains are separated according to confidence. The amplitude $|\psi_g|$ of each grain's reflections is high on its own side of the boundary and low on the other (**Figs. 4c2, c3**). Assigning every pixel to the lattice with the highest confidence therefore partitions the field into domains without prior knowledge of the boundary. A separate tracked reference is then used for each domain (**Figs. 4c4, c5**), and the two references are stitched into one (**Fig. 4c6**). The boundary itself is a wall of edge dislocations, whose positions are obtained from the curl of the stitched wave-vector field. Each core is written into the topological reference as a phase vortex, in the same way as the isolated core of **Fig. 3p**.

In a final iteration, the elastic field of both grains is measured together with the dislocations that form the boundary (**Fig. 4c7**). The misorientation is recovered as 14.7° for a true 15°, and the seven cores inside the scored area are placed within about one lattice period of their true positions. The strain error in the grains is 1.5%, which is much lower than 12% for GPA in **Fig. 1c**. The same procedure is used to separate mirror-twin variants whose reflections overlap within one GPA aperture (**Supplementary Fig. S6**). Every step of the algorithm on a second boundary is shown in **Supplementary Fig. S24** (Supplementary Note 6).

**All reflections are fused.** In **Fig. 1d**, one of the two reflections used by GPA was extinguished by the envelope, and using only two reflections leaves no redundancy with which the loss could be detected. Confidence-weighted fusion of six reflections in the inversion gives both the redundancy and the detection (**Fig. 4d1**). Every reflection of the lattice lies at $m\boldsymbol{g}_1 + n\boldsymbol{g}_2$, which is a linear combination of the two basis vectors. The higher-order reflections are therefore located automatically from the basis and included once their confidence exceeds the threshold. Inside the reversal band, the phase of the suppressed (1,0) family is noise, but its amplitude (i.e., confidence) falls to near zero at the same place (**Figs. 4d4, d5**). Its weight in Eq. (3) decreases continuously to zero as the reflection approaches extinction, and the lattice basis is then reconstructed from the remaining reflections. The diagonal and higher-order reflections constrain the same basis and give the component lost when the reflection was extinguished. The strain error inside the band falls from 1.9% (two-reflection) to 0.44% (fused solution) (**Fig. 4d6, d7**), matching the value outside the band. The same procedure works for a bent foil in which different reflections vanish in different regions (**Supplementary Fig. S7**) and an envelope that would otherwise nucleate false dislocation cores (**Supplementary Fig. S9**). With the four components, the fixed settings of GPA are replaced by quantities determined from the image at every pixel. The analysis adapts to the specimen, removing the requirement for the specimen to fit the analysis.

## Comparison with existing methods

We compare GPTA with GPA, windowed GPA (S-GPA, **Supplementary Fig. S25**) and peak-pair analysis (PPA, **Supplementary Fig. S26**) on our benchmark of 13,795 simulated HRTEM images with known strain fields. The implementations are described in Methods and Supplementary Note 7. The images fall into ten families, each containing image types that share one physical origin of the lattice distortion (smooth elastic deformation, foil thickness and geometry, dislocations, planar defects, grain boundaries, heterointerfaces and cracks, continuous lattice rotation, lattice modulation, real crystal structures and mixed conditions). Dose is varied within every family under one fixed scoring protocol. Accuracy is scored from the error of the recovered local wave vectors relative to the ground truth, $|\boldsymbol{k} - \boldsymbol{k}_{\mathrm{GT}}|/|\boldsymbol{g}|$. A map fails when more than 20% of the scored pixels exceed an error of 0.02. For the maps that pass, $k_{50}$ is the median error over the scored area (Methods and Supplementary Note 8). In **Fig. 5**, we first select the four conditions of **Fig. 1** from this benchmark and compare the methods under them.

**Figure 5a** compares the four methods on four images that represent the four conditions of **Fig. 1** (large strain, low dose, a bicrystal and reflection extinction). Maps from all four methods for each image are shown in **Supplementary Fig. S27**. These comparisons show three patterns, with GPTA gaining a factor of six over GPA when large strain displaces a reflection. Under this condition, the sliding window of S-GPA gains little, and PPA loses accuracy as the distortion degrades its column localization. When nothing displaces the reflection and the dose is low, the three reciprocal-space methods perform similarly because the narrow fixed aperture of GPA already rejects most of the noise. PPA falls behind because the scatter of individual column positions enters its strain directly. The gain of GPTA reaches two orders of magnitude on the second grain of a bicrystal and under reflection extinction, conditions under which GPA breaks down. PPA is the most accurate method on the high-dose bicrystal because it measures individual columns without assuming a shared lattice. Under extinction, S-GPA falls below GPA because a sliding window cannot recover a reflection that is gone.

In the series in **Fig. 5b–e**, each condition becomes progressively more challenging, giving a quantitative comparison and showing where each method fails. As the peak strain rises (**Fig. 5b**), the fixed apertures of GPA and S-GPA truncate the displaced reflection and the distortion degrades column localization in PPA. GPTA tracks the reflection, and its error rises by less than a factor of three. With increasing dose (**Fig. 5c**), the GPTA error continues to fall with counting statistics over two decades, whereas GPA and S-GPA reach a plateau set by their apertures. GPTA recovers the step in $\theta$ across the boundary as grain misorientation varies from 1° to 45° (**Fig. 5d**). GPA and S-GPA lose the step once the reflections of the second grain leave their fixed apertures, at about 15° and 10°. PPA also recovers the step at this high dose because it does not assume a common lattice, but its precision deteriorates as columns become indistinct at low dose (**Supplementary Figs. S28–S30** and Supplementary Note 9). The methods differ most under reflection extinction (**Fig. 5e**), when the loss of one of two reflections leaves GPA and S-GPA without a measurement. PPA degrades by more than an order of magnitude because it locates the columns in the Bragg-filtered image. When one fringe family has vanished, the columns form ridges along the surviving family instead of isolated maxima, leaving their positions along the lost direction undetermined. GPTA retains its overdetermined basis through the fusion, with an error nearly two orders of magnitude below GPA at the deepest reversal.

In **Fig. 5a**, we also determine how the components of GPTA (tracking with the narrowing aperture, the confidence-weighted fusion and the domain-aware algorithm) contribute to its gain over GPA by switching them on in sequence (Supplementary Note 10 and Supplementary Table 1). This identifies which component resolves each condition. Under large strain, tracking and fusion contribute about equally, since the displaced reflection is recovered using the tracked reference and the noise of the narrowed aperture is reduced by fusion. At low dose on an undisplaced lattice, no component contributes much because the fixed aperture of GPA is already narrow. On the bicrystal, the gain results almost entirely from the domain-aware algorithm, since tracking cannot help until the second grain has its own reference. Under extinction, the gain results mainly from fusion, in which the remaining reflections are used in place of the lost reflection, with tracking adding a smaller factor. With all components switched off, GPTA is identical to GPA, showing that the gains result from the components. There is no difference in implementation that contributes to these gains.

The full benchmark extends the comparison to the mixed and complex conditions encountered in practice, where several distortions and low dose coincide. Across all 13,795 images, GPTA reduces the failure rate from 27% for GPA and 24–27% for the three comparison methods to 11.5% and halves the median error of the passing maps (Table 1 and Supplementary Tables 2 and 3). It has the lowest failure rate in nine of the ten families. The difference is largest in the families of grain boundaries and mosaics, heterointerfaces and crack tips, and planar defects, in which the conditions of **Fig. 1** combine and the fixed reference and the fixed pair of reflections of GPA fail together. PPA depends on reliable column positions, which explains both its accuracy on the high-dose bicrystal and its large deficit in the grain-boundary family, which spans the full dose range. Continuous rotation with kink bands and slow lattice modulations remain the hardest families for every method. In the first, GPTA and PPA lower the failure rate of GPA by a third, with no further reduction achieved by any method. The sliding window of S-GPA, designed for gradual spatial variation, has the lowest failure rate in the second. Tracking takes about an order of magnitude more computing time than GPA and little more than S-GPA, requiring a few seconds for a $256^2$ image on a laptop processor (Supplementary Note 11).

## Experimental HAADF-STEM images

**Figure 6** compares the performance of different methods on real HAADF-STEM frame stacks of three Pt specimens. Each map is computed from one image, either a single frame or the registered average of sixteen frames of the same region. The two doses therefore differ by a factor of sixteen (Supplementary Note 12). The specimens are, respectively, a [110] polycrystal with more than six grains (**Fig. 6a**), a [111] bicrystal with a 23.5° tilt boundary (**Fig. 6b**), and a [001] region crossed by a low-angle wall of eight dislocation cores (**Fig. 6c**).

The behavior seen in the simulations recurs in all three specimens, with GPTA resolving all six grains in the polycrystal (**Fig. 6a**) at both doses, each with its own reference, while GPA measures only the grain containing its reference and returns the others as noise. PPA also resolves the grains since the columns can be clearly localized (**Supplementary Fig. S31**). In the bicrystal (**Fig. 6b**), GPA again measures only one grain while GPTA resolves the boundary at both doses, with low noise in both grains. Column localization limits PPA in single-frame data. In the sixteen-frame average, PPA resolves the boundary as sharply as GPTA (**Supplementary Fig. S32**). In the dislocation region (**Fig. 6c**), which contains a single lattice, GPA remains valid but noisier than GPTA, whereas column displacements and spurious peaks degrade PPA near the cores (**Supplementary Fig. S33**). Quantitative comparison shows that the core positions detected by GPTA from a single frame lie within one lattice period of their reference positions, whereas those detected by GPA, S-GPA and PPA lie more than one period away at every dose (Supplementary Note 12). A dislocation core is thus localized from one frame more finely than with the real-space method. The single-frame maps show the dose efficiency of the reciprocal-space measurement, with GPTA recovering the six grains, the boundary and the eight cores from one frame. PPA recovers the grains at both doses and the boundary after sixteen frames but not the cores, and GPA recovers none of them at low dose (**Supplementary Figs. S31–S33**).

## Discussion

GPTA changes the source of the reference while retaining the phase measurement established by GPA. GPA measures the lattice phase relative to a reference that contains no information about how the lattice varies across the specimen. Every departure of the lattice from that reference, whether strain, rotation, a second grain or a lost reflection, is measured through one fixed aperture. With this aperture, the departure that should be measured cannot be distinguished from the one that moves the reflection out of its range. GPTA measures the phase relative to a reference derived from the specimen itself, and after three iterations that reference is the local lattice at every pixel. A small residual remains to be measured throughout the field, allowing the aperture to narrow around it without losing signal. The measurement remains in reciprocal space and retains the dose efficiency of a reflection containing the signal of thousands of columns. It yields the reciprocal-lattice basis at each pixel. As a further consequence of this change, the amplitude of a demodulated reflection, which GPA discards, becomes a confidence measure because it measures how much lattice information the reflection contains at that pixel. The confidence is used to weight the joint inversion and is included with every strain map.

These results bear on a long-standing division in strain mapping, in which a microscopist locates an atomic column using two kinds of information at once. These are the local appearance of the image, which constrains the column's position, and the periodicity of the lattice, which constrains where the column is expected. Real-space methods use the first kind of information, and GPA uses the second. The two have been converging, as the Bragg filter applied before column localization in peak-pair analysis is a periodicity prior. Consequently, the detected peaks belong to the periodic component of the image and not to the raw image. A convolutional network for column localization[28,29] applies a learned filter over a receptive field of several lattice periods. This operation has finite spatial resolution, and its low-dose performance is limited by the same balance between field size and averaged noise. The full lattice GPTA combines these kinds of information from the reciprocal-space side, and the basis tracked at each pixel localizes the measurement as far as the signal allows. In the benchmark, peak-pair analysis matches or exceeds GPTA when individual columns are dependable, as on the high-dose bicrystal (**Fig. 5a**), while GPTA performs better when the lattice is present and its columns are not clear. An optimal low-dose strain estimator will use both kinds of information explicitly, with the confidence map of GPTA giving the quantity needed to weight this combination.

## Conclusion

We have introduced geometric phase tracking analysis, which retains the phase measurement of GPA while tracking its reference lattice from the image. As a result, the Fourier aperture follows each reflection and narrows around the residual, each lattice is treated separately, and all visible reflections are fused according to confidence. On 13,795 simulated images and three experimental HAADF-STEM stacks, this lowers the failure rate from 27% to 11.5% and improves the accuracy by up to two orders of magnitude when a reflection is displaced, extinguished or belongs to a second grain. The measurement remains in reciprocal space and retains its dose efficiency, which extends atomic-scale strain mapping to single low-dose frames of beam-sensitive materials.

## Methods

**GPTA algorithm.** The iteration in **Fig. 3** is run with the following settings. Before demodulation, the image is detrended using a least-squares plane, tapered with an 8% Hann window and masked to the crystalline area. The basis reflections are found from the diffractogram. For every reflection, the noise level $N_0$ is the median spectral power in an annulus at $|\boldsymbol{g}|$ from which all reflections are excluded (**Supplementary Figs. S17 and S18**). Reflections whose median confidence $|\psi_g|^2/N_0$ on the mask falls below 3 are excluded from the inversion (**Supplementary Fig. S18**). The wave-vector gradient is the $|\psi_g|^2$-weighted mean over an averaging window of 2.5 lattice periods by default. Its dependence on dose is mapped in Supplementary Note 10.4. The fused basis is smoothed by an edge-preserving median filter before least-squares integration into the next reference, and dislocation cores are detected as integer windings of the fused field and inserted into the reference as phase vortices during the first two iterations (**Supplementary Fig. S21**). If the diffractogram contains more than one lattice, a candidate lattice is accepted when its reciprocal cell matches an observed set of first-order reflections and its orientation differs by more than 2° from every accepted lattice. Each pixel is assigned to the candidate with the highest demodulated confidence, and a separate reference is tracked in each domain. The references are stitched, and the boundary dislocations located from the curl of the stitched field are injected as phase vortices before a final iteration (**Supplementary Fig. S24**). Three iterations are run with the aperture schedule 0.25, 0.125 and $0.09|\boldsymbol{g}|$, which is the software default used in **Fig. 3**, the benchmark and the experimental maps. In **Figs. 2 and 4a, b**, three iterations are run with 0.25, 0.125 and $0.05|\boldsymbol{g}|$, no averaging window and a curl-free reference. With these settings, iteration 1 in these panels is GPA at $0.25|\boldsymbol{g}|$. The formulation and the equivalence to GPA are given in Supplementary Note 1, and the full procedure with every intermediate array is given in Supplementary Note 5.

**Comparison methods.** Standard GPA[10] uses a Gaussian aperture of width $\sigma|\boldsymbol{g}|$ ($\sigma$ is the Gaussian width). In **Figs. 1, 4 and 5** and Table 1, we set $\sigma = 0.10$, which minimizes the error on undistorted benchmark lattices. Supplementary Note 10.5 gives the **Fig. 5a** cases at the textbook setting $\sigma = 0.25$, which is also used for the GPA panels in **Fig. 2**. S-GPA follows the windowed Fourier-transform method of Zhang et al.[36] with a two-lattice-period window. It uses a $0.10|\boldsymbol{g}|$ aperture, five frequency samples and an amplitude threshold of 0.15. PPA implements the peak-pairs construction of Galindo et al.[20] with an updated peak-finding step comprising a Bragg filter limited to first-order reflections, non-maximum suppression at the lattice spacing, centroid localization on the unfiltered image and clustering of bonds into grains. This implementation is used in **Fig. 5**, Table 1 and **Supplementary Fig. S27**. In **Fig. 6**, we use the classic construction with one global reference cell because grain clustering assigns the 23.5° bicrystal to a single grain at every dose up to sixteen frames (Supplementary Note 12). We use the same reference basis for all methods, obtained by lattice fitting to the image. With basis finding held constant, the comparison concerns phase extraction. The basis finder selects the strongest first-order reflections. If a weak reflection is selected beside a much stronger one, demodulation draws it to the neighbor, and this condition is reported in the confidence map (**Supplementary Fig. S34** and Supplementary Note 13).

**Simulated images.** The rotating lattice in **Figs. 1a, 2 and 4a** is a $256^2$ px benchmark image whose lattice has a period of 9.3 px and turns by 28° across the field, recorded at 200 e/px. Its low-dose versions in **Figs. 1b, 2 and 4b** are obtained by binomial thinning of the same counts to 20 and 8 e/px. The boundary in **Figs. 1c and 4c** is a 15° tilt wall of edge dislocations with the isotropic Hirth–Lothe field plus a 3% smooth elastic field, at 50 e/px, with seven of its nine cores inside the scored area. For the reflection-loss image in **Figs. 1d and 4d**, we multiply the amplitude of the (1,0) fringe family by the envelope

$$A(x) = 1 - m + m\cos(2\pi x/\lambda) \quad (4)$$

with a period $\lambda$ = 96 px at 200 e/px. The contrast-reversal depth $m$ gives a minimum amplitude of $1 - 2m$, which causes the contrast to reverse for $m > 0.5$. The reversal band is the region where $|A| < 0.15$. The dislocation in **Fig. 3** is a $512^2$ lattice with a period of 8 px and one edge dislocation with a Burgers vector of one period, at 120 e/px. For **Fig. 5**, the strain series is the smooth displacement field $u_x = A\sin(2\pi x/N)\cos(\pi y/N)$ on a $256^2$ lattice with peak strains of 0.5–25% at 200 e/px. For the dose series, we record one smooth-strain image at 3–300 e/Å$^2$. For the misorientation series, we use a bicrystal with a tilt boundary of 1–45°. We use Eq. (4) with depths of 0.6–1.2 for the extinction series. The four images in **Fig. 5a** are a strained layer at 3 e/px, the 25% image of the strain series, a 30° bicrystal at 200 e/px and the $m$ = 1.2 image of the extinction series.

**Benchmark data generation.** To generate each simulated image, we start with a lattice of Gaussian columns and a prescribed displacement, thickness or envelope field and add dislocations where specified and Poisson noise at doses from 1 to 500 electrons per pixel. The lattice period is 8 px and the reference basis is known exactly. The ten physical families and their sizes are smooth elastic fields (6,950 images), thickness and foil geometry (1,266), dislocations and dipoles (1,535), planar defects (521), grain boundaries and mosaics (1,239), heterointerfaces and crack tips (310), continuous rotation and kink bands (342), lattice modulations (756), real crystal structures (314), and mixed or masked images (562). The two image sets contain 10,248 images of $256^2$ px and 3,547 images of $1024^2$ px. We use multislice simulation[53–55] to generate the thickness, bending and tilt cases of the thickness and foil-geometry family.

**Scoring.** For the wave-vector field ($\boldsymbol{k}_1$, $\boldsymbol{k}_2$) obtained with each method, we calculate the error metric $k_{50}$ as the median of $|\boldsymbol{k} - \boldsymbol{k}_{\mathrm{GT}}|/|\boldsymbol{g}|$ over the scored area after global re-indexing of the basis by an integer transformation. For images with several grains, the ground truth is assigned per grain, and a 24 px margin is omitted from the scored area. The bad area is the fraction of this area at which $|\boldsymbol{k} - \boldsymbol{k}_{\mathrm{GT}}|$ exceeds $0.02|\boldsymbol{g}|$. A map fails when the bad area exceeds 0.20. The median of three noise realizations is plotted for every curve in **Fig. 5**. In the rotation maps in **Figs. 1, 2 and 4**, we plot $\theta = \arg(\boldsymbol{k}_1/\boldsymbol{g}_1)$ relative to the fitted $\boldsymbol{g}_1$, with the ground truth re-indexed onto the same basis. We calculate their RMS error over the assessed field, 40 px inside the frame.

**Software.** Lattice Studio is the open-source desktop application distributed at https://github.com/YTianUCI/LatticeStudio and shown in **Supplementary Fig. S35** and Supplementary Note 14. GPA, S-GPA, PPA and GPTA are implemented in one Python code base and use the same reference basis. The user loads a frame, selects or accepts the basis reflections in the diffractogram, inspects the amplitude previews and starts an extraction. The GPTA reference type and map resolution are available in the application. Users can substitute a pinned basis with the topological reference for the default domain-aware analysis. The Compare view runs the four methods on the same frame and basis and displays the selected strain component on one color scale. The release includes the benchmark image generator, the scoring code and the $256^2$ image set. The $1024^2$ set is available from the corresponding author.

## Data availability

The $256^2$ px simulated benchmark set with its ground-truth fields, the per-scene results of all methods on both simulated sets underlying Table 1 and **Figs. 4–6**, and the experimental HAADF-STEM image stacks are deposited at Science Data Bank (https://doi.org/10.57760/sciencedb.010xy). The $1024^2$ px simulated set (about 100 GB) is available from the corresponding author on request. Source data for the figures are provided with this paper.

## Code availability

Lattice Studio (the open-source application implementing GPTA together with the GPA, S-GPA and PPA comparison methods) and the benchmark generator, scene library and scoring code are available at https://github.com/YTianUCI/LatticeStudio under an open-source license.

## Acknowledgements

Y.H. acknowledges support from the National Natural Science Foundation of China (Grant Nos. 52501055 and 130-0401260014).

## Author contributions

Y.T. conceived the tracked-reference approach and supervised the project. Y.H. and Y.T. developed the method and the software, designed and generated the simulated benchmark, acquired and analyzed the experimental data, and wrote the manuscript.

## Competing interests

The authors declare no competing interests.

# Table

**Table 1 | Strain-mapping benchmark over 13,795 simulated images.** Image families are grouped by physical content from the $256^2$ and $1024^2$ px image sets. Columns are GPA, windowed GPA (S-GPA), peak-pair analysis (PPA) and GPTA. Each cell gives the failure rate in % / median $k_{50}$ of the passing images in $10^{-3}|\boldsymbol{g}|$ (criteria in Methods). GPTA is the domain-aware default, and the lowest failure rate is in bold. Full family-level results are in Supplementary Table 2, and image sets and GPTA variants are presented separately in Supplementary Table 3.

| **Image family** | **n** | **GPA** | **S-GPA** | **PPA** | **GPTA** |
|---|---|---|---|---|---|
| Pristine and smooth elastic fields | 6,950 | 14.2 / 2.53 | 15.0 / 2.51 | 16.6 / 2.50 | **5.6 / 1.12** |
| Thickness, tilt and foil geometry | 1,266 | 14.9 / 2.16 | 15.3 / 2.22 | 16.0 / 2.35 | **4.7 / 1.00** |
| Dislocations and dipoles | 1,535 | 19.1 / 6.15 | 17.9 / 5.35 | 19.5 / 4.89 | **7.9 / 4.70** |
| Planar defects (antiphase, twin) | 521 | 38.2 / 3.52 | 39.3 / 3.46 | 24.0 / 3.04 | **7.3 / 1.29** |
| Grain boundaries and mosaics | 1,239 | 67.2 / 5.25 | 64.1 / 5.72 | 37.2 / 3.89 | **8.2 / 3.00** |
| Heterointerfaces and crack tips | 310 | 58.1 / 4.12 | 57.4 / 4.08 | 36.8 / 3.15 | **14.2 / 1.93** |
| Continuous rotation and kink bands | 342 | 81.9 / 9.46 | 81.6 / 6.16 | 56.4 / 7.74 | **56.1 / 7.39** |
| Lattice modulations | 756 | 64.9 / 10.76 | **52.0 / 9.49** | 53.6 / 8.92 | 58.5 / 9.03 |
| Real crystal structures (Au, Si, $SrTiO_3$, graphene, $MoS_2$, MOF) | 314 | 43.6 / 3.68 | 47.8 / 3.84 | 54.1 / 4.11 | **31.5 / 2.49** |
| Mixed and masked images | 562 | 31.1 / 4.20 | 31.1 / 4.38 | 33.1 / 4.19 | **17.4 / 2.97** |
| All images | 13,795 | 27.3 / 3.24 | 26.7 / 3.23 | 24.0 / 3.19 | **11.5 / 1.68** |

Failure = bad area > 0.20. The $k_{50}$ of each method is taken over its own passing images. GPA and S-GPA use a Gaussian aperture of $0.10|\boldsymbol{g}|$, and GPTA is the domain-aware default.

## Figure legends

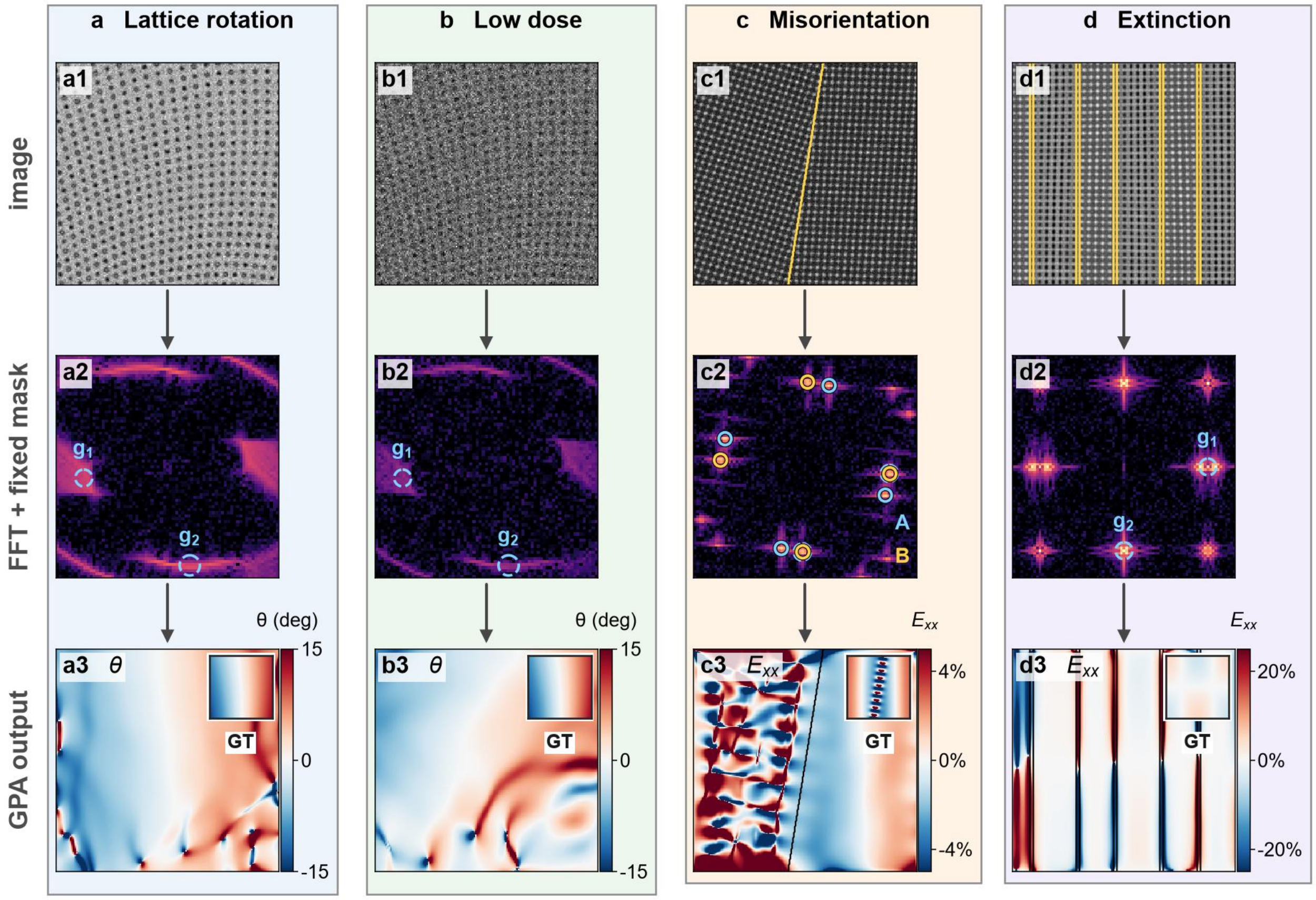

**Fig. 1 | Four experimental conditions at the limits of GPA.** Simulated HRTEM images (top), their diffractograms with the basis reflections $\boldsymbol{g}_1$ and $\boldsymbol{g}_2$ and the fixed GPA aperture (dashed, $\sigma = 0.10|\boldsymbol{g}|$) (middle), and the GPA maps with the ground truth (GT) inset (bottom). **a** Continuous lattice rotation. The lattice turns by 28° across the field, and the reflections are smeared into arcs. The aperture passes 30% of their power, and the rotation map $\theta$ has a RMS error of 4.4°. **b** Low dose. The same field at one tenth of the dose, where the error rises to 6.3°. **c** Misorientation. A 15° tilt boundary (yellow line) separates grains A (cyan) and B (yellow), whose reflections lie 2.6 aperture widths apart. The map of the normal strain $E_{xx}$ has an error of 12% RMS against a 3% signal. **d** Reflection extinction. One fringe family is suppressed inside the reversal bands (yellow lines), where the map shows 16% RMS strain although the true strain is zero. Related cases are in **Supplementary Figs. S3–S8**, and the GPA workflow is in **Supplementary Fig. S1**.

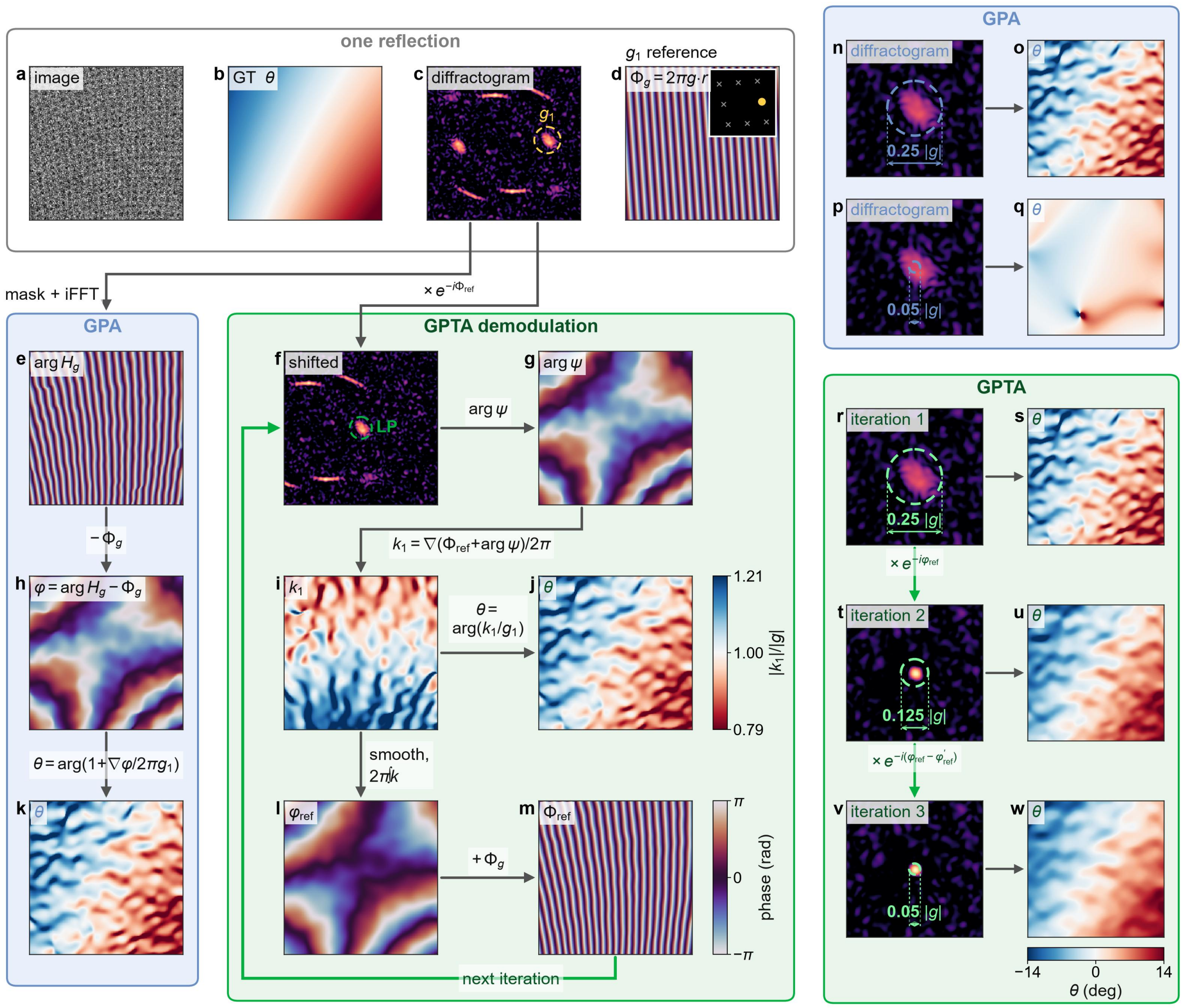

**Fig. 2 | From a fixed to a tracked reference.** One reflection of the rotating lattice in **Fig. 1a** at 8 e/px. **a–d** Image, ground-truth rotation $\theta$, diffractogram with $\boldsymbol{g}_1$ ringed, and the reference phase $\Phi_g = 2\pi\boldsymbol{g}\cdot\boldsymbol{r}$ of the mean lattice, which is one point in reciprocal space (inset). **e, h, k** GPA. Masking $\boldsymbol{g}_1$ and transforming back gives the Bragg-filtered image $H_g$, shown by its phase. Subtracting $\Phi_g$ gives the geometric phase $\varphi_g$, whose gradient gives $\theta$. **f, g, i, j** The same operations expressed as demodulation. Multiplying the image by $\exp(-i\Phi_{\text{ref}})$ shifts the reflection to the origin, and the aperture LP gives the residual phase $\varphi_{\text{res}} = \arg\,\psi_g$, the local wave vector $\boldsymbol{k}_1 = \nabla(\Phi_{\text{ref}} + \varphi_{\text{res}})/2\pi$ and $\theta$. With $\Phi_{\text{ref}} = \Phi_g$, the result is GPA. **l, m** GPTA integrates the smoothed $\boldsymbol{k}_1$ into the correction $\varphi_{\text{ref}}$, and $\Phi_{\text{ref}} = \Phi_g + \varphi_{\text{ref}}$ is the reference for the next iteration. **n–q** GPA with apertures of $0.25|\boldsymbol{g}|$ and $0.05|\boldsymbol{g}|$, with errors of 4.2° and 5.1°. The narrow aperture removes the arc together with the noise. **r–w** GPTA iterations 1–3. The demodulated diffractogram contracts around the origin as the reference tracks the lattice, the aperture narrows from $0.25|\boldsymbol{g}|$ to $0.05|\boldsymbol{g}|$, and the error falls from 4.2° to 1.6°. The demodulation and the tracked reference are developed step by step in **Supplementary Figs. S11–S16**.

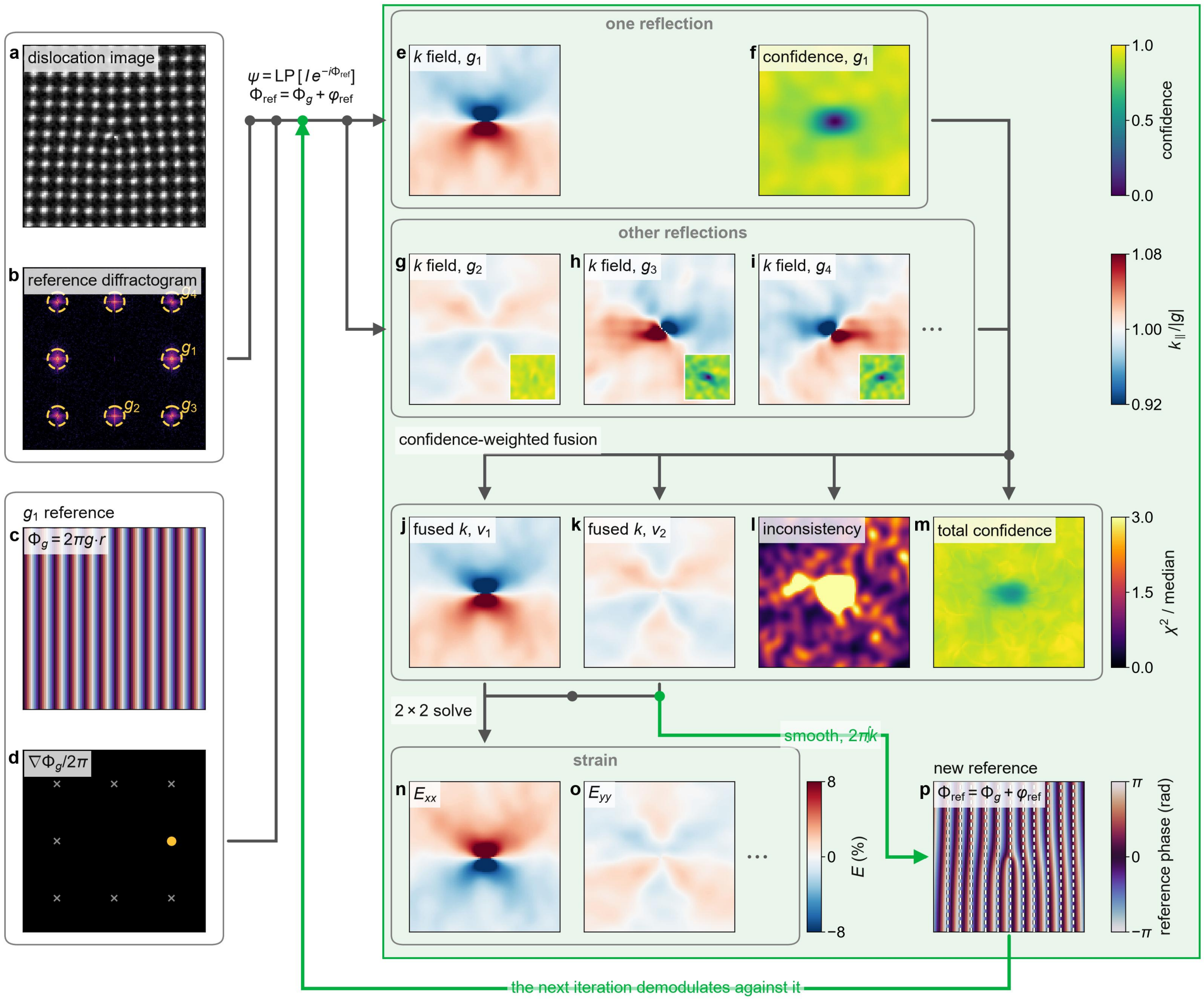

**Fig. 3 | One iteration of GPTA step by step.** A simulated edge dislocation. **a, b** Image and its diffractogram with the four reflections $\boldsymbol{g}_1$–$\boldsymbol{g}_4$ used in the analysis. **c, d** The $\boldsymbol{g}_1$ reference, shown as the phase $\Phi_{\boldsymbol{g}} = 2\pi\boldsymbol{g}\cdot\boldsymbol{r}$ and its wave vector, one point per reflection. **e, f** Demodulation of $\boldsymbol{g}_1$ against the reference gives the local wave-vector field $\boldsymbol{k}_1$ and the confidence $|\psi_1|^2/N_0$, where $N_0$ is the noise level of the reflection. The confidence falls at the core. **g–i** The $\boldsymbol{k}$ fields of the other reflections with their confidences (insets). **j–m** Confidence-weighted fusion of all reflections into one lattice basis, giving the fused $\boldsymbol{k}$ of $\boldsymbol{v}_1$ and $\boldsymbol{v}_2$, the inconsistency $\chi^2$ between the reflections and the total confidence. **n, o** The strain components $E_{xx}$ and $E_{yy}$. **p** The new reference $\Phi_{\rm ref} = \Phi_{\boldsymbol{g}} + \varphi_{\rm ref}$ from the smoothed and integrated fused $\boldsymbol{k}$, which contains the dislocation as a phase vortex. The next iteration demodulates against it. Over three iterations, the $E_{xx}$ error falls from 0.23% to 0.13%. The complete workflow is in **Supplementary Fig. S23**.

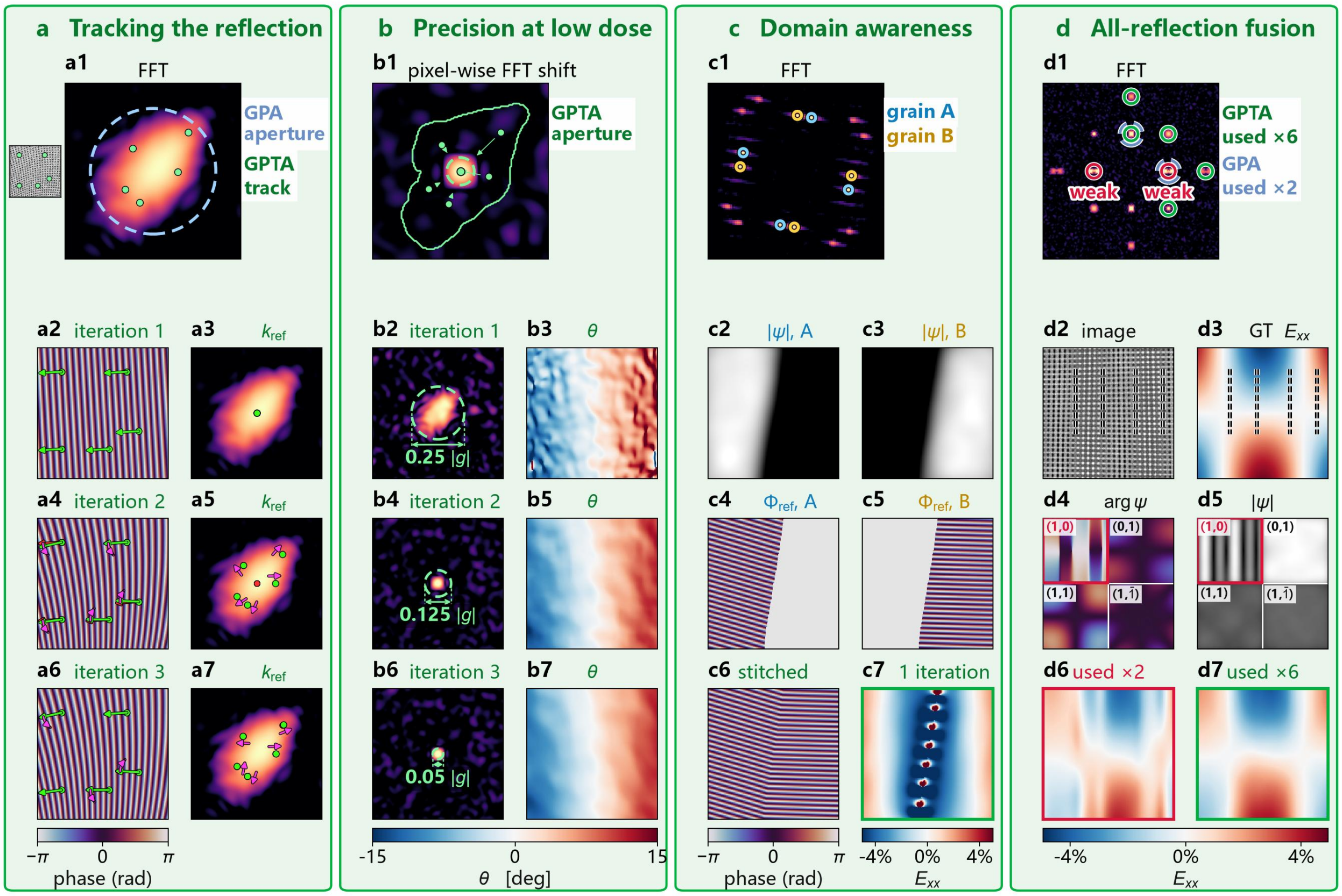

**Fig. 4 | Four components of the iteration at work.** Each column uses the image from the corresponding column of **Fig. 1**. **a Tracking the reflection.** a1, diffractogram of **Fig. 1a** about $\boldsymbol{g}_1$ with the iteration-1 aperture (blue, $0.25|\boldsymbol{g}|$) and the final wave vectors of five regions of the field (green), which GPTA tracks to five points on the arc. a2–a7, the reference phase of each iteration with the local reference vectors of the five regions, and their reference wave vectors $\boldsymbol{k}_{ref}$ on the spectrum (red the preceding iteration, magenta the move). The rotation error falls to 0.26° after three iterations, compared with 4.4° in **Fig. 1a**. **b Precision at low dose.** b1, the demodulated diffractogram of the final iteration for **Fig. 1b**. The reflections of the five regions are brought to the origin by their own references (arrows), and the last aperture ($0.05|\boldsymbol{g}|$, dashed) closes around them. b2–b7, the demodulated diffractogram with its aperture and the rotation map for iterations 1–3, with a final error of 0.67° compared with 6.3° in **Fig. 1b**. **c Domain awareness.** c1, diffractogram of **Fig. 1c** with the reflections of grain A (cyan) and grain B (yellow). c2, c3, the confidence of each grain's reflections, which marks its domain. c4–c6, the reference of each domain after its own iterations and the stitched reference. c7, $E_{xx}$ after the boundary dislocations have been injected into the reference, with a measured misorientation of 14.7° for a true 15° and a strain error of 1.5% in the grains compared with 12% in **Fig. 1c** (**Supplementary Fig. S24**). **d All-reflection fusion.** d1, diffractogram of **Fig. 1d** with the six reflections used by GPTA (green) and the two used in the two-reflection solve (blue). d2, d3, image with the reversal bands (yellow lines) and ground-truth $E_{xx}$. d4, d5, the phase arg $\psi_g$ and the amplitude $|\psi_g|$ of four reflections. The suppressed (1,0) family loses both inside the bands. d6, d7, $E_{xx}$ from the two basis reflections (error inside the bands 1.9%) and from all six (0.44%, **Supplementary Fig. S19**).

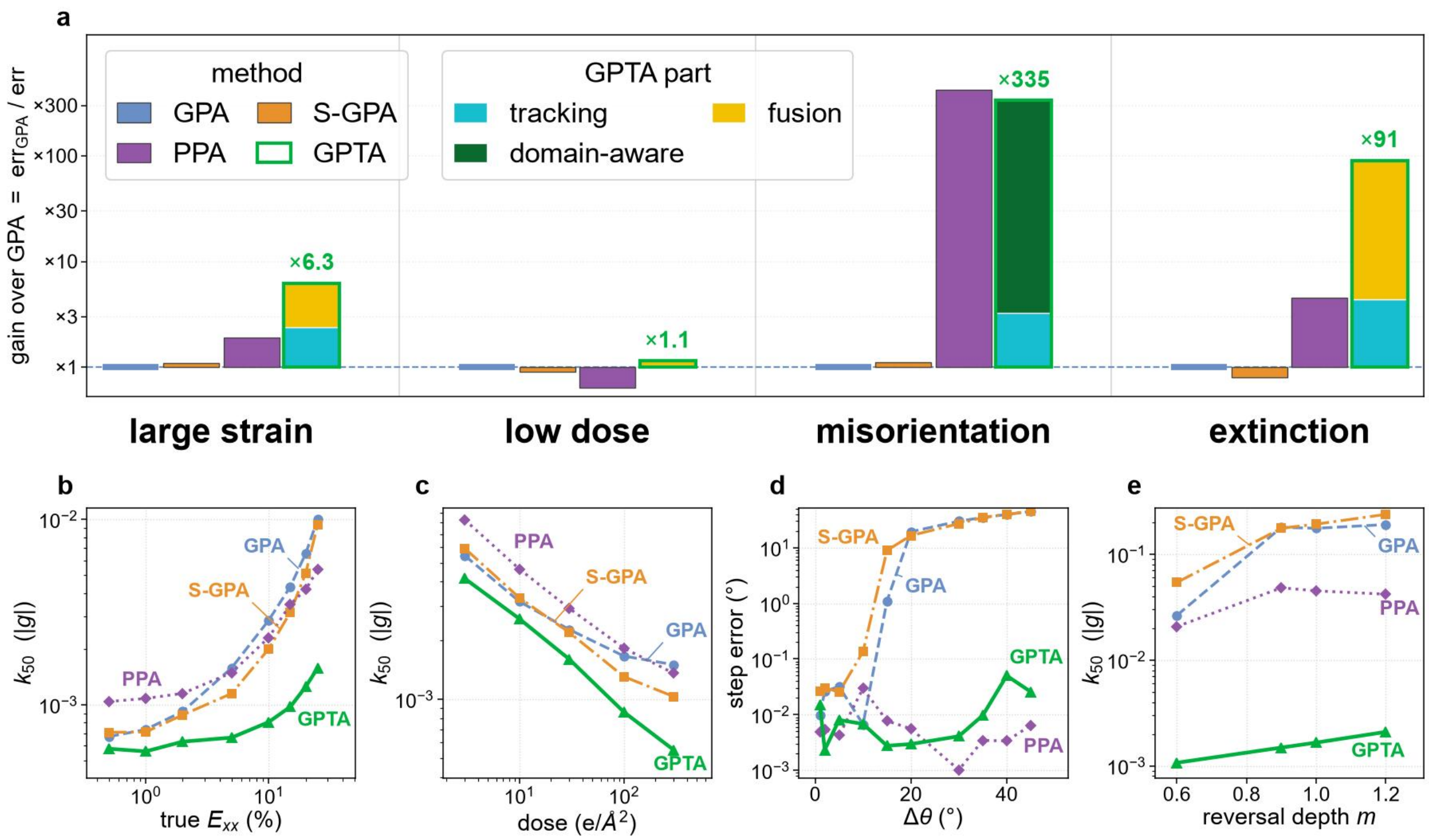

**Fig. 5 | Comparison with existing methods.** **a** Accuracy gain over GPA, defined as the GPA error divided by the error of each method, for GPA, S-GPA, PPA and GPTA on four images representing the conditions of **Fig. 1**, namely large smooth strain, low dose, a 30° bicrystal and reflection extinction. GPA at $\sigma = 0.10|\boldsymbol{g}|$ is ×1, and bars below the line are worse than GPA. The GPTA bar is built by switching on its components in sequence (tracking, confidence-weighted fusion and the domain-aware algorithm), and each colored segment is the factor of one component (Supplementary Table 1). The GPTA totals are ×6.3, ×1.1, ×335 and ×91, and PPA is best on the bicrystal. **b–e** Error against the driving parameter for the four methods, as the median of three noise realizations. The error is $k_{50}$, the median wave-vector error $|\boldsymbol{k} - \boldsymbol{k}_{\mathrm{GT}}|/|\boldsymbol{g}|$ over the scored area (Methods). **b** Peak strain of 0.5–25%. **c** Dose of 3–300 e/Å$^2$, where the error of GPTA continues to fall while GPA and S-GPA settle at the plateau set by their apertures. **d** Misorientation of 1–45°, where GPA loses the boundary step above 15° and S-GPA above 10°. **e** Contrast-reversal depth of 0.6–1.2. The principles of S-GPA and PPA are in **Supplementary Figs. S25 and S26**, and maps of all four methods on one image of each regime are in **Supplementary Fig. S27**.

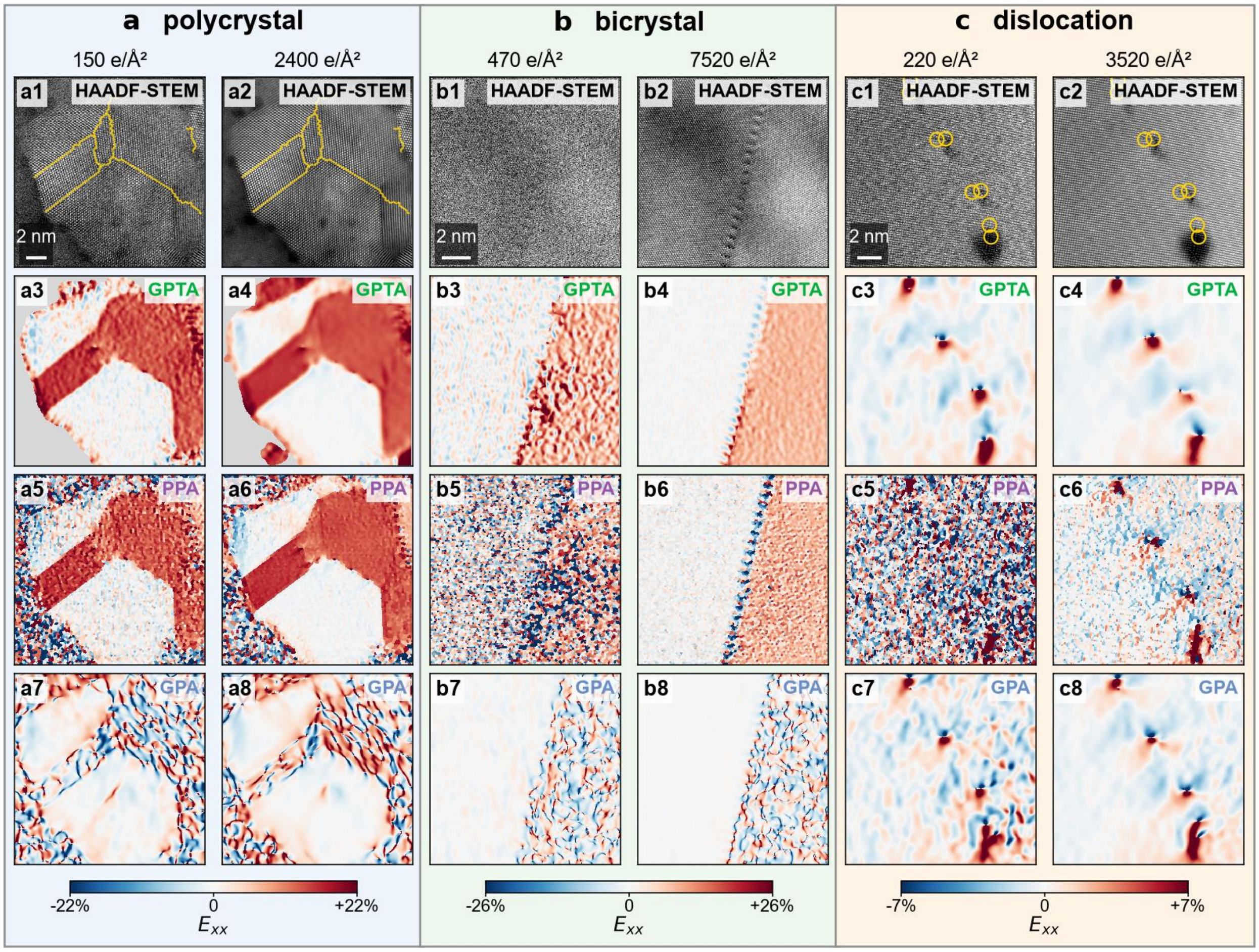

**Fig. 6 | Strain maps of three Pt specimens at two doses.** HAADF-STEM images of each specimen from a single frame (left) and from an average of sixteen frames of the same field (right). Rows show the image and the $E_{xx}$ maps from GPTA, PPA and GPA, on one color scale per specimen. **a** [110] polycrystal with six grains at 150 e/Å$^2$ per frame, with the grain boundaries from the 40-frame average overlaid. **b** [111] bicrystal with a 23.5° tilt boundary at 470 e/Å$^2$ per frame. **c** [001] region with eight dislocation cores at 220 e/Å$^2$ per frame. The marked core positions are the phase singularities of the Bragg-filtered 24-frame average and are independent of the methods compared. GPTA places the cores within one lattice period of these positions at both doses, whereas the other methods place them more than one period away. Gray in the GPTA maps indicates the area outside the image mask. A rotated grain appears as a uniform offset under the single global reference used for display, so the comparison concerns noise and boundary sharpness. Scale bars, 2 nm. The full dose series with S-GPA are in **Supplementary Figs. S31–S33**.

# Supplementary Information for

## From phase analysis to phase tracking: a robust approach to atomic-scale strain mapping

Ying Han[1] and Yuan Tian[2]*

[1] School of Materials Science and Engineering, Tianjin University, Tianjin, China.

[2] Department of Materials Science and Engineering, Southern University of Science and Technology, Shenzhen, China.

*Corresponding author: Yuan Tian (tiany3@sustech.edu.cn).

ORCID: Y.H. 0000-0002-6621-2351; Y.T. 0000-0002-7074-5468.

## Supplementary Note 1 | Mathematical formulation of GPTA

This note presents the equations behind geometric phase tracking analysis (GPTA) in the order used in one iteration. The sequence runs from the image model through the demodulated field, the local wave vector and its confidence, the tracked reference, the joint inversion over reflections, and the topological reference to the resulting displacement and strain. The final section states the conditions under which one iteration reduces exactly to geometric phase analysis (GPA). Boldface $\boldsymbol{r}$ denotes position, $\boldsymbol{g}$ a reciprocal-lattice vector, and $\boldsymbol{u}(\boldsymbol{r})$ the displacement of the lattice from its reference.

**1.1 Image model.** An atomic-resolution image of a distorted lattice is modeled as

$$I(\boldsymbol{r}) \ = \ \sum_{\boldsymbol{g}} A_{\boldsymbol{g}}(\boldsymbol{r}) \exp\left[2\pi i \boldsymbol{g}\cdot\left(\boldsymbol{r} \ - \ \boldsymbol{u}(\boldsymbol{r})\right)\right] \ + \ n(\boldsymbol{r}) \tag{S1}$$

a sum over reflections with slowly varying amplitudes $A_{\boldsymbol{g}}$ and a noise term $n$. The same displacement field is represented in the phase of every reflection, $\varphi_g(\boldsymbol{r}) = -2\pi\, \boldsymbol{g}\cdot\boldsymbol{u}(\boldsymbol{r})$, which is the quantity estimated by all phase methods. The amplitude $A_{\boldsymbol{g}}$ corresponds to the envelope (thickness, bending, defocus) and is discarded by GPA.

**1.2 Demodulation.** Before Fourier transformation, the image is detrended and apodized with an 8% Hann taper (**Supplementary Fig. S17**). For a reflection $\boldsymbol{g}$ and a reference phase $\varphi_{\mathrm{ref}}$, the demodulated field is

$$\psi_{\boldsymbol{g}}(\boldsymbol{r}) \ = \ \mathrm{LP}_b \left[I(\boldsymbol{r}) \exp\left[-i\left(2\pi\boldsymbol{g}\cdot\boldsymbol{r} \ + \ \varphi_{\mathrm{ref}}(\boldsymbol{r})\right)\right]\right] \tag{S2}$$

Here, $\mathrm{LP}_b$ is a Gaussian aperture of width $b|\boldsymbol{g}|$ applied around the origin. The first iteration uses the mean-lattice plane wave, for which $\varphi_{\mathrm{ref}} = 0$ in Eq. (S2), the total reference is $\Phi_{\boldsymbol{g}} = 2\pi\boldsymbol{g}\cdot\boldsymbol{r}$, and $b = 0.25$. By the shift theorem, demodulating with this plane reference and applying the aperture around the origin is identical to masking the Fourier transform with a Gaussian aperture centered on $\boldsymbol{g}$, which is the GPA construction (**Supplementary Figs. S11 and S12**). Thus, $\psi_g = H_{\boldsymbol{g}} \exp(-2\pi i \mathrm{g}\cdot\boldsymbol{r})$ and arg $\psi_g$ is the geometric phase $\varphi_g$ of Eq. (1). The parameter $b$ and the GPA aperture width σ are therefore the same. The argument of $\psi_g$ is the residual phase $\varphi_{\mathrm{res}} = \arg\psi_g$, which is the phase of the lattice relative to the total reference $\Phi_{\mathrm{ref}} = \Phi_{\boldsymbol{g}} + \varphi_{\mathrm{ref}}$. Its modulus $|\psi_g|$ is the local amplitude of the reflection after filtering.

**1.3 The local wave vector and its confidence-weighted estimate.** The local wave vector of the reflection is obtained from the phase gradient in the complex domain,

$$\boldsymbol{k}_{\boldsymbol{g}}(\boldsymbol{r}) \ = \ \boldsymbol{k}_{\mathrm{ref}}(\boldsymbol{r}) \ + \ 1/2\pi\, \mathrm{Im}\left\langle\psi_{\boldsymbol{g}}{}^{*} \ \nabla\psi_{\boldsymbol{g}}\right\rangle_{\sigma\mathrm{res}} \Big/ \left\langle\left|\psi_{\boldsymbol{g}}\right|^2\right\rangle_{\sigma\mathrm{res}} \tag{S3}$$

Here, $\boldsymbol{k}_{\text{ref}} = \nabla \Phi_{\text{ref}}/2\pi$ is the wave vector of the reference and $\langle \cdot \rangle_{\sigma_{\text{res}}}$ is a Gaussian average of width $\sigma_{\text{res}}$. For $\sigma_{\text{res}} \to 0$, the quotient is identically $\nabla \arg \psi_g/2\pi$ (the phase gradient of GPA). No unwrapping is required because the gradient is computed before any phase is taken. The main text writes this case as $\boldsymbol{k}_1 = \nabla(\Phi_{\text{ref}} + \varphi_{\text{res}})/2\pi$. For $\sigma_{\text{res}} > 0$, Eq. (S3) is the $|\psi|^2$-weighted local mean of the phase gradient. This weighting has a statistical basis, since for shot noise the variance of the phase of a demodulated pixel is $N_0/2|\psi_g|^2$, where $N_0$ is the noise power passed by the aperture. Thus, $|\psi_g|^2/N_0$ is proportional to the inverse variance of the observation, and Eq. (S3) is the minimum-variance local average. The proportionality between confidence and the inverse variance of the phase measurement in the main text is defined in this sense. A pixel in which the reflection has vanished makes no contribution to its neighbors, including no noise contribution. This distinguishes the $\sigma_{\text{res}}$ average from the GPA aperture (which averages signal and noise alike, before the nonlinearity) and from any smoothing applied to a finished map (which is unweighted).

**1.4 The tracked reference.** The reference for the next iteration is the integral of the current estimate,

$$\varphi_{\text{ref},n+1}(\boldsymbol{r}) = \operatorname{argmin}_{\varphi} \textstyle\sum_{\boldsymbol{r}} w(\boldsymbol{r}) \left| \nabla \varphi(\boldsymbol{r})/2\pi - \left(\boldsymbol{k_g}(\boldsymbol{r}) - \boldsymbol{g}\right) \right|^2 \tag{S4}$$

a weighted least-squares integration (a Poisson equation in $\nabla\cdot(w\nabla\varphi)$), preceded by edge-preserving median smoothing of $\boldsymbol{k}_g$. As a result, a discontinuity in the lattice, such as a boundary, is represented in the reference as a step rather than a ramp. The iteration is completed by substituting $\varphi_{\text{ref},n+1}$ into Eq. (S2). Convergence is governed by the spread of the residual, since after demodulation with $\varphi_{\text{ref},n}$, the spectrum of $\psi_g$ is centered on the part of the lattice distortion not yet included in the reference. The aperture for iteration $n$+1 therefore need only cover that residual. Narrowing the aperture is meaningful provided that $b_{n+1}$ exceeds the root-mean-square (RMS) width of the distribution of $\boldsymbol{k}_g - \boldsymbol{g} - \nabla\varphi_{\text{ref},n}/2\pi$ over the field. The aperture can close only as fast as lattice variation is incorporated into the reference, and without tracking ($\varphi_{\text{ref}}$ fixed) it cannot close below the spread of the signal itself (**Supplementary Figs. S13–S16**).

**1.5 Joint inversion over reflections.** The wave vectors of all demodulated reflections $\boldsymbol{g}_{mn} = m\boldsymbol{g}_1 + n\boldsymbol{g}_2$ are combined per pixel into one lattice basis ($\boldsymbol{v}_1$, $\boldsymbol{v}_2$),

$$(\boldsymbol{v}_1, \boldsymbol{v}_2)(\boldsymbol{r}) = \operatorname{argmin} \textstyle\sum_{mn} w_{mn}(\boldsymbol{r}) \| \boldsymbol{k}_{mn}(\boldsymbol{r}) - (m\boldsymbol{v}_1 + n\boldsymbol{v}_2) \|^2, \quad w_{mn} = |\psi_{mn}|^2 / N_{0,mn} \tag{S5}$$

a $2N \times 4$ weighted linear system. Every reflection whose median signal-to-noise ratio (SNR) exceeds 3 is included, with as many as twelve reflections for a square lattice. For each reflection, $N_0$ is the median spectral power in an annulus at $|\boldsymbol{g}_{mn}|$ after excluding the reflections (**Supplementary Fig. S18**). The basis is obtained from the normal equations, and its $4 \times 4$ covariance gives the per-pixel uncertainty. The weighted sum of squared residuals gives the per-pixel $\chi^2$ consistency statistic with $2N - 4$ degrees of freedom, which is plotted in the inconsistency map of **Fig. 3l**. Two passes of robust reweighting reduce the weight of a reflection whose residual is inconsistent with the others at that pixel. For $N = 2$, the system is exactly determined, the solution is independent of the weights, and the residual is zero. Eq. (S5) is then the $2 \times 2$ inversion of GPA. Redundancy begins at $N = 3$, allowing tolerance of a lost reflection and detection of an inconsistency.

**1.6 The topological reference.** Around a dislocation core, the phase $\varphi_g$ is multi-valued. Its circulation along a closed path enclosing the core is $2\pi\, \boldsymbol{g}\cdot\boldsymbol{b}$, with $\boldsymbol{b}$ denoting the Burgers vector. Thus, $\boldsymbol{k}_g - \boldsymbol{g}$ contains a delta function of curl at the core and cannot be written as a gradient. The least-squares integration of Eq. (S4) projects this part out, which is why the entire singular phase remains in the residual when the reference is curl-free (i.e., without phase vortices). To separate the two parts explicitly in the topological reference, integer circulations are detected on the fused field before integration. Each core contributes an analytic vortex term $\varphi_{\text{vort}}(\boldsymbol{r}) = q\cdot\operatorname{atan2}(y - y_0, x - x_0)$ with integer $q = \boldsymbol{g}\cdot\boldsymbol{b}$, and the reference is the sum of the integrated smooth part and the vortex terms. The residual around the core is then single-valued and small, and the zero of $\psi_g$ is pinned to the core. The winding numbers $q$ are reported as the dislocation content of the field (**Supplementary Fig. S21** compares the available integrals). The same vortex term represents the dislocations of a boundary once they have been located (Note 6.3).

**1.7 Displacement and strain.** Because the bulk of the phase is included in the tracked reference, the residual phase $\varphi_{res} = \arg \psi_g$ does not wrap and the total phase $\varphi_{ref} + \varphi_{res}$ is available without unwrapping. The displacement field is obtained from Eq. (S1) as the weighted least-squares solution of $\varphi_g = -2\pi\, \boldsymbol{g}\cdot\boldsymbol{u}$ over all reflections. The strain tensor is the symmetric part of its gradient $\nabla\boldsymbol{u}$. Equivalently, the distortion is obtained from the reference and measured reciprocal bases. The main text maps the linearized normal strain $E_{xx}$ and the in-plane rotation $\theta = \arg(\boldsymbol{k}_1/\boldsymbol{g}_1)$.

**1.8 Equivalence to GPA.** Combining 1.2 (a fixed plane reference corresponds to a Gaussian aperture), 1.3 ($\sigma_{res} = 0$ corresponds to the pointwise phase gradient), and 1.5 ($N = 2$ gives a weight-independent exact solve) shows that one iteration of GPTA with $b = 0.25$, a fixed plane reference, $\sigma_{res} = 0$, two reflections, and no reweighting is equivalent to the GPA estimator in every operation (**Fig. 2e, h, k** compared with **f, g, i, j**). Numerically, the two agree to a median $|\Delta\boldsymbol{k}|$ of $2.5 \times 10^{-5}|\boldsymbol{g}|$, or $4.3 \times 10^{-7}|\boldsymbol{g}|$ at $b = 0.09$. The residual comes from the finite-difference stencil of the gradient. GPA is therefore the anchor point of the family, with each component of GPTA relaxing one of these fixed choices. Switching all of them off returns the original method exactly (Note 10.1).

**1.9 Domain-aware analysis, resolution and averaging.** When several distinct reciprocal lattices occur in the diffractogram, the domain-aware algorithm identifies candidate lattices and assigns each pixel to the candidate with the highest demodulated confidence. GPTA is applied to each domain with that domain's reference, and the results are joined at the maximum-confidence boundary. If the boundary is a wall of dislocations, the cores are located from the curl of the stitched wave-vector field and injected into the reference as phase vortices before a final iteration (Note 6.3 and **Supplementary Fig. S24**). A candidate is accepted only when its reciprocal cell area and period match an observed collection of first-order reflections and its orientation differs by more than 2° from those of all accepted lattices. For a single lattice, the domain-aware algorithm remains inactive and the output is identical to a single-reference run. The final aperture b and the averaging window $\sigma_{res}$ are separate settings. $\sigma_{res}$ is reported in lattice periods (2.5 by default), and the normalized map-resolution setting of the software is $\sigma_{res}$ in units of that default. For experimental data, the map resolution is the full width at half maximum of the kernel that combines the final aperture and $\sigma_{res}$. The first iteration of the formulation with fixed settings is therefore GPA, and those settings are revised from the image itself in every later iteration.

## Supplementary Note 2 | GPA as an estimator

This note follows GPA through its intermediate quantities on a simulated image (**Supplementary Fig. S1**) and shows where the strain signal lies in reciprocal space (**Supplementary Fig. S2**). Both points are used in the main text to explain why a fixed aperture and a single reference limit the method.

**2.1 Workflow (Supplementary Fig. S1).** GPA is examined at each step on a simulated $512^2$ lattice (period 8 px, 500 e/px) containing two edge dislocations with Burgers vectors along $x$ (core A) and along $y$ (core B). The two reflections $\boldsymbol{g}_1$ and $\boldsymbol{g}_2$ are chosen such that $\boldsymbol{g}_1\cdot\boldsymbol{b} = 1$ at A and 0 at B, and $\boldsymbol{g}_2\cdot\boldsymbol{b}$ has the reverse values. Each chain of the analysis therefore contains one core, and the two chains are not redundant. For each reflection, the analysis consists of four quantities. The first is one patch of the transform, isolated by a Gaussian aperture of width $\sigma = 0.25|\boldsymbol{g}|$, whose inverse transform is the Bragg-filtered image $H_g$ of Eq. (1). The second is one point of the transform (the reflection $\boldsymbol{g}$ itself), which is the global reference. The difference between $\arg H_g$ and the reference phase $2\pi\boldsymbol{g}\cdot\boldsymbol{r}$ is the third quantity, the geometric phase $\varphi_g$, and its gradient gives the fourth, the local wave vector $\boldsymbol{k}$. The results from the two chains are combined in a 2 × 2 solve that yields the four components of the distortion tensor. The main text uses two properties of this workflow. The first is the computation and subsequent discarding of the amplitude $|H_g|$, which equals $|\psi_g|$ of Eq. (3). This amplitude is the local strength of the reflection and would be the natural measure of confidence in the phase, but GPA does not use it. The other property is the use of one point as the reference, with the entire specimen measured relative to a single wave vector. Every departure of the lattice from that wave vector, whether strain or a second grain, is expected to lie inside the aperture.

**2.2 Where the strain signal lives (Supplementary Fig. S2).** The aperture determines what GPA extracts and is therefore not an implementation detail. A perfect lattice transforms to a point at $\boldsymbol{g}$, and a single edge dislocation adds only a diffuse skirt around that point, which is the strain signal. Its second moment (the excess spread of the reflection) has a one-to-one correspondence with the RMS strain of the field (**Supplementary Fig. S2i**, window floor $0.0054|\boldsymbol{g}_1|$ compared with $0.0473|\boldsymbol{g}_1|$ for the dislocation). All displayed transforms and spread values use the periodic component of a periodic-plus-smooth decomposition of the image, which removes the cross-shaped streaks of the finite window. The second block of the figure varies the aperture alone ($\sigma = 0.25$, 0.10 and $0.04|\boldsymbol{g}_1|$) on the same noiseless image. The $E_{xx}$ error rises from 0.044% to 0.179% to 0.347% as the aperture cuts into the skirt. The trade-off between resolution and noise in GPA arises because a narrow aperture excludes both noise and the part of the strain signal that lies outside it. These cannot be separated because the aperture is applied before the phase is measured. In GPTA, the two settings corresponding to these properties (a discarded amplitude and a single global reference) are replaced.

## Supplementary Note 3 | Failure survey of the fixed reference

This note follows each of the four failures in **Fig. 1** through the intermediate quantities in which it develops, which are the per-reflection amplitude $|\psi|$, the phase, the aperture transmittance and the aperture-width series. These quantities establish that the map is wrong and identify the step at which it became wrong. The following analyses correspond to the columns of **Fig. 1**. The rotation column is expanded using a 37° rotation with three regions sampled along the arc (**Supplementary Fig. S3**), and the low-dose column is expanded using an inclusion over aperture widths and apertures (**Supplementary Fig. S4**). For the misorientation column, the analysis covers two severities, which are a small-angle boundary (**Supplementary Fig. S5**) and a mirror twin (**Supplementary Fig. S6**). The extinction column is expanded across physical mechanisms, which are an envelope null (**Supplementary Fig. S9**), a bent foil (**Supplementary Fig. S7**) and three multislice mechanisms (**Supplementary Fig. S8**), and includes the false cores produced by one of them (random thickness) (**Supplementary Fig. S10**). The simulated images are drawn from the benchmark image set.

**3.1 Low dose (Supplementary Fig. S4).** For an Eshelby inclusion at 10 e/Å$^2$, the SNR within the aperture of $\boldsymbol{g}_1$ is 10. Varying the GPA aperture gives a U-shaped curve of bad area as a function of width, with a minimum of 0.06 for $\sigma = 0.03|\boldsymbol{g}|$ and a value of 0.62 at the fixed aperture of $\sigma = 0.10|\boldsymbol{g}|$. Noise is excluded by a narrow aperture, and the signal is retained by a wide aperture. No single width achieves both, and GPTA narrows its aperture over three iterations on the tracked reference. The bad area is 0.27 from the base pair alone. With weighted inversion over (0,1), (1,0), (1,1) and (2,0), whose SNRs within the aperture are 5, 5, 18 and 15, the bad area is 0.00. At this dose, the higher-order reflections contain more signal than the base pair, although GPA does not use them.

**3.2 Small-angle boundary (Supplementary Fig. S5).** At 7.2°, the two grains' reflections are split by $0.126|\boldsymbol{g}|$, placing grain B on the flank of the fixed aperture with a transmittance of 0.46 and $|\psi_B|/|\psi_A| = 0.47$. B remains measurable at a degraded SNR, and no aperture width is suitable for both grains at once (bad area 0.27 at best). GPTA narrows its aperture from 0.25 to $0.09|\boldsymbol{g}|$ over three iterations on each grain's own reflection. The bad area is 0.07 with a single tracked reference and 0.05 with the domain-aware algorithm, consistent with the regime of **Fig. 4a** in which tracking alone suffices.

**3.3 Mirror twin (Supplementary Fig. S6).** The reflections of the two variants are split by $0.48|\boldsymbol{g}|$. The aperture transmittance at variant B is $1.4 \times 10^{-5}$, its SNR within the aperture is 0.00 and $|\psi_B|/|\psi_A| = 0.02$. GPA measures variant B as noise (bad area 0.51). The domain-aware algorithm finds the two domains, places the aperture on each variant's own reflections and stitches the result (bad area 0.00).

**3.4 Continuous rotation (Supplementary Fig. S3).** In a dislocation-free domain whose lattice rotates by 37° across the field, the reflection is spread into an arc. Along the arc, the aperture transmittance falls 1.00 → 0.17 → 0.00. GPA $E_{xx}$ has a bad area of 0.69, whereas GPTA gives a bad area of 0.13 as the tracked reference follows the arc with a total drift of 0.39$|\boldsymbol{g}|$ in the tail. The residual bad area is at the tail, where the arc of $\boldsymbol{g}_1$ approaches the neighboring reflection. The domain-aware algorithm does not apply because there is no boundary in this image. The 28° rotation of **Figs. 1a, 2 and 4a** is in the same regime at a smaller angle, for which tracking recovers the whole field.

**3.5 Vignette envelope (Supplementary Fig. S9).** A dislocation array under a vignette envelope looks intact to the eye, but the envelope reduces $|\psi(\boldsymbol{g}_2)|$ to 0.10 in its null, compared with a noise level within the aperture of 0.36. The phase there is noise, which nucleates 16 spurious phase vortices, and the GPA $E_{xx}$ has a bad area of 0.76. In GPTA, the weight falls to zero in the null, and the tracked reference is used to determine the phase across it. The bad area is 0.01 with no spurious vortices. With the tracked reference alone, without the confidence weight, the bad area is 0.40, indicating that both mechanisms are needed here.

**3.6 Physical mechanisms of extinction (Supplementary Figs. S7, S8 and S10).** In a bent foil (multislice, 50 e/Å$^2$), different reflections are extinguished in different parts of the field. The two hard-wired reflections of GPA collapse over different regions, and their phase is noise at positions where their amplitude vanishes. The fixed-pair $E_{xx}$ has a bad area of 0.90. The GPTA base pair alone has a bad area of 0.96, and the weighted inversion over the surviving reflections gives 0.00 (**Supplementary Fig. S7**, mechanism in Note 6.6). In three further multislice images, a reflection is suppressed by a thickness wedge, by random bending and by a random height field. The ground truth is strain-free, so every feature of a strain map comes from the envelope. GPA gives $E_{xx}$ RMS errors of 4.92, 5.54 and 5.27%, compared with 0.021, 0.44 and 0.43% for GPTA, reductions by factors of 229, 12.5 and 12.1 (**Supplementary Fig. S8**). The thickness wedge smoothly suppresses individual reflections in sequence, and its effect is removed almost completely. The two disordered mechanisms weaken several reflections at once and leave a residual of about 0.44%. **Supplementary Fig. S10** shows the consequence for dislocation detection by the transform, mask and inverse-transform construction. In a random-thickness image, the envelope alone causes the fringe amplitude to cross zero away from the core, and each zero is interpreted as a phase singularity. The fitted reflection is stable (0.0994–0.1004 cycles per pixel), and the true core is found in every image. Spurious cores arise from the envelope independently of the mask width, and **Supplementary Fig. S9** shows that the confidence weight removes them. In every case, the map fails at the step where a fixed setting no longer matches the image and is restored by the corresponding component of GPTA. This pattern is developed in **Fig. 4** and Note 6.

## Supplementary Note 4 | Demodulation and the tracked reference step by step

This note develops the demodulation and the tracked reference using the simplest signals that illustrate them (a one-dimensional fringe pattern and a single dislocation), before returning to the rotating lattice of **Fig. 2**. Each subsection corresponds to one supplementary figure.

**4.1 Demodulation in one dimension (Supplementary Fig. S11).** A fringe pattern with a period of 14.9 px is modulated in phase by a Gaussian bump of 3.4 rad on a ramp of 0.6 rad, giving a local-frequency swing of 0.18 $\boldsymbol{g}$, and buried in Gaussian noise at 0.7 of its amplitude. Its spectrum has the reflection at +$\boldsymbol{g}$ with a skirt that represents the modulation. Multiplying the signal by exp(−i2πgx) shifts the reflection to zero frequency. A mask of half-width 0.35 $\boldsymbol{g}$, wide enough to pass the whole swing, and an inverse transform give the complex field $\psi_g$, whose two quadratures Re $\psi_g$ and Im $\psi_g$ are the two components needed to determine the phase. The wrapped phase arg $\psi_g$ = atan2(Im $\psi_g$, Re $\psi_g$) is the modulation up to multiples of 2π, and unwrapping recovers it together with the noise passed by the 0.35 $\boldsymbol{g}$ mask. The mask cannot be narrower because the modulation itself spans 0.18 $\boldsymbol{g}$ of local frequency, and a narrower mask would cut it.

**4.2 Demodulation in two dimensions (Supplementary Fig. S12).** The same operations are applied to a plain edge dislocation (lattice period 9.85 px, 500 counts). Masking $\boldsymbol{g}_1$ in the diffractogram and transforming back is the Bragg filter of textbook GPA and gives the Bragg-filtered image $H_g$ of Eq. (1), in which the extra half-plane is visible. Multiplying the image by $\exp(-i2\pi \boldsymbol{g}_1 \cdot \boldsymbol{r})$ instead shifts $\boldsymbol{g}_1$ to zero frequency. The mask of half-width $0.30|\boldsymbol{g}_1|$ with the inverse transform gives Re $\psi_g$ and Im $\psi_g$, whose argument is the displacement as a phase and whose derivative is the strain. The gradient of the same phase gives the wave-vector field of $\boldsymbol{g}_1$ (i.e., $\boldsymbol{k}_1 = \boldsymbol{g}_1 + \nabla\varphi/2\pi$). The field of $\boldsymbol{g}_2$ is obtained by demodulating the other reflection, and the two fields are needed to solve the 2 × 2 system for the strain tensor. Of the two details that matter for a quantitative result, the first is removal of the image mean before the shift because the $\boldsymbol{k} = 0$ term of the image otherwise lies exactly $|\boldsymbol{g}_1|$ from the center of the aperture and leaks into $\psi_g$ as a ripple at the lattice frequency. The second is that the dislocation is synthesized on a larger grid than is displayed, because a single dislocation winds the phase by $2\pi$ and its image cannot be periodic.

**4.3 The tracked reference in one dimension (Supplementary Fig. S13).** For the signal of 4.1, the modulation recovered in iteration 1 is filtered in the complex field (a super-Gaussian with a cutoff at 0.24 $\boldsymbol{g}$, above the modulation's own bandwidth and below the 0.35 $\boldsymbol{g}$ mask that iteration 1 required) and used as the reference phase $\varphi_{\mathrm{ref}}$. Multiplying the signal by $\exp(-i2\pi gx)$ and by $\exp(-i\varphi_{\mathrm{ref}}(x))$ removes the modulation's frequency swing from the spectrum. The residual is a narrow peak at zero, allowing the mask to narrow to 0.12 $\boldsymbol{g}$ in iteration 2 and 0.09 $\boldsymbol{g}$ in iteration 3. The residual phase $\varphi_{\mathrm{res}} = \arg \psi_g$ is within ±0.2 rad. The recovered modulation is $\varphi_{\mathrm{ref}} + \arg \psi_g$, and noise in the deviation of the local wave vector from $\boldsymbol{g}$, $(1/2\pi)d\varphi/dx$, decreases with successive iterations while the modulation is preserved. Using the tracked reference rejects noise without increasing sharpness because the reference contains the signal's own bandwidth and the mask only has to pass the residual.

**4.4 The tracked reference in two dimensions (Supplementary Fig. S14).** For the dislocation of 4.2, the iteration-1 wave-vector field is averaged in the complex field over 12 px and integrated into the reference phase $\varphi_{\mathrm{ref}}$. In this averaging, every pixel is weighted by its own amplitude, preserving the winding of the core. After multiplication by $\exp(-i\varphi_{\mathrm{ref}})$, the baseband contracts by a factor of 2.5 and the mask narrows from 0.30 to $0.12|\boldsymbol{g}_1|$. The residual phase is flat. $|\psi_g|$ is unity except at the core, where no single-valued reference can follow the winding (Note 1.6).

**4.5 The reference as a function of position (Supplementary Fig. S15).** In the rotating lattice of **Fig. 1a**, three points 68 px apart have reflections spread over $0.28|\boldsymbol{g}|$ in the diffractogram. After two iterations of the default schedule (0.25, 0.125 and $0.09|\boldsymbol{g}|$), the reference phase $\Phi_{\mathrm{ref}}(x, y)$ has fringes that turn with the lattice, and its local wave vector $\boldsymbol{k}_{\mathrm{ref}}$ therefore varies with position. At each point, multiplying the image patch by the local reference wave removes the fringes in real space. Equivalently, convolving the patch's reflection with the point $\boldsymbol{k}_{\mathrm{ref}}$ shifts it to the origin in reciprocal space. The residual $|\boldsymbol{k} - \boldsymbol{k}_{\mathrm{ref}}|$ is 0.020, 0.008 and $0.004|\boldsymbol{g}|$, inside the $0.09|\boldsymbol{g}|$ aperture. This operation is performed simultaneously at every pixel in **Fig. 2f**, allowing the reference to track the reflection. A single reference phase evaluated at different positions of the same image corresponds to different positions of the same reflection.

**4.6 Iteration by iteration on the rotating lattice (Supplementary Fig. S16).** The iteration in **Fig. 2** is shown with every intermediate array at 200 e/px, including the demodulated diffractogram with its aperture, $|\psi_g|$ and $\arg \psi_g$, the wave-vector field $\boldsymbol{k}_1 = \boldsymbol{k}_{\mathrm{ref}} + \nabla \arg \psi_g/2\pi$, the reference $\varphi_{\mathrm{ref}}$ obtained in the iteration by smoothing and integrating $\boldsymbol{k}_1$, and the output of the 2 × 2 solve ($E_{xx}$, $E_{yy}$, $E_{xy}$ and $\theta$). The reflection contracts to a point as the reference tracks the lattice. The aperture narrows from 0.25 to 0.125 and $0.05|\boldsymbol{g}|$, and the $\theta$ error falls from 0.94° to 0.30° and 0.25°. Iteration 1 is GPA at $0.25|\boldsymbol{g}|$ with the plane-wave reference. The arrows between iterations indicate the increment of the correction ($\varphi_{\mathrm{ref}} - \varphi'_{\mathrm{ref}}$) because the diffractogram of an iteration already contains the correction from the preceding iteration. The sequence from **Supplementary Figs. S11–S16** follows the main text in condensed form. Demodulation is used in place of the Bragg filter, the lattice variation is included in the tracked reference, and the aperture narrows around the residual.

## Supplementary Note 5 | The GPTA iteration in full

This note documents the settings and intermediate quantities of the GPTA iteration summarized in **Fig. 3** of the main text. These are the noise level and the weights, the pre-treatment of the image, the signal-to-noise threshold, the confidence-weighted fusion, the integrals used to construct the reference, and the schedule of the three iterations.

**5.1 Workflow with noise level and weights (Supplementary Fig. S23).** We use a simulated edge dislocation with an 18% smooth strain at 20 e/px, an averaging window of 8 px and four reflections. The noise level $N_0$ of each reflection is the median power of the spectrum in an annulus at $|\boldsymbol{g}_{mn}|$ from which disks around all reflections are excluded (measured values of 4158, 4158, 4000 and 4000 for (1,0), (0,1), (1,1) and (1,−1)). The confidence is the demodulated power $|\psi_g|^2$ of each field divided by its $N_0$. Reflections whose median confidence falls below 3 are dropped from the inversion (here all four pass, with medians of 5416, 5543, 987 and 992). The weights $\mathrm{w} = |\psi_g|^2/N_0$ are normalized across reflections. Per-reflection inconsistencies are obtained from the robust inversion, and the fused basis $v_1$ is integrated into the next reference. The curl of the fused field is computed before integration because it is projected out during integration. One vortex of charge −1 is found on $\boldsymbol{g}_1$ at the dislocation and included in the rebuilt reference. The demodulated spectrum contracts from iteration 1 to iteration 3. The final outputs are the strain components, the rotation and the per-pixel uncertainty from the covariance of the weighted solve.

**5.2 Pre-treatment (Supplementary Fig. S17).** Before demodulation, the image is detrended by a least-squares plane and tapered by an 8% Hann window, which suppresses the edge power by a factor of 3.1 and removes the window cross from the diffractogram. The basis $\boldsymbol{g}_1, \boldsymbol{g}_2$ is found from the diffractogram (exactly, on the synthetic image). If the diffractogram contains more than one lattice, each lattice's reflections are collected separately for the domain-aware algorithm. The noise level $N_0$ of every reflection is the median of $|\mathrm{F}|^2$ in an annulus at $|\boldsymbol{g}|$ with a width of $0.15|\boldsymbol{g}|$, from which every reflection and its Friedel mate are excluded by $0.35|\boldsymbol{g}|$. It is 1.23 times the Poisson expectation on this image. An image mask separates the crystal from the vacuum, the amorphous rim and the apodized border. The evidence $|\psi_g|^2$ (the largest demodulated power over the four first-order families) is thresholded, closed, opened, hole-filled and eroded by 7 px. The mask is then apodized over half a lattice period (4.9 px on this 9.85 px lattice). On an image with a vacuum corner covering 16% of the field, the mask excludes 27% of the frame, 78% of the vacuum pixels and all of the vacuum more than 20 px from its edge. Noise added only where the mask is zero changes the measured $\boldsymbol{k}$ by zero with the mask and by $0.006|\boldsymbol{g}|$ without it. The reflection thresholds are evaluated on the mask.

**5.3 Noise level and SNR threshold (Supplementary Fig. S18).** The per-reflection SNR on the mask is $|\psi_i|^2/N_0$. The demodulated power $|\psi_i|^2$ is measured through the $0.25|\boldsymbol{g}|$ aperture, and $N_0$ is referred to the same aperture. On a dislocation at 10 counts with a contrast-reversal band on the (1,0) family, the median SNR inside the band falls to 9.2 for (1,0) and 2.7 for (2,0), below the threshold drawn at 12 in the figure. The (0,1), (0,2), (1,1) and (1,−1) reflections retain SNRs of 353, 70, 129 and 132. Outside the band, all six reflections pass. The threshold is drawn at 12 in **Supplementary Figs. S18 and S19**, which makes the decision inside the reversal band visible. The software default is 3. At either value, no reflection in this image is dropped over the whole mask because the band covers 31% of the mask, and the result of **Supplementary Fig. S19** therefore does not depend on the threshold. The suppressed family is removed inside the band by the per-pixel weight, which follows the confidence continuously. The threshold does not remove this family inside the band.

**5.4 Confidence-weighted fusion (Supplementary Fig. S19).** On the same image, the confidence and the phase of each of the six reflections are shown as tiles. The (1,0) and (2,0) phases are noise inside the band, where their confidence is near zero and their weights are correspondingly low. $E_{xx}$ from the two basis reflections alone has an RMS error of 2.31% inside the reversal band. The weighted inversion over all reflections gives 0.36%, a gain of 6.4 inside the band and 1.3 outside it. The residual $\chi^2$ of the inversion is 1.55 inside the band compared with 1 outside. The uncertainty of $E_{xx}$ from the covariance is 0.25% compared with 0.14%, indicating where the map is less reliable. The information that would have been obtained from the (1,0) reflection is obtained from (1,1) and (1,−1), whose sum is (2,0) and whose difference is (0,2). In the least-squares sense, $v_1$ is overdetermined by the remaining reflections, and the extinguished reflection is not needed for the solution.

**5.5 The integrals of the iteration (Supplementary Fig. S21).** The reference in Eq. (S4) is an integral of a wave-vector field, for which six integrals are available. The first is least squares, which is the default. Three more are applied once after the iteration to the resulting phase (weighted by confidence, weighted by curl, and a branch cut). Two more are applied to the reference at every iteration and include the circulation of a dislocation as an integer vortex (the topological reference) or as a fractional curl. On the fused field of an edge dislocation at 20 counts, the winding of the core remains in the residual with the four single-valued integrals (median $0.018|\boldsymbol{g}|$), and $|\psi_g|$ at the core stays at 0.01–0.02 of its far-field value. With the two integrals that include circulation, the residual decreases to $0.0003|\boldsymbol{g}|$ and $|\psi_g|$ at the core is restored to 0.96. All six agree outside the core. The vortex integral is the topological reference used in every dislocation result in **Figs. 3–6**.

**5.6 Implementation of the three iterations.** The 13 px median filter (1.6 periods of the 8 px benchmark lattice) is the only smoothing applied between the fused basis and its integration. It preserves edges. A discontinuity in $\boldsymbol{k}$, such as a grain boundary, is therefore retained in the reference as a step rather than a ramp. At the end of each iteration, the fused lattice basis is passed through this filter on a downsampled grid and interpolated back before least-squares integration gives $\varphi_{\mathrm{ref}}$ for the next iteration. Within each iteration, the same integration is applied to the wave vector of each reflection. Each reflection is redemodulated with the narrower aperture using the correction implied by its own $\boldsymbol{g}$ for the common lattice reference. The apertures are 0.25, 0.125 and $0.09|\boldsymbol{g}|$ by default in the software and in **Fig. 3**, and 0.25, 0.125 and $0.05|\boldsymbol{g}|$ in **Figs. 2 and 4a, b**, where only two reflections are demodulated. Three iterations are sufficient, with the $E_{xx}$ error for the dislocation in **Fig. 3** decreasing to 0.23, 0.17 and 0.13% after one, two and three iterations. For the rotating lattice in **Fig. 2**, the $\theta$ error is 4.2, 2.2 and 1.6° at 8 e/px and 0.94, 0.30 and 0.25° at 200 e/px. The reference has converged after the second iteration, and only the aperture is narrowed in the third. Cores are recorded in the topological reference during the first two iterations only, since a core recorded in the last iteration cannot influence the field because the reference is not used again. These settings fully specify the iteration and are used, unless stated otherwise, in every figure and in the benchmark.

## Supplementary Note 6 | The four components of the tracked reference in detail

This note describes the settings and additional cases for each component of GPTA corresponding to the four columns of **Fig. 4**, together with the two additional outputs obtained from the topological reference and the joint inversion. The failures in each image are described in Note 3. The GPTA results for each image are described in this note.

**6.1 Tracking the reflection (Fig. 4a, Supplementary Figs. S3, S5, S15 and S16).** The five regions of **Fig. 4a** are chosen in five directions around the centroid of the smeared $\boldsymbol{g}_1$ reflection. Each is the farthest region in its direction whose final wave vector lies within $0.03|\boldsymbol{g}|$ of the ground truth. The iteration starts from the centroid of the smeared $\boldsymbol{g}_1$ reflection. The iteration-1 reference is the center of the arc, where the GPA aperture sits. Over the three iterations, the reference wave vector of each region moves outward from the center along the arc (**Fig. 4a3, a5, a7**), and the $\theta$ error falls from 0.89° to 0.30° to 0.26° with no averaging window. The gain therefore comes from tracking with the narrowing aperture, with no contribution from averaging. In **Supplementary Fig. S16**, the same iteration is run on the same image from the fitted $\boldsymbol{g}_1$ instead of the arc center, which accounts for the errors of 0.94° in iteration 1 and 0.25° in iteration 3. Tracking is also used in the small-angle boundary of **Supplementary Fig. S5** and in the 37° rotation of **Supplementary Fig. S3** (Notes 3.2 and 3.4). **Supplementary Fig. S15** shows that the tracked reference is one phase field whose local wave vector varies with position. Convergence is governed by the inequality in Note 1.4. The aperture of iteration $n$+1 need only cover the residual spread left by the reference of iteration $n$.

**6.2 The narrowing aperture at low dose (Fig. 4b, Supplementary Fig. S4).** At 20 e/px, the arc in **Fig. 1b** is the same as in **Fig. 1a**, with the aperture noise now visible. Using each region's own reference shifts its reflection to the origin of the demodulated diffractogram. The five regions therefore coincide there, allowing the aperture to close around the residual (0.25 → 0.125 → 0.05|***g***|). The $\theta$ error falls from 4.0° to 0.88° to 0.67° compared with 6.3° for the fixed 0.10|***g***| aperture. The gains and costs of narrowing the aperture are shown for the strained-layer case at 3 e/px (Note 10.4 gives the aperture–window map). For a fixed reference, noise and resolution are coupled through the aperture width along one curve, and GPTA falls outside this curve. The low-dose inclusion in **Supplementary Fig. S4** shows the second component of the low-dose gain, which comes from the higher-order reflections included in the joint inversion (Note 3.1).

**6.3 Domain awareness (Fig. 4c, Supplementary Figs. S6 and S24).** *Per-domain iterations and stitching (***Supplementary Fig. S24***).* On a benchmark bicrystal at 50 e/px with two lattices at a misorientation of 7.2°, the confidence of each lattice's reflections is high on its own side of the boundary and is used to identify its domain. Each pixel is assigned to the lattice with the highest confidence (55% and 45% of the field). Three iterations are run in each domain with its own basis, and the two $\theta$ maps are stitched at the boundary. This gives a ***k*** error of 0.0043|***g***| (median), a bad area of 0.001 and an $E_{xx}$ RMS of 0.57% in the grains. *Injected dislocations.* The boundary is then treated as a wall of edge dislocations and localized using the curl of the stitched wave-vector field. The cores are identified by the holes in $|\psi|$ obtained in one demodulation using the stitched reference. An even comb of sites is placed along the boundary, with a count of 15 determined from the sheet charge of the curl and a spacing of 41.4 px that matches Frank's rule (42.5 px expected). It is aligned with the holes in $|\psi|$ (median distance 4.7 px) and added to the reference as phase vortices. Both grains and the boundary are then measured together in one final iteration. The holes in $|\psi|$ on the boundary shrink from 48% to 11% of the boundary pixels, and the cores appear in $E_{xx}$ as the singular fields of edge dislocations (13% within 8 px of the boundary, where the cores' own fields dominate). Measuring the boundary in the same map raises the grain RMS from 0.57% to 0.92% in the final iteration because the singular fields of the injected cores now extend into the grains beyond the 8 px disks excluded from the grain mask. This bicrystal (lattice period 5.3 px, from Frank's rule) is not the image of **Supplementary Fig. S5**. Applying the same sequence to the 15° boundary of **Fig. 1c** produces **Fig. 4c**. The two domains agree with the ground-truth assignment on 99.9% of the pixels, the measured misorientation is 14.7°, and seven cores inside the scored area are placed a median of 8.5 px from their true positions on an 8 px lattice. The $E_{xx}$ RMS in the grains is 1.5% after the final iteration. The analysis is not applied when one lattice is found, and the result is then identical to the single-reference run. The mirror twin of **Supplementary Fig. S6**, whose variants share one GPA aperture, is resolved by the same assignment (Note 3.3).

**6.4 Topological reference and inconsistency map (Supplementary Figs. S20 and S21).** On a wall of three same-sign edge dislocations at 10 e/px, the measured local ***k*** vectors circulate around each core. With a curl-free reference, this circulation must be represented by a smooth field during integration, spreading each core over the averaging window and shifting it from its position. With the topological reference, the integer winding of each core is recorded as a phase vortex and included in the reference. The residual around the core is therefore small, and the core is located at the topologically pinned zero of $\psi$. In this image, the reference type contributes a factor 1.13 to the full-field error of the step sequence (Supplementary Table 1, L3 → L4), and the three cores are located at their ground-truth positions using the topological reference. For a dissociated dislocation (partials of a/3 and 2a/3 at (132, 192) and (252, 192)), the $\chi^2$ statistic of the weighted solve rises on the ribbon between the partials. There, the displacement error is 0.286 px compared with 0.020 px in the far field, and $\chi^2$ is 19.5 times its far-field value ($4.19 \times 10^{-8}$ compared with $2.15 \times 10^{-9}$). A single-valued phase field cannot represent the fractional winding of a partial, so no phase method, GPTA included, measures the partials individually. GPTA also yields a $\chi^2$ map that identifies regions in which the model is invalid. The curl-free and topological references give the same far-field error and differ at the cores. With the curl-free reference, the zero of $|\psi|$ at the core never closes (**Supplementary Fig. S21**). On the experimental dislocation stack (Note 12.4), the core position is determined by the reference type and is independent of the map resolution.

**6.5 Deliverables (Supplementary Fig. S22).** Because the low-frequency part of the lattice distortion is removed using the tracked reference before the phase is measured, the residual phase is small everywhere and the integrated displacement field is continuous without unwrapping. On an analytic dislocation dipole (50 e/Å$^2$), GPA unwraps the single-valued phase of each reflection and solves the two phases for $\boldsymbol{u}$. The resulting RMS error in $u_x$ is 1.43 px, compared with 0.031 px for GPTA ($u_y$ 0.032 px). The winding density of the fused field is −1.00 and +1.00 at the two cores. The recorded Burgers vectors are 1.00 a in magnitude, and the core positions are within 0.85 px. On an analytic 10° bicrystal with a boundary normal at 12°, ten dislocations are recorded at both 200 and 10 e/Å$^2$, with median spacings of 47.3 and 47.4 px, a total wall charge of 10.008 and 9.998 compared with 10.0, and a $\theta$ step of 10.001° and 9.990°. Frank's rule predicts 10.34 dislocations inside the trimmed window. The dislocation content of a boundary is thus obtained from the phase topology with the same precision at both doses.

**6.6 All-reflection fusion under extinction (Fig. 4d, Supplementary Figs. S7, S9 and S19).** *Mechanism.* Extinction is imposed as a contrast-reversal envelope of depth $m$ = 1.2 at 200 e/px in **Fig. 4d** and as a reversal band at 10 counts in **Supplementary Fig. S19**. Six reflections are demodulated, namely (1,0), (0,1), (1,1), (1,−1), (2,0) and (0,2). Inside the reversal band, the (1,0) family loses its phase and its confidence. Because the weights depend on the confidence, this family contributes nothing there, and the basis is reconstructed from the remaining reflections. In **Fig. 4d**, a two-reflection solve restricted to (1,0) and (0,1) has an $E_{xx}$ RMS error of 1.86% inside the band compared with 1.20% outside. The weighted inversion over all six gives 0.44% inside and 0.43% outside, and the band therefore disappears from the error. *Why the weights are confidence and not a threshold.* The weights are $|\psi|^2/N_0$ per pixel. A strong reflection has a weight close to its normal value, whereas a reflection driven to zero by the envelope has a weight close to zero. Intermediate reflection strengths have intermediate weights, and this continuous transition leaves no step in the fused error at any depth. In a reversal-depth series on the image of **Fig. 4d** (not shown), the fused error stays between 0.10% and 0.34% as the depth $m$ runs from 0 to 0.8 because the basis is reconstructed from the other five reflections. The two-reflection error rises from 0.11% to 4.1%. A threshold would produce a step at the depth at which the reflection is dropped, and the map would change discontinuously there. *Region-by-region reflection loss (***Supplementary Fig. S7***).* In the bent foil of Note 3.6, the SNR within the aperture of the reflections available to GPTA is 4, 2, 25 and 23 for (0,1), (1,0), (2,0) and (2,1). (1,0) falls below the threshold and is dropped. The field is then obtained from the two higher-order reflections in regions in which it cannot be obtained from the base pair. For $E_{xx}$, the bad area is 0.96 with the base pair alone and 0.00 with the weighted inversion over the remaining reflections. The joint inversion is a different estimator and should not be regarded as an improvement on the two-reflection solve. It uses the reflections available from the specimen at each pixel. *Envelope null (***Supplementary Fig. S9***).* The vignette null of Note 3.5 requires both the confidence weight and the tracked reference (bad area 0.01 with both, 0.40 with the tracked reference alone). Each component addresses one failure in **Fig. 1** through a mechanism that can be followed in the intermediate quantities. In all cases, the weights depend on the same confidence quantity.

## Supplementary Note 7 | Comparison methods and fairness of the comparison

This note describes the implementation and tuning of the three comparison methods, GPA, windowed GPA (S-GPA) and peak-pair analysis (PPA), which allow the comparisons in **Fig. 5** and Table 1 to test phase extraction under equal conditions.

**7.1 Peak pairs (Supplementary Fig. S26).** PPA is shown on data suited to it (a noiseless simulated edge dislocation). A Bragg filter ($\sigma$ = 0.30$|\boldsymbol{g}|$, three reflections) is applied, and the local maxima of the filtered image are found and located to subpixel precision by a quadratic fit (4103 peaks, 3975 paired at ×4 Fourier upsampling). A reference basis is obtained from a 64 px box far from the core ($\boldsymbol{a}$ = (7.969, 0.028) px, $\boldsymbol{b}$ = (−0.035, 8.015) px). Each peak is paired with its neighbors along the basis vectors, and the strain per peak is obtained from the pair vectors. In the affine space of the reference basis, a perfect lattice is an integer grid and the dislocation appears as an extra half-plane. The interpolated $E_{xx}$ has an RMS error of 0.130% outside $\boldsymbol{r}$ < 25 px on this noiseless image. In the updated peak-finding step used in the benchmark, the Bragg filter is restricted to first-order reflections (second-order reflections produce weak maxima at the interstitial sites of a three-beam reconstruction, which the peak finder returns as atoms). The maxima are located by non-maximum suppression at the lattice spacing, each position is determined as a centroid on the unfiltered image, and bonds are clustered into grains. In **Fig. 6** and **Supplementary Figs. S31–S33**, the classic construction with one global reference cell is used instead, for the reason given in Note 12.5.

**7.2 Windowed GPA (Supplementary Fig. S25).** S-GPA uses a bank of windowed transforms in place of the single global transform of GPA. On a sinusoidal grating with a peak strain of 25% (window $\sigma$ = 12 px, aperture 0.30|$\boldsymbol{g}$|, nine frequencies, threshold 0.12 of the bank maximum), each window follows its own local frequency. At the three marked positions, the offsets from $\boldsymbol{g}_1$ are −0.22, 0.00 and +0.23 of the aperture. The filtered terms are stitched by the inverse windowed transform, and $E_{xx}$ is obtained from the resulting phase with an RMS error of 0.233%. S-GPA thus follows slow changes in the lattice at the cost of the resolution set by its window, with a fixed reference within each window and a two-reflection inversion. In the benchmark, it is run with a window of two lattice periods, an aperture of 0.10|$\boldsymbol{g}$|, five frequency samples and an amplitude threshold of 0.15. For the large-strain case of **Fig. 5**, opening its aperture to 0.40 with thirteen frequency samples gives an error of 1.65 × $10^{-3}$|$\boldsymbol{g}$|, equal to GPTA (Note 10.5).

**7.3 Settings and the basis (Supplementary Fig. S27).** The same reference basis $\boldsymbol{g}_1$, $\boldsymbol{g}_2$, obtained by lattice fitting on the simulated image, is used for every method. This makes phase extraction the subject of the comparison and excludes basis finding. The exceptions are the domain-aware algorithm of GPTA and the grain clustering of PPA, which are allowed to find further lattices as part of their specified analysis. In **Figs. 1, 4 and 5** and Table 1, GPA is run at $\sigma$ = 0.10|$\boldsymbol{g}$|, which gives the lowest error on undistorted lattices in the benchmark. Note 10.5 gives **Fig. 5a** at the textbook 0.25|$\boldsymbol{g}$|. **Supplementary Fig. S27** shows ground truth and all four methods on one further image from each regime in **Fig. 5** (a low-dose image at 2 e/px, a 25% strain image, a 30° bicrystal and a bent foil with reflection loss). None of these images appears in **Fig. 5a**. PPA is the most accurate method on the 30° bicrystal there (1.29 compared with 1.76 × $10^{-3}$|$\boldsymbol{g}$| for GPTA), and GPTA is the most accurate in the other three regimes, as in **Fig. 5a**. The order of the three comparison methods differs from image to image. Under these settings, the same basis and images are used for every method. The differences in **Fig. 5** and Table 1 therefore arise from the estimators, with no differences in their inputs.

## Supplementary Note 8 | Benchmark image set, scoring and Supplementary Tables 2 and 3

This note defines the benchmark image set and the scoring behind Table 1 and Supplementary Tables 2 and 3.

**8.1 Image set.** 13,795 simulated images in two image sets (10,248 at $256^2$ px and 3,547 at $1024^2$ px) were generated as lattices of Gaussian columns with prescribed displacement, thickness or envelope fields and Poisson noise at doses from 1 to 500 e/px. The extinction cases involving thickness, bending and tilt are multislice simulations. For Table 1, the 41 generators are grouped into ten families according to the physical origin of lattice distortion shared by the image types. These families are smooth elastic fields (pristine lattices, smooth Fourier strain, Eshelby inclusions, device lobes, 6,950 images), thickness and foil geometry (thickness fields, tilt, wedges, torsion, 1,266), dislocations and dipoles (1,535), planar defects (antiphase boundaries and twins, 521), grain boundaries and mosaics (tilt bicrystals from 1° to 45° and multi-grain mosaics, 1,239), heterointerfaces and crack tips (310), continuous rotation and kink bands (342), lattice modulations (756), real crystal structures (Au, Si, $SrTiO_3$, graphene, $MoS_2$, a metal–organic framework (MOF), 314) and mixed or masked images (562).

**8.2 Scoring.** Each method returns ($\boldsymbol{k}_1$, $\boldsymbol{k}_2$), from which $k_{50}$ is calculated as the median over the scored area of $|\boldsymbol{k} - \boldsymbol{k}_{\mathrm{GT}}|/|\boldsymbol{g}|$ after re-indexing the basis by an integer transformation. The ground-truth basis is taken per grain, and a 24 px margin is excluded. The bad area is the fraction of the scored area at which $|\boldsymbol{k} - \boldsymbol{k}_{\mathrm{GT}}| > 0.02|\boldsymbol{g}|$. A map fails when the bad area exceeds 0.20, and a missing result is also counted as a failure. The median $k_{50}$ of a method is taken over its own passing images, so a method that passes more difficult ones has a more difficult pool. For the kink-band and single-mode-modulation generators of Supplementary Table 2 (a5knk and a4qa), GPTA's median $k_{50}$ is above GPA's (14.7 compared with 12.3 and 11.9 compared with $11.3 \times 10^{-3}|\boldsymbol{g}|$), although it passes 2.8 and 1.2 times as many images there.

Supplementary Table 2 lists the failure rate and median $k_{50}$ of each of the 41 generators, and Supplementary Table 3 separates the two image sets and the GPTA variants. In the lattice-modulation family of Table 1, S-GPA has the lowest failure rate, 52.0% against 58.5% for GPTA, and GPTA has the lowest in the other nine families.

The domain-aware default is within 0.5 percentage points of the per-image best, indicating that the gain over the single-reference iteration comes from segmentation and is not explained by the choice of variant. Table 1 therefore reports one configuration (the domain-aware default), scored by one protocol on one image set. Supplementary Tables 2 and 3 give the per-generator and per-variant detail behind it.

## Supplementary Note 9 | Failure survey of the real-space methods

In this note, we document the failure of the real-space method at low dose and on two superimposed lattices. We also compare the reciprocal-space windows through which PPA and GPTA measure the same image.

**9.1 Low-dose limit of a peak method (Supplementary Fig. S28).** At 1 e/Å$^2$, the PPA peak-finding step returns 1015 maxima for 1017 cells and indexes all detected maxima, but the precision of each column is insufficient. The scatter of the peaks about the best single lattice grows from 0.009 cell edges at 100 e/Å$^2$ to 0.058 at 1 e/Å$^2$, a factor of 6.4 over two decades of dose, compared with 10 for pure $1/\sqrt{\text{dose}}$ scaling. Because PPA measures the lattice at each column, that scatter contributes directly to the strain, giving a bad area of 0.24 compared with 0.02 for GPA and 0.00 for GPTA. A phase method estimates the same displacement from the reflection, which contains information from approximately $10^3$ columns. Across the image set, the PPA failure curve rises above the GPA curve at the lowest doses, with PPA failing on 68% of the images and GPA on 56% at 1 e/Å$^2$. The second block of the figure shows the other limit of a single reference cell, which occurs on two superimposed lattices. The detected peaks belong to both lattices, and their nearest-neighbor spacing splits into two populations. No single cell fits them, and PPA with one reference cell converts the beating of the two lattices into stripes of false strain. The two lattices can be separated only by their reflections, which are used by the domain-aware algorithm of GPTA.

**9.2 PPA and GPTA at 1 e/px on synthetic images (Supplementary Figs. S29 and S30).** Two simulated images were constructed to compare the individual reciprocal-space windows of PPA and GPTA. On a synthetic bicrystal (9.68°, Frank wall of five dislocations spaced 46 px apart), PPA's Bragg filter is a union of 33 Gaussians at $0.30|\boldsymbol{g}|$ and gives a region SNR of 2.4. GPTA's final aperture at $0.09|\boldsymbol{g}|$ on the tracked reference gives 20.4. The difference between the two windows, which contains what PPA passes and GPTA excludes, has an SNR of −0.1, that is, it is noise. The peak displacement caused by that noise is 0.093 of a period RMS, and PPA's $\theta_1$ has a bad area of 0.32 with a profile spread of 1.44°. GPTA with the domain-aware algorithm of **Fig. 4c** gives 0.02 and 0.89°. On a circular bend (radius 978 px, 15° of rotation across the field, arc strain ±13.1%, no boundary), the reflection is smeared over a patch 4.8 aperture widths wide. PPA's fixed masks give SNR 2.5, GPTA's tracked aperture 11.9, and the bad areas are 0.28 and 0.04. In both images, the advantage is confined to low dose. At 200 counts per pixel, both methods reach a bad area of 0.000, and on the bicrystal, PPA is then the more precise of the two ($k_{50}$ 1.69 compared with 1.87 × $10^{-3}|\boldsymbol{g}|$). The low-dose behavior of a peak method is determined by the window through which its peaks are measured. GPTA's window follows the reference, and PPA's window is fixed. Both limits arise because a peak method measures each column individually through a fixed window, whereas a phase method measures the lattice through a window that follows the reference.

## Supplementary Note 10 | Attribution, or where each factor of accuracy comes from

In this note, we explain how the gain of GPTA over GPA is attributed to its components in **Fig. 5a**. GPTA with every component switched off is first shown to be GPA. We then add the components in a fixed order and record the error at each step. Finally, we document how the comparison methods were tuned.

**10.1 Reduction to GPA.** Six settings separate GPTA from GPA, which are the number of iterations, the aperture schedule, the averaging window $\sigma_{\text{res}}$, the SNR threshold on reflections, the robust reweighting and the set of reflections. Using iterations = 1, a single aperture, $\sigma_{\text{res}} = 0$, a threshold that accepts all reflections and the reflection set (1,0), (0,1) switches off five of them, and the sixth is inactive by construction. With two reflections, the weighted solve is exactly determined and the weights cancel. Robust reweighting is skipped for fewer than three reflections. In a comparison at each point with GPA on the same image using the same Gaussian aperture, the median $|\Delta\boldsymbol{k}|$ is $2.5 \times 10^{-5}|\boldsymbol{g}|$ at $0.25|\boldsymbol{g}|$ and $4.3 \times 10^{-7}|\boldsymbol{g}|$ at $0.09|\boldsymbol{g}|$, compared with a signal of $10^{-2}|\boldsymbol{g}|$. The residual comes from the finite-difference stencil (two implementations of the same gradient). The equivalence follows from three identities, the first of which is that demodulating against a plane-wave reference and then applying the aperture around the origin is the same operation as masking the transform around $\boldsymbol{g}$ (the shift theorem). $\text{Im}(\psi_g^* \nabla \psi_g)/|\psi_g|^2$ is identically $\nabla \arg \psi_g$ when $\sigma_{\text{res}} = 0$. With two reflections, the joint inversion is the $2 \times 2$ solve of GPA. L0 in the step sequence of Supplementary Table 1 is the GPA-equivalent configuration. **Fig. 5a** plots the sequence's positive increments rescaled to the gain over GPA at $0.10|\boldsymbol{g}|$, which is the aperture of **Figs. 1, 4 and 5**.

The default averaging window $\sigma_{\text{res}}$ requires care because it is 2.5 lattice periods. This corresponds to 20 px on an 8 px lattice. A GPA aperture of $0.25|\boldsymbol{g}|$ corresponds to a real-space kernel of 5.1 px. A single-iteration GPTA run at the default $\sigma_{\text{res}}$ corresponds to GPA at an effective aperture of $0.062|\boldsymbol{g}|$, four times narrower than $0.25|\boldsymbol{g}|$. The full three-iteration run has an effective kernel of 17 px, equivalent to $0.074|\boldsymbol{g}|$. Any improvement over GPA in a single-iteration GPTA run would be due to $\sigma_{\text{res}}$ and could not be attributed to tracking. The comparisons in this paper therefore either set $\sigma_{\text{res}} = 0$ (**Figs. 2 and 4a, b**, **Supplementary Fig. S16**, the GPA-equivalent configuration L0) or compare with GPA at its best aperture (**Figs. 1, 4 and 5**, Table 1, Supplementary Note 10.5). The one property of the $\sigma_{\text{res}}$ average that GPA cannot reproduce at any aperture is the $|\psi_g|^2$ weighting. With this weighting, which is added in step L2, a low-amplitude pixel does not contaminate its neighbors in the weighted mean. The aperture averages phase and noise alike.

**10.2 The step sequence.** The components of GPTA are added in a fixed order, which is L1 tracking the reflection, L2 confidence averaging ($\sigma_{\text{res}} = 1.5$ periods), L3 all-reflection fusion, L4 topological reference and L5 domain-aware algorithm. At each step, the settings of the preceding step are used, and Supplementary Table 1 lists the error at each step. The factor for a step is the ratio of the errors of consecutive steps. These factors multiply to give the total gain over the GPA-equivalent configuration L0, which is why **Fig. 5a** plots the gain on a logarithmic axis, on which the segments add. In **Fig. 5a**, the positive factors are rescaled to make the stack span the gain over GPA at $0.10|\boldsymbol{g}|$. L2 and L3, which are both parts of the confidence-weighted inversion, are drawn as one "fusion" segment. The raw factors quoted below refer to the step sequence itself. On the bicrystal, the second grain is measured on the wrong lattice at each step before L5, so only L1 and L5 are shown as segments there. One step dominates in every case, with the low-dose gain coming from L2 (3.3×) and L1 contributing 1.9× because the narrowed aperture excludes noise. The higher-order reflections of L3 add 1.05× because at 3 e/px they are themselves close to the noise level. At large strain, the gain comes from L1 (1.9×) and L2 (2.0×). On the bicrystal, the gain comes from L5 because tracking cannot recover the second grain until it has its own reference. L1 gives 4.2× on grain

B because the tracked reference moves part of the way toward grain B's reflections, as on the flank of a fixed aperture in **Supplementary Fig. S5**, without reaching them at 30°. L2 and L3 increase the error in the second grain (×3.3 and ×2.3) because averaging and fusion are performed on the wrong lattice. L5 reduces the error 288-fold from L4, or 38-fold from L1. Extinction is the one case in which three steps each give large gains (L1 4.2×, L2 4.9× and L3 3.7×). The fixed aperture, the noise inside the reversal band and the loss of a reflection are three separate mechanisms there.

Three steps reduce accuracy in a single-lattice case, each by less than 10%, and each reduction has a physical reason. For L3 on large strain (−9%), the higher-order reflections have a tail twice as wide as that of the fundamental at 25% strain and add noise where no reflection has been lost. For L4 on large strain (−7%), the case has no dislocation. The vortex detector records a small number of spurious windings in the tail region, which are then included in the reference. For L5 on the dislocation wall (−2.8%), the segmentation finds one lattice and the step only reruns the iteration. None of these costs is visible in a case that needs the step, which supports using one configuration with everything on.

**10.3 Core disk against far field.** The dislocation cases in the step sequence are scored over the full valid field. When scored inside a disk of 16 px around the core of an isolated dislocation (the edge dislocation with 12% smooth strain in Note 10.4), the step sequence reverses, with L0 giving 0.021|***g***| and L4 0.051|***g***|. The reversal is due to resolution and does not reflect attribution. L0 does no averaging and is therefore the sharpest estimator of a singularity by construction, while every averaging step smooths the core field over its window. In the far field ($\boldsymbol{r}$ > 25 px), the same step sequence gives 4.8 → 2.8 → 1.5 → 1.25 → 1.05 × $10^{-3}$|***g***| from L0 to L4. On a wall of three edge dislocations spaced 64 px (eight periods) apart, the GPTA $\theta$ map reproduces all three cores at the ground-truth positions (**Supplementary Fig. S20a**). At 32 px, the averaging windows of every method merge neighboring cores.

**10.4 Aperture and averaging window against dose.** For the strained-layer case at 3 e/px, the $E_{xx}$ error was mapped over 8 apertures (0.04 to 0.35|***g***|) and 8 windows (0.25 to 5 periods). Along the aperture axis at a fixed window, the error is flat below 0.10|***g***| and rises above it because a wide aperture passes noise that must then be removed by averaging over the window. Along the window axis at a fixed narrow aperture, the error falls to a floor of 0.377% at 1.6 periods and rises slowly beyond it as the window begins to smooth the 6 px interface. In GPA, the single aperture width determines the averaging window, confining GPA to the curve window = 1/(2$\pi$·aperture width). Its best point on that curve is 0.445% ($\sigma$ = 0.10|***g***|). The optimal window depends on dose. For an edge dislocation with a 12% smooth strain at 50 e/px, the error outside the core, relative to the best GPA (0.222%), is 1.29× better at $\sigma_{res}$ = 2.5 periods, 1.75× at 1.5 (the optimum), 1.59× at 1.0, 1.10× at 0.5, and 0.95×, which is worse than GPA, at 0.25. At $\sigma_{res}$ = 0, the ratio is 0.90×. For a noiseless version of the same case, GPA at 0.25|***g***| is 3.5× better than the three-iteration GPTA because there is no noise to average and the window only smooths. At 500 e/px, the two are equal, and at 100 e/px, 30 e/px and 10 e/px, GPTA is 2.4×, 3.9× and 5.5× better. The window is a dose-dependent setting, and the 1.5 periods used in the step sequence are the optimum at 50 e/px on an 8 px lattice. The software default of 2.5 periods is used in **Fig. 4c**, and the experimental maps are obtained with the per-map setting in Note 12.3. On the rotating lattice of **Figs. 2 and 4a, b**, the gain comes from tracking with the narrowing aperture, with no window used at all.

**10.5 Tuning of the comparison methods.** *GPA aperture.* Across the four series in **Fig. 5b–e**, the aperture that minimizes the error on undistorted lattices is 0.10|***g***|. This is the setting used in **Figs. 1, 4 and 5**, Table 1 and **Supplementary Figs. S3–S10**. At the textbook aperture of 0.25|***g***|, GPA is worse in the low-dose and dislocation-wall cases and better with large strain and extinction (with errors of 18.5, 5.13, 44.4, 11.0 and 163 × $10^{-3}$|***g***| compared with 3.27, 10.2, 520, 3.03 and 195 at 0.10|***g***|). The two bicrystal values are not comparable, since the 0.25|***g***| value is the field-wide median and the 0.10|***g***| value is the median over grain B. The GPTA gains over 0.25|***g***| are 6.5, 3.1, 29, 4.4 and 76. *S-GPA.* The benchmark setting (window two periods, aperture 0.10|***g***|, five frequencies) is that of the series in **Fig. 5b–e**. In the large-strain case of **Fig. 5**, the strain sideband extends to 0.24|***g***|, whereas the 0.10|***g***| aperture with five samples covers 0.18|***g***|. S-

GPA is worse than GPA in this case. With an aperture of 0.40|$\boldsymbol{g}$| and thirteen frequencies, its error is 1.65 × $10^{-3}$|$\boldsymbol{g}$|, equal to that of GPTA (1.63), and the best GPA aperture (0.175|$\boldsymbol{g}$|) gives 3.74. For large smooth strain, a windowed transform with an aperture opened to match the strain is as accurate as tracking, and GPTA reaches the same accuracy without requiring the aperture to be chosen. *PPA.* Three peak-finding approaches were run on the images of **Supplementary Fig. S27**, namely the original construction, a Voronoi bond-clustering construction and the updated peak-finding step in Methods. Their $k_{50}$ (× $10^{-3}$|$\boldsymbol{g}$|) are 20.8, 98.0 and 21.2 at low dose, 9.31, not available and 9.08 at large strain, 230, 5.16 and 1.29 on the bicrystal, and 46.2, 140 and 15.4 under extinction. The updated peak-finding step is the best of the three on the bicrystal and under extinction and within 2% of the best at low dose and large strain. This is the PPA used in **Fig. 5** and Table 1. Across the $256^2$ image set, its introduction lowered the PPA failure rate from 50% to 25%. The attribution is therefore based on a GPA-equivalent starting point, a fixed order of steps and comparison methods tuned to their best settings. The factors in **Fig. 5a** measure the components, with tuning excluded as a source of the differences.

## Supplementary Note 11 | Computational cost

The computing cost of the four methods on the benchmark was measured as wall-clock time per simulated image (median / 90th percentile, Apple M5 Pro, 18 cores, 64 GB, one process per image). On the $256^2$ image set (10,248 images), GPA takes 0.15 / 0.29 s, S-GPA 1.3 / 1.5 s, PPA 0.52 / 0.80 s, single-reference GPTA 1.3 / 1.6 s and domain-aware GPTA 1.4 / 2.0 s. On the $1024^2$ image set (3,547 images), the times are 2.2 / 3.4 s, 21.4 / 25.4 s, 7.6 / 10.6 s, 25.2 / 29.6 s and 28.6 / 49.2 s.

GPTA requires three demodulations per reflection instead of one, up to twelve reflections instead of two, and one reference integration per iteration. The cost scales as the number of reflections times the number of iterations times one pair of fast Fourier transforms (FFT). The cost is within a factor of 1.4 of S-GPA at both sizes and 8–13 times that of GPA. The domain-aware algorithm requires segmentation and one run per domain, and all timings include the basis search shared by the four methods. GPTA therefore costs about an order of magnitude more than GPA and about the same as S-GPA, with a computation time of a few seconds per image on a laptop processor.

## Supplementary Note 12 | Experimental data

This note documents the three Pt specimens in **Fig. 6**, their calibration, the quantities by which maps are evaluated without ground truth, and the results for each specimen.

**12.1 Specimens and calibration.** Three Pt specimens were imaged by high-angle annular dark-field scanning transmission electron microscopy (HAADF-STEM) as frame stacks. The specimens were a [110] polycrystal of six grains ($512^2$, period 5.49 px), a [001] region with eight dislocation cores ($512^2$, period 5.81 px), and a [111] bicrystal ($896^2$, period 8.64 px). The detector noise is Poisson to within 30%. The local variance-to-mean ratio is 2.4–3.2 analog-to-digital units (ADU) across the signal range, compared with the nominal gain of 2.52 ADU per electron. The two detector settings had gains of 2.52 and 6.77 ADU per electron, giving 25.4, 25.8 and 12.2 e/px per frame for the three stacks. A Poisson injection from the 40-frame average reproduces the single frame's spectral noise level to within 2.5%, and the 0.25-frame column of **Supplementary Figs. S31–S33** is such an injection. Pixel sizes are derived from the measured reciprocal basis of the dominant grain indexed against the Pt lattice, giving 0.4077 Å/px ([110], {111}+{002}, residual 1.23%), 0.3414 Å/px ([001], {200}+{020}, 1.13%) and 0.1606 Å/px ([111], {2−20}+{−220}, 1.01%). Same-field stacks of 40 frames (polycrystal), 30 (dislocation region) and 150 (bicrystal) were recorded, and registered averages of 1, 4 and 16 frames were formed from each, with an additional 24-frame average for the dislocation stack. With these pixel sizes, the frame doses are 150, 220 and 470 e/Å$^2$, respectively.

**12.2 Evaluation without ground truth.** We report four quantities, which are the change in one method's map with dose (self-consistency), the split-half noise level between odd and even frames, the disagreement between methods at one dose, and, for the dislocation stack, the core offset relative to the method-independent reference positions of Note 12.4. The disagreement between methods locates their differences without deciding which is right. Any count of which method is best is made on one shared mask, since GPA and S-GPA return a value at every pixel and GPTA and PPA do not. The map resolution of GPTA is the full width at half maximum of the kernel that combines the final aperture and the averaging window. For each map, we choose the finest setting that returns a stable lattice (Note 12.3). The map noise is the robust standard deviation of $E_{xx}$, calculated as 1.4826 times its median absolute deviation. Cores are scored over the full field, and inside a core disk, the unaveraged estimator is the sharpest (Note 10.3).

**12.3 Polycrystal (Supplementary Fig. S31).** We choose the GPTA map resolution for each dose using the finest averaging window that returns a stable lattice. At one frame, this corresponds to a map resolution (full width at half maximum, FWHM) of 0.35 nm ($\sigma(E_{xx})$ 0.0117), and at sixteen frames it corresponds to 0.54 nm (0.0015). The finest stable setting is thus coarser at the higher dose because at 0.28 nm on the sixteen-frame average the reference converges to a different lattice (median $|\Delta \boldsymbol{k}|$ 0.20$|\boldsymbol{g}|$ compared with the 0.54 nm map), which is the condition described in Note 13. We compute the grain boundaries once on the 40-frame average from 7810 detected columns, of which 5018 pass a lattice test, using a nematic order parameter built from the two long bonds of the pseudo-hexagonal [110] projection (the six-fold bond-orientational order parameter alone folds the 55.7° grains onto 4.3°). The boundaries are drawn identically at every dose. Under the single global re-indexing used for display, a misoriented grain appears as large uniform strain, with the color representing orientation.

**12.4 Dislocation stack (Supplementary Fig. S33).** The eight reference cores are the phase singularities of the Bragg-filtered 24-frame average. They were obtained without contribution from any method and move by ≤ 2 px as the mask width is varied. Cores are scored for every method using one estimator, which is the Burgers-vector density $\alpha = \nabla \times \boldsymbol{k}$ thresholded into sign-connected components whose charge centroids are matched to the reference within four periods. The median core offset is 2.5 px for GPTA at one frame and 1.6–4.4 px over 2–24 frames, within the 5.81 px period. The offsets are 6.5–9.6 px for GPA, 7.3–10.1 px for S-GPA and 6.8–12.6 px for PPA. The GPTA value is determined by the reference rather than the map resolution, with a curl-free reference giving 8.2–11.8 px at every map resolution at 24 frames and the topological reference giving 2.4–2.6 px. Changing the map resolution from 5.9 to 1.9 periods with a curl-free reference improves the median offset from 11.8 to 8.2 px at a cost of 23% in far-field noise. The topological reference at the same map resolution gives 2.4 px at a cost of 2%. GPA and S-GPA improve slightly at lower doses (9.6 → 7.3 px) because their detection threshold scales with the noise and the curl components fragment. Their high-dose value is the representative one. PPA detects one to eight spurious components at 1–16 frames.

**12.5 Bicrystal (Supplementary Fig. S32).** The measured step across the boundary is 0.409$|\boldsymbol{g}|$ = 23.5°. PPA uses the classic peak-pairs construction here because the segmenting peak-finding step assigns the two grains to one domain at 1, 2, 4, 8 and 16 frames and returns uniform blocks of the wrong lattice (median $|\boldsymbol{F} - \boldsymbol{I}|$ 0.12–0.58 compared with 0.017 when it segments correctly, with F as the local deformation matrix and I as the identity). These blocks result from a segmentation failure and do not constitute a measurement. The classic construction gives a per-atom error at the limit of localization $\sqrt{2}\cdot\sigma_{atom}/|a|$, predicted to be 0.135 at one frame and 0.027 at 25 and measured as 0.12 and 0.025–0.036. At sixteen frames, the two grains are 60 $\sigma$ apart in per-atom orientation and PPA resolves the boundary as sharply as GPTA. The comparison in **Fig. 6b** concerns the low-dose columns, where the fixed apertures of GPA and S-GPA mix the two lattices across the boundary and GPTA's domain-aware algorithm separates them.

**12.6 Self-consistency across dose (all stacks).** For every stack, each method's $\boldsymbol{k}$ field at N frames is compared with the same method's field at the full frame count (relative median $|\Delta \boldsymbol{k}|/|\boldsymbol{k}|$). The split-half noise level is the difference between the odd-frame and even-frame averages divided by two, that is, the noise at full dose. On the Pt bicrystal stack, the deviation from the full-dose field falls monotonically for GPA (0.018 → 0.0026), S-GPA (0.020 → 0.0016) and GPTA (0.015 → 0.0017), each reaching its split-half noise level (0.0037, 0.0024, 0.0022). This corresponds to a 6.8- to 12.5-fold improvement from one frame to sixteen. PPA on the same stack starts at 0.25 at one frame because its grain clustering assigns the two grains differently at low dose. Its split-half noise level of 0.15 shows that the two half-averages are clustered differently as well. Its error is determined by segmentation rather than noise, which is why we use the classic construction in **Fig. 6b**. The same behavior is observed on the Pt polycrystal stack, where the grain count for PPA with the updated peak-finding step is 1, 1, 1, 1, 1, 2, 4, 4, 2, 3, 3 across eleven dose levels. Its deviation does not fall with dose (0.9×), while the three phase methods improve 10- to 13-fold. Segmentation is decided before the measurement as a discrete choice that does not average down. Across the three stacks, the phase methods behave as the simulations predict. PPA departs from this behavior with a dose-independent error caused by segmentation rather than noise.

## Supplementary Note 13 | Where GPTA fails

This note records the conditions under which a default setting of GPTA gives a wrong map and the setting that corrects each of them.

The reference can follow a reflection to its position in the local lattice, but this freedom also allows it to settle on a neighboring reflection when the nominated one is much weaker (**Supplementary Fig. S34**). In a dislocation-free single crystal denoised by a neural network, the reflection nominated by the basis finder (1,1) is 136 times weaker than the adjacent (1,0) fundamental. After iterations 1, 2 and 3, the tracked reference lies 0.42, 0.55 and 0.63$|\boldsymbol{g}|$ from the nominated reflection, toward the fundamental at a distance of $|\boldsymbol{g}|$. Its normalized confidence (the weight w in the figure) is 0.61, 0.17 and 0.19. The GPTA map is built on the wrong lattice (bad area 1.00, correlation 0.00 with the ground truth), while PPA measures the columns and is correct (bad area 0.00, correlation 0.78). GPA uses a fixed aperture and passes the criterion (bad area 0.16), but its $E_{xx}$ correlates 0.04 with the ground truth, so it does not resolve the field either. The confidence indicates the failure and is the quantity an automated implementation uses as its SNR threshold. In the benchmark, this condition occurs when a weak reflection is nominated as a basis beside a strong one. The basis finder avoids this on undenoised images by nominating the strongest first-order reflections.

These notes record three further conditions in which a default setting gives a wrong map, each with the setting that corrects it. The vortex detector records spurious windings on a dislocation-free image at 25% strain, costing 7% (Note 10.2), and these are removed by using the curl-free reference. With a map-resolution setting finer than the finest stable one, the reference settles on a different lattice on the sixteen-frame polycrystal average (Note 12.3). This is avoided by using the per-map setting. The domain-aware default segments a 3° dislocation wall into two domains (Note 14), while measurements using the pinned basis with the topological reference identify it as one lattice. In each case, the failure is visible in the confidence or in the stability of the reference and is removed by using a documented setting.

## Supplementary Note 14 | Software

This note describes Lattice Studio, which is the desktop application in which every map in this paper was produced and in which GPA, S-GPA, PPA and GPTA are implemented in one code base. The application is shown in **Supplementary Fig. S35**. The Input card loads a frame, displays the diffractogram, lets the user pick or accept the basis reflections, and previews $|\psi|$ for each. The Compare card runs the four methods on the same frame and basis and displays any strain component in a quad view on one color scale. GPTA in the screenshots uses the topological-reference variant of Supplementary Table 3 with a pinned basis at a map-resolution setting of 0.5 (the setting of **Supplementary Fig. S33**), on the 16-frame average of that dislocation stack. The domain-aware default accepts a second lattice at 2° (Note 1.9) and segments this 3° wall into two domains. On a low-angle wall, the pinned basis with the topological reference is the setting to use (Note 13). Code, the image generator and the scoring scripts are released at https://github.com/YTianUCI/LatticeStudio, and the $256^2$ px image set and the per-image results are deposited at https://doi.org/10.57760/sciencedb.010xy. With the released code and data, every comparison in the paper can therefore be repeated on any image.

## Supplementary Tables

**Supplementary Table 1 | Error after each added component of GPTA for the Fig. 5a cases.**

| case | L0 | L1 | L2 | L3 | L4 | L5 | S-GPA | PPA |
|---|---|---|---|---|---|---|---|---|
| low dose (3 e/px) | 18.5 | 9.79 | 2.97 | 2.84 | 2.84 | 2.84 | 3.61 | 5.12 |
| large strain (25%) | 5.13 | 2.73 | 1.38 | 1.52 | 1.63 | 1.63 | 9.37 | 5.36 |
| bicrystal 30° (grain B) | 247 | 58.5 | 194 | 447 | 447 | 1.55 | 467 | 1.24 |
| dislocation wall | 11.0 | 6.26 | 2.85 | 2.79 | 2.46 | 2.53 | 2.95 | 3.73 |
| extinction (m = 1.2) | 163 | 39.1 | 7.90 | 2.15 | 2.15 | 2.15 | 242 | 43.1 |

Error is the median $|\Delta\boldsymbol{k}|/|\boldsymbol{g}|$ over the field in units of $10^{-3}|\boldsymbol{g}|$, then the median over noise realizations 1–3. L0 = GPA-equivalent configuration at the aperture $0.25|\boldsymbol{g}|$, L1 + tracking the reflection, L2 + confidence averaging, L3 + all-reflection fusion, L4 + topological reference, L5 + domain-aware algorithm. The dislocation wall of **Supplementary Fig. S20a** is listed for completeness and is not one of the four **Fig. 5a** cases.

**Supplementary Table 2 | Benchmark results for the 41 generators.**

| generator | content | *n* | GPA fail % | GPA $k_{50}$ | S-GPA fail % | S-GPA $k_{50}$ | PPA fail % | PPA $k_{50}$ | GPTA fail % | GPTA $k_{50}$ |
|---|---|---|---|---|---|---|---|---|---|---|
| g4s | twisted bilayer moiré, twist 4–18°, dose series | 100 | 78.0 | 9.03 | 75.0 | 5.65 | 100.0 | – | 64.0 | 2.14 |
| b2gra | monolayer graphene | 77 | 81.8 | 10.25 | 83.1 | 11.22 | 88.3 | 9.77 | 62.3 | 9.63 |
| a5spn | continuous lattice rotation across the field, 10–37° | 150 | 85.3 | 2.75 | 84.0 | 2.87 | 67.3 | 4.58 | 74.0 | 2.92 |
| b4het | heterointerface between two lattices | 155 | 99.4 | 5.23 | 97.4 | 4.20 | 51.6 | 3.63 | 18.1 | 1.60 |
| lhs | pristine, dislocation, twin or mosaic lattice with 3–6 stacked imaging errors | 68 | 66.2 | 4.73 | 70.6 | 5.09 | 77.9 | 6.72 | 39.7 | 4.07 |
| a5knk | kink bands, 2–6 walls with steps of 2–8° | 192 | 79.2 | 12.25 | 79.7 | 11.55 | 47.9 | 10.92 | 42.2 | 14.65 |
| a4qa | single-mode lattice modulation, period 18–45 Å | 567 | 68.8 | 11.30 | 52.0 | 10.27 | 47.8 | 9.13 | 63.1 | 11.85 |
| b2mos | monolayer $MoS_2$ | 65 | 43.1 | 5.67 | 50.8 | 6.09 | 67.7 | 5.93 | 41.5 | 3.00 |
| a3mos | mosaic of grains misoriented by up to 5–37° | 669 | 68.6 | 9.74 | 65.2 | 10.51 | 42.5 | 5.63 | 10.3 | 4.84 |
| bt | hard cases (high-angle boundaries, semicoherent interface, strained wedge, modulated film, extreme low dose, defocus and time series) | 104 | 51.9 | 2.56 | 43.3 | 2.73 | 40.4 | 2.91 | 37.5 | 1.48 |
| bxt | bicrystal with a tilt boundary, 1–45° | 570 | 65.4 | 2.85 | 62.8 | 3.03 | 31.1 | 3.05 | 5.8 | 1.78 |
| a2twn | twin boundary | 273 | 59.7 | 3.80 | 60.8 | 3.80 | 30.8 | 2.77 | 9.5 | 1.80 |
| b1au | Au crystal, several zone axes | 91 | 33.0 | 3.15 | 36.3 | 3.54 | 38.5 | 3.45 | 17.6 | 2.49 |
| a5par | nanoparticle with radial surface relaxation, 4–9% at the surface, R = 20–34 Å | 89 | 25.8 | 7.00 | 25.8 | 8.54 | 38.2 | 8.22 | 22.5 | 4.41 |
| v5s3tilt | bent plate with local off-axis tilt up to 4° | 304 | 26.6 | 2.41 | 27.6 | 2.60 | 27.6 | 3.13 | 6.2 | 1.63 |
| b1sr | $SrTiO_3$ crystal, several zone axes | 76 | 21.1 | 1.94 | 26.3 | 2.03 | 28.9 | 3.86 | 10.5 | 1.28 |
| a2msk | irregular field boundary with 1–6% vacancies | 149 | 19.5 | 3.19 | 21.5 | 3.36 | 26.2 | 4.79 | 8.1 | 1.42 |
| b1sithk | Si crystal, 60–90 Å thick | 4 | 0.0 | 0.53 | 0.0 | 0.59 | 75.0 | 1.51 | 0.0 | 0.45 |
| a5esh | Eshelby inclusion | 130 | 24.6 | 3.84 | 26.2 | 3.08 | 19.2 | 4.15 | 3.8 | 1.70 |
| a3mix | smooth strain and a dislocation dipole, some with mask or vacancies | 241 | 19.5 | 4.91 | 20.7 | 4.67 | 21.6 | 4.31 | 8.3 | 3.48 |
| g2s | smooth strain, amplitude 0.3–4%, dose grid | 100 | 18.0 | 4.00 | 24.0 | 3.56 | 19.0 | 3.20 | 6.0 | 2.44 |
| a5crk | mode-I crack tip (linear elastic fracture mechanics) | 155 | 16.8 | 4.08 | 17.4 | 4.08 | 21.9 | 2.97 | 10.3 | 2.21 |
| v5s2 | one dislocation dipole with smooth strain of 1–2% | 1,108 | 19.1 | 6.28 | 18.1 | 5.47 | 20.4 | 4.87 | 8.5 | 4.74 |
| v5s3wedge | thickness wedge | 190 | 17.9 | 2.59 | 17.4 | 2.53 | 20.0 | 2.71 | 7.9 | 0.94 |
| a2dis | 1–3 dislocation dipoles | 427 | 19.0 | 5.88 | 17.6 | 4.96 | 17.1 | 5.17 | 6.6 | 4.48 |
| v5s1 | smooth strain, 1–2%, wavelength 60–200 Å | 3,954 | 15.7 | 2.68 | 16.3 | 2.73 | 17.3 | 2.61 | 5.6 | 1.14 |
| v1s | pristine lattice with one imaging error at a time | 1,681 | 13.7 | 1.82 | 14.7 | 1.79 | 18.0 | 2.02 | 7.1 | 0.80 |
| a2apb | antiphase boundary in a two-sublattice lattice | 248 | 14.5 | 3.32 | 15.7 | 3.32 | 16.5 | 3.26 | 4.8 | 0.87 |
| b3mof | metal–organic framework, square, honeycomb or kagome net, cell 16–26 Å | 2 | 0.0 | 5.02 | 0.0 | 5.63 | 50.0 | 4.14 | 0.0 | 1.70 |
| v5s3tors | elastic torsion, rotation span 1.5° | 302 | 15.2 | 2.67 | 15.6 | 2.67 | 12.6 | 2.59 | 5.6 | 1.08 |
| v3d | perfect lattice, fine sampling of 11–27 px per period | 78 | 7.7 | 1.14 | 9.0 | 1.10 | 16.7 | 1.13 | 5.1 | 0.52 |
| a5dev | device strain lobe, 0.5–2.5% | 88 | 9.1 | 2.76 | 10.2 | 2.84 | 10.2 | 2.61 | 5.7 | 1.18 |
| g1s | smooth strain, 0.5–2%, dose and defocus grid | 100 | 11.0 | 2.64 | 12.0 | 2.78 | 6.0 | 2.36 | 3.0 | 1.51 |
| a2thk | thickness field (wedge, dome or step), some with tilt | 226 | 8.8 | 2.23 | 10.2 | 2.26 | 8.8 | 2.56 | 4.0 | 1.33 |
| a2str | smooth Fourier strain, 0.3–3% | 517 | 7.5 | 3.38 | 8.3 | 3.30 | 12.4 | 2.99 | 2.9 | 2.55 |
| v3s | perfect lattice | 296 | 7.4 | 2.18 | 7.8 | 2.26 | 10.5 | 2.51 | 3.4 | 0.80 |
| g3s | perfect lattice, thickness 10–110 Å, defocus grid | 100 | 4.0 | 1.64 | 5.0 | 1.47 | 10.0 | 1.98 | 0.0 | 0.48 |
| v1t | pristine lattice, thickness series | 100 | 3.0 | 1.58 | 2.0 | 1.65 | 9.0 | 1.57 | 0.0 | 0.47 |
| v1x | pristine lattice, beam-tilt series | 40 | 2.5 | 1.32 | 0.0 | 1.20 | 0.0 | 1.58 | 0.0 | 0.53 |
| v5s3 | demonstration set (tilt, torsion, wedge) | 6 | 0.0 | 2.83 | 0.0 | 2.64 | 0.0 | 2.33 | 0.0 | 1.40 |
| b1si | Si crystal, several zone axes | 3 | 0.0 | 2.56 | 0.0 | 3.00 | 0.0 | 4.13 | 0.0 | 1.88 |
| **all images** | **both image sets** | **13,795** | **27.3** | **3.24** | **26.7** | **3.23** | **24.0** | **3.19** | **11.5** | **1.68** |

Failure rate (%) and median $k_{50}$ of passing images ($\times 10^{-3}|\boldsymbol{g}|$), both image sets merged. GPTA is the domain-aware default. Failure = bad area > 0.20. –, no passing image.

**Supplementary Table 3 | GPTA variants and image sets.**

| image set | *n* | GPA | S-GPA | PPA | domain-aware (Table 1) | single-reference | topological | best of the three |
|---|---|---|---|---|---|---|---|---|
| $256^2$ | 10,248 | 29.2 / 3.72 | 28.7 / 3.59 | 24.8 / 3.53 | 12.3 / 2.05 | 17.4 / 2.03 | 17.4 / 1.97 | 11.9 / 2.01 |
| $1024^2$ | 3,547 | 21.7 / 2.38 | 21.0 / 2.50 | 21.5 / 2.50 | 9.2 / 1.06 | 13.4 / 1.05 | 13.4 / 1.04 | 8.6 / 1.04 |
| both | 13,795 | 27.3 / 3.24 | 26.7 / 3.23 | 24.0 / 3.19 | 11.5 / 1.68 | 16.4 / 1.65 | 16.4 / 1.62 | 11.0 / 1.64 |

Failure rate (%) / median $k_{50}$ of passing images ($\times 10^{-3}|\boldsymbol{g}|$). Single-reference and topological are the same iteration without the segmentation. Best of the three is the per-image best, the upper bound for any variant selection.

## Supplementary Figures

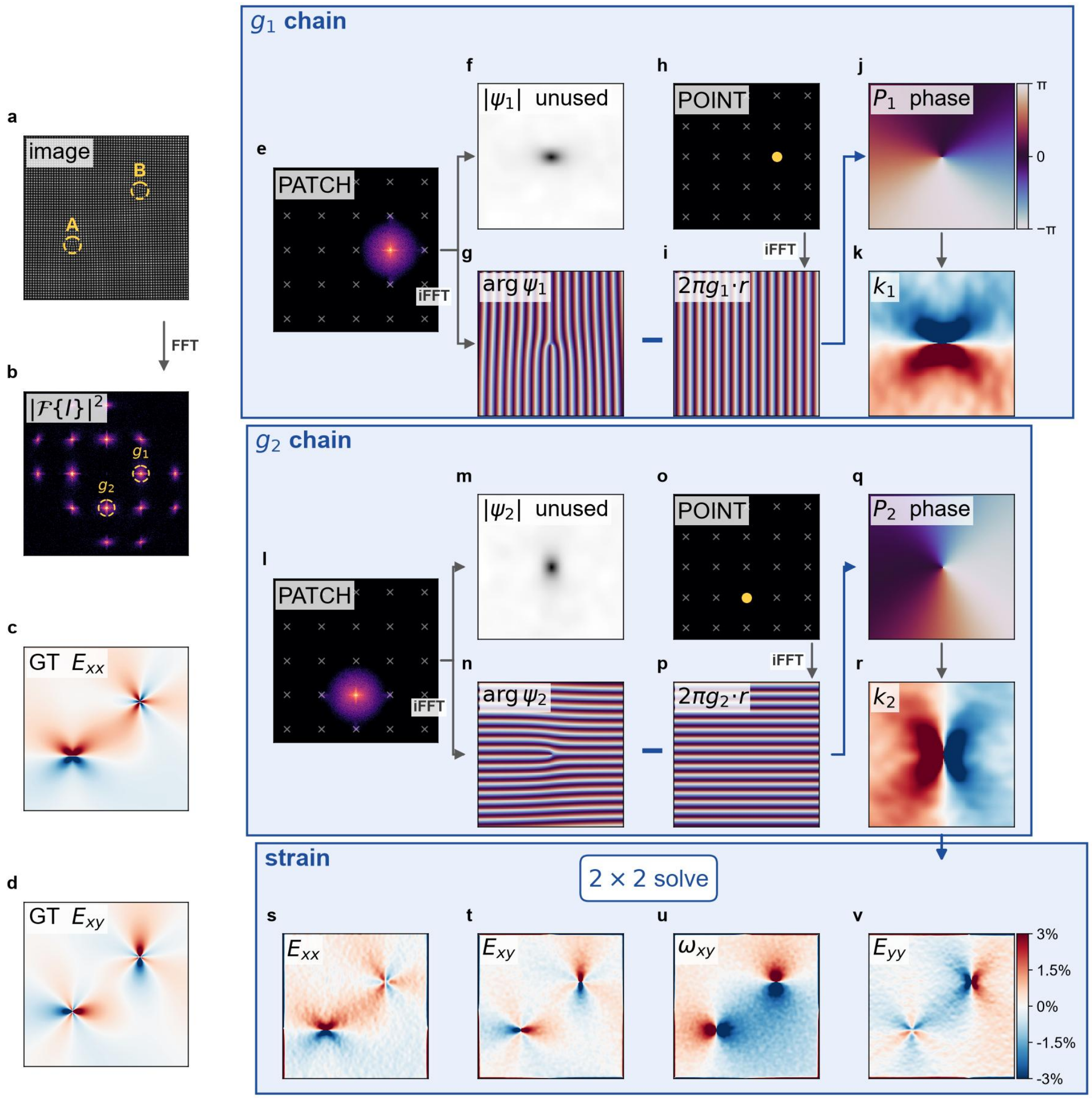

**Supplementary Fig. S1 | Geometric phase analysis workflow for two edge dislocations.** A simulated high-resolution transmission electron microscopy (HRTEM) image with two edge dislocations, core A with its Burgers vector along $x$ and core B with its Burgers vector along $y$, analyzed by GPA with an aperture of $\sigma = 0.25|\boldsymbol{g}|$. **a** Image. **b** Diffractogram with the basis reflections $\boldsymbol{g}_1$ and $\boldsymbol{g}_2$ ringed. **c, d** Ground-truth normal strain $E_{xx}$ and shear strain $E_{xy}$. **e–k** The $\boldsymbol{g}_1$ chain, comprising the aperture applied to the transform (PATCH), the amplitude and the phase of the Bragg-filtered image (labeled $|\psi_1|$ and arg $\psi_1$), the single point of the transform that serves as the global reference (POINT), the reference phase $2\pi\boldsymbol{g}_1\cdot\boldsymbol{r}$, the geometric phase $\varphi_1$ and the local wave vector $\boldsymbol{k}_1$. GPA discards the amplitude. **l–r** The same seven steps for $\boldsymbol{g}_2$. **s–v** The 2 × 2 solve, giving $E_{xx}$, $E_{xy}$, the rotation $\theta$ (labeled $\omega_{xy}$) and $E_{yy}$. Arrows denote the fast Fourier transform (FFT), its inverse (iFFT) and the 2 × 2 solve.

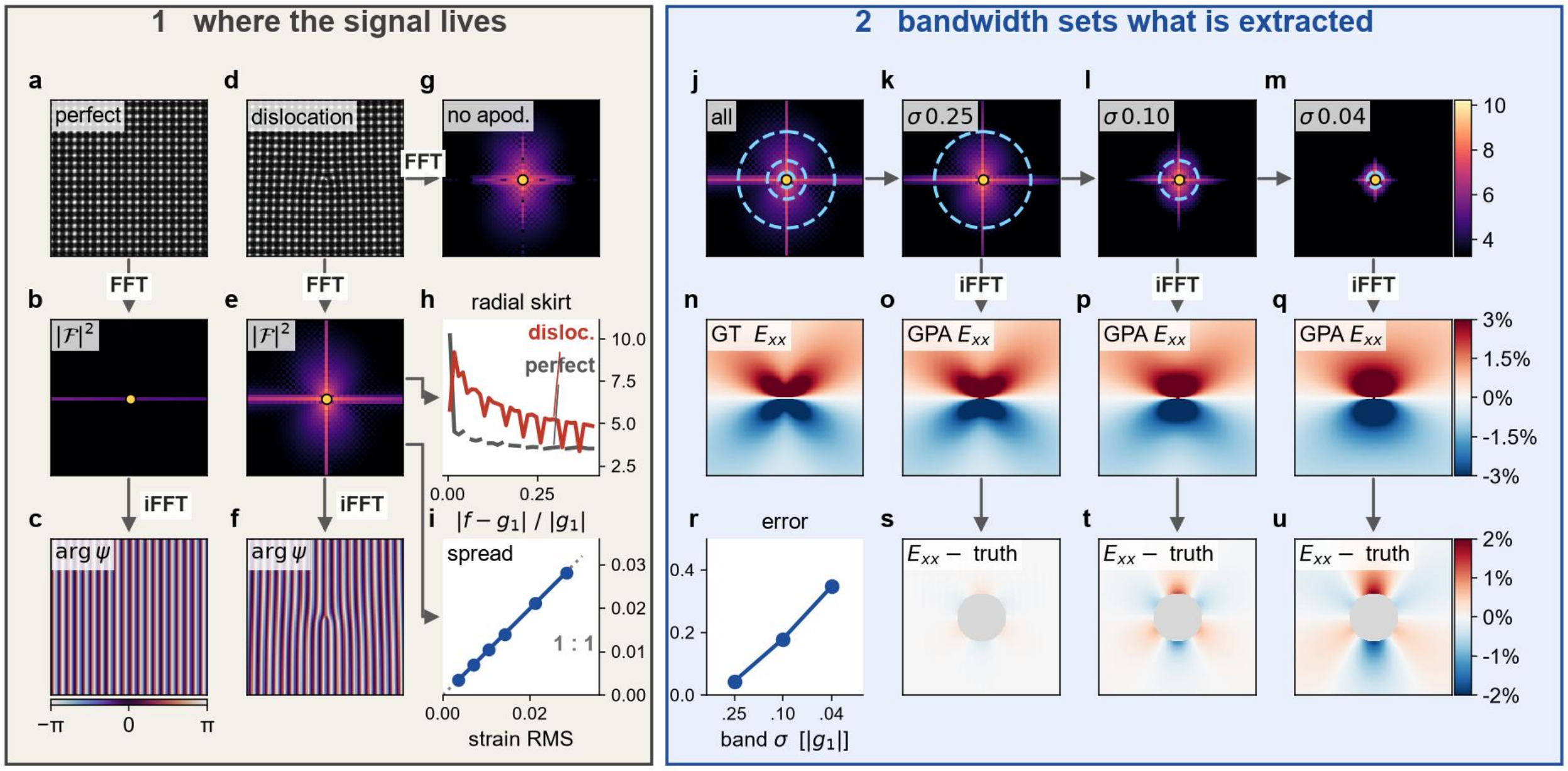

**Supplementary Fig. S2 | Reciprocal-space strain signal and aperture width.** A noiseless simulated HRTEM image of a lattice with one edge dislocation at the center. **a–c** The perfect lattice, its transform, which is a single point, and the phase arg $\psi$ of the demodulated field. **d–f** The same for the crystal with the dislocation, whose transform has a diffuse skirt along the Burgers vector $\boldsymbol{b}$. **g** The transform without apodization, whose window cross is the artifact removed in the other panels. **h** The radial power profile through $\boldsymbol{g}_1$. **i** The excess spread of the reflection against the RMS strain, which follows the 1:1 line. **j–u** The same image analyzed by GPA with the aperture alone changed. No aperture (j), $\sigma$ = 0.25, 0.10 and 0.04$|\boldsymbol{g}_1|$ (k–m), ground truth (n), the three $E_{xx}$ maps (o–q), the error against $\sigma$ (r) and the three difference maps (s–u), with RMS errors of 0.044, 0.179 and 0.347%.

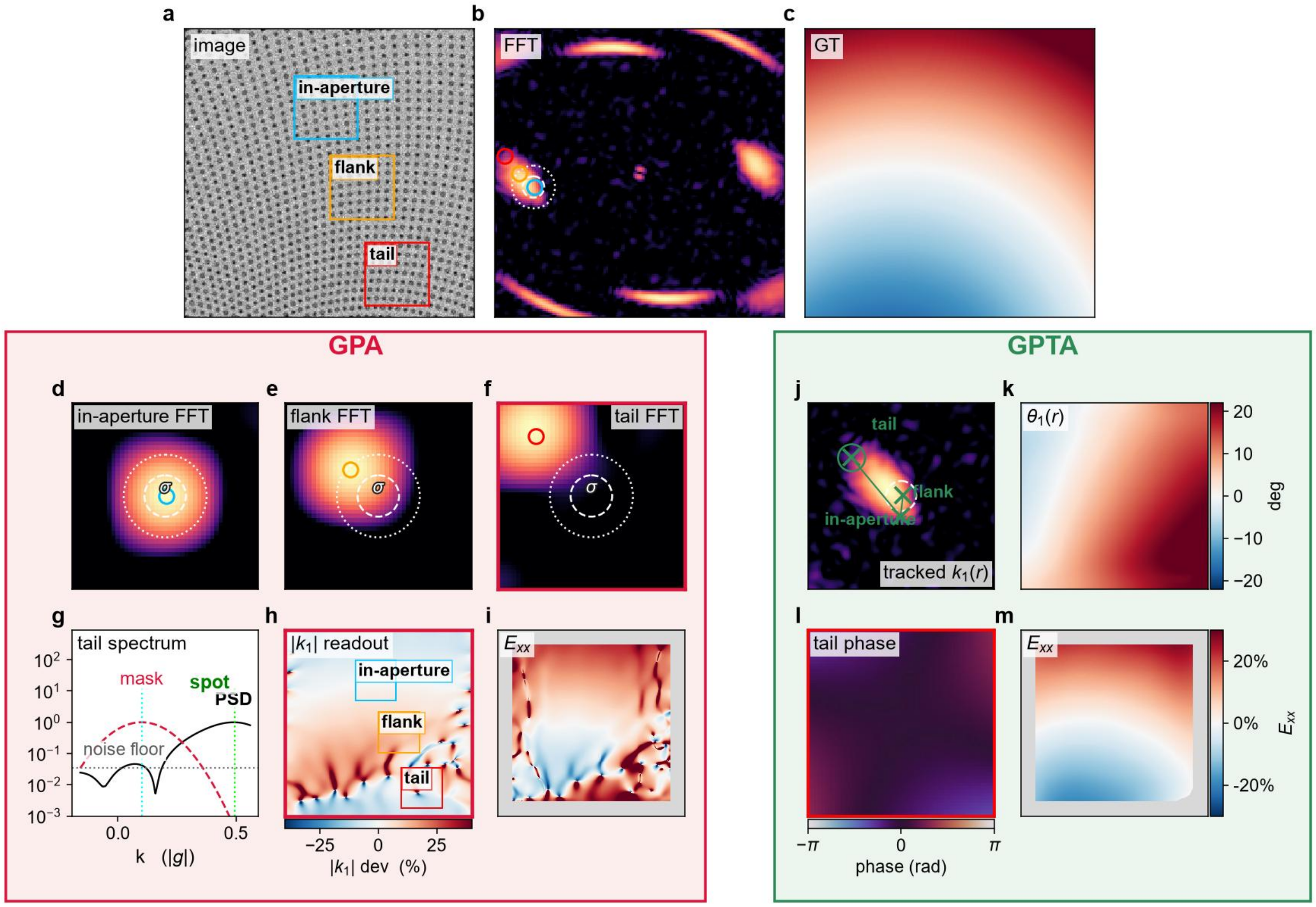

**Supplementary Fig. S3 | Strain mapping under continuous lattice rotation.** A simulated HRTEM image of one dislocation-free domain whose lattice turns by 37° across the field, which smears the reflection into an arc. Three regions are sampled along the arc, inside the aperture, on its flank and in the tail. **a** Image with the three regions marked. **b** Diffractogram with the fixed aperture ($\sigma = 0.10|\boldsymbol{g}|$). **c** Ground-truth $E_{xx}$. **d–f** The local transform of each region, in which the aperture transmittance falls from 1.00 to 0.17 and 0.00. **g** The spectrum in the tail against the noise level. **h** The deviation of the measured $|\boldsymbol{k}_1|$ over the field. **i** $E_{xx}$ from GPA, bad area 0.69. **j–l** The tracked reference $\boldsymbol{k}_1(\boldsymbol{r})$ following the arc, the measured lattice orientation $\theta_1(\boldsymbol{r})$ and the tracked phase in the tail. **m** $E_{xx}$ from GPTA, bad area 0.13.

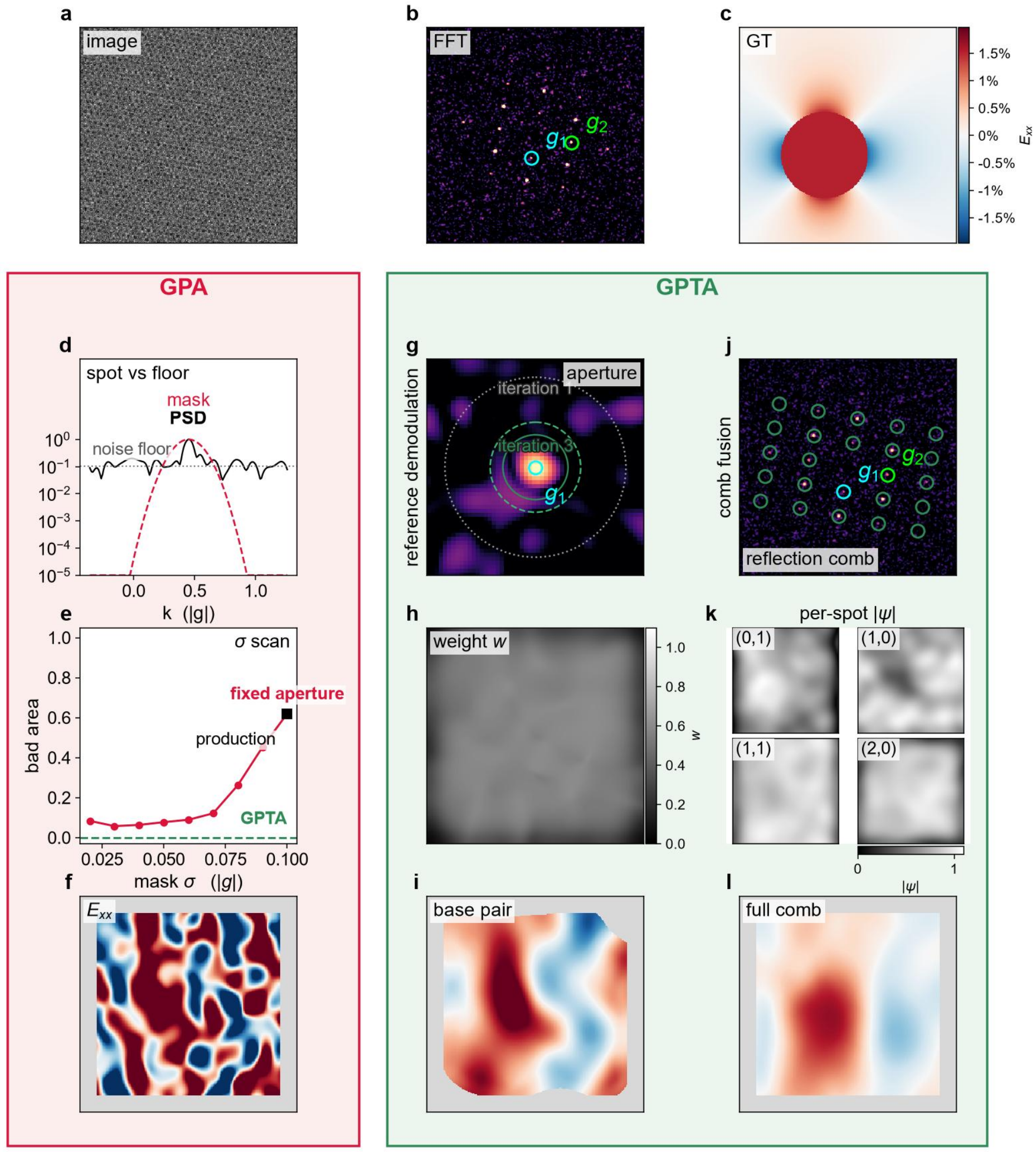

**Supplementary Fig. S4 | Low-dose strain maps across reciprocal-space settings.** A simulated HRTEM image of an Eshelby inclusion at 10 e/Å$^2$. **a–c** Image, diffractogram with $\boldsymbol{g}_1$ and $\boldsymbol{g}_2$, and ground-truth normal strain $E_{xx}$. **d–f** GPA with a fixed aperture of $\sigma = 0.10|\boldsymbol{g}|$. The power spectral density of $\boldsymbol{g}_1$ against the noise level (d), the bad area against the aperture width (e), whose minimum is 0.06 at $\sigma = 0.03|\boldsymbol{g}|$ compared with 0.62 at the fixed aperture, and the resulting $E_{xx}$ (f). **g–l** GPTA. The aperture narrowing over three iterations on the tracked reference (g), the normalized inversion weight $w$ (h), $E_{xx}$ from the base pair alone (i, bad area 0.27), the full set of reflections (j), their $|\psi|$ maps (k) and $E_{xx}$ from the weighted inversion over all of them (l, bad area 0.00).

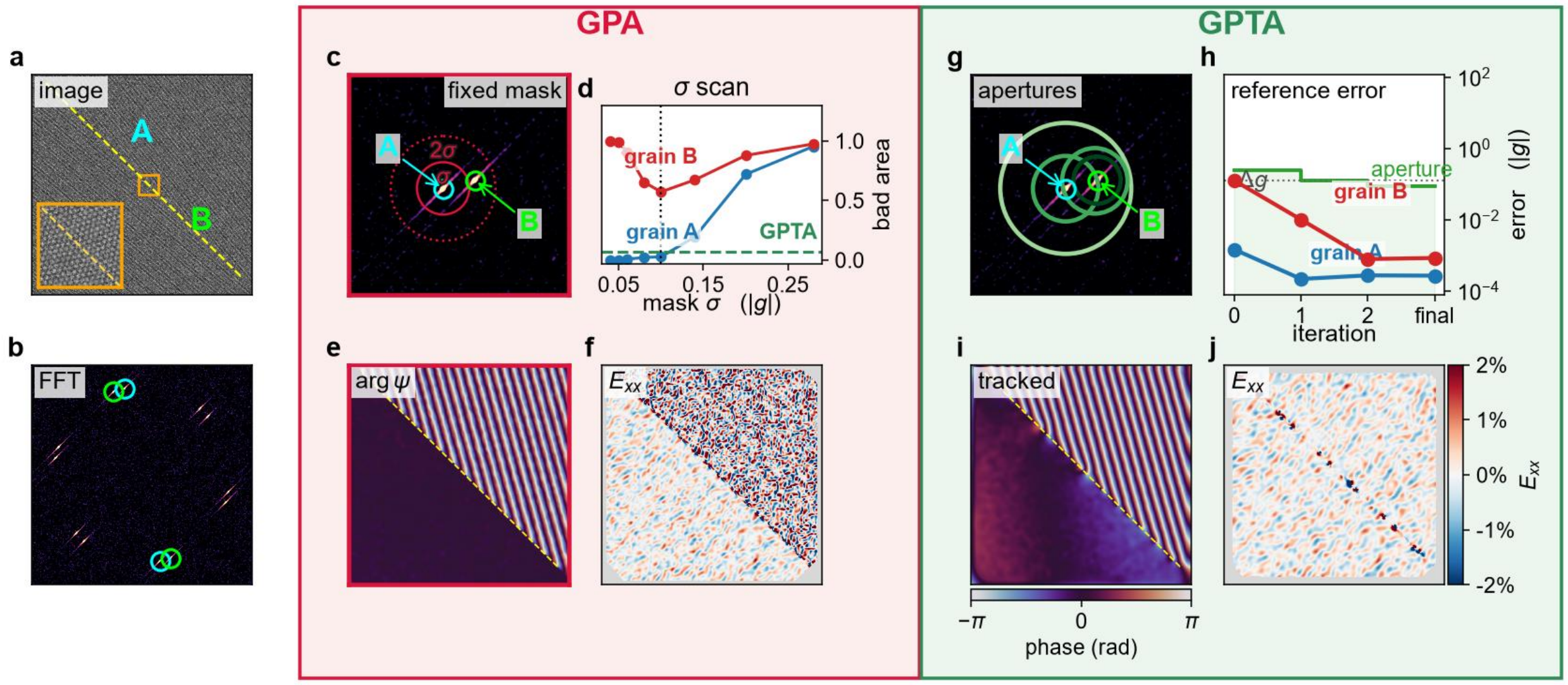

**Supplementary Fig. S5 | Strain mapping across a 7.2° boundary.** A simulated HRTEM image of a 7.2° bicrystal, in which the reflections of grains A and B are split by 0.126|$\boldsymbol{g}$|. **a, b** Image with the two grains marked and its diffractogram. **c–f** GPA with a fixed aperture of $\sigma$ = 0.10|$\boldsymbol{g}$|. The aperture over the two reflections (c), where the transmittance at B is 0.46, so that B is measured with a degraded signal. The bad area against the aperture width for each grain (d), in which no width serves both grains. The phase (e) and $E_{xx}$ (f, bad area 0.27). **g–j** GPTA. The aperture closing over iterations 1–3 on the reflection of each grain (g), the reference error per iteration (h), the tracked phase (i) and $E_{xx}$ (j), with a bad area of 0.07 for a single tracked reference and 0.05 for the domain-aware algorithm.

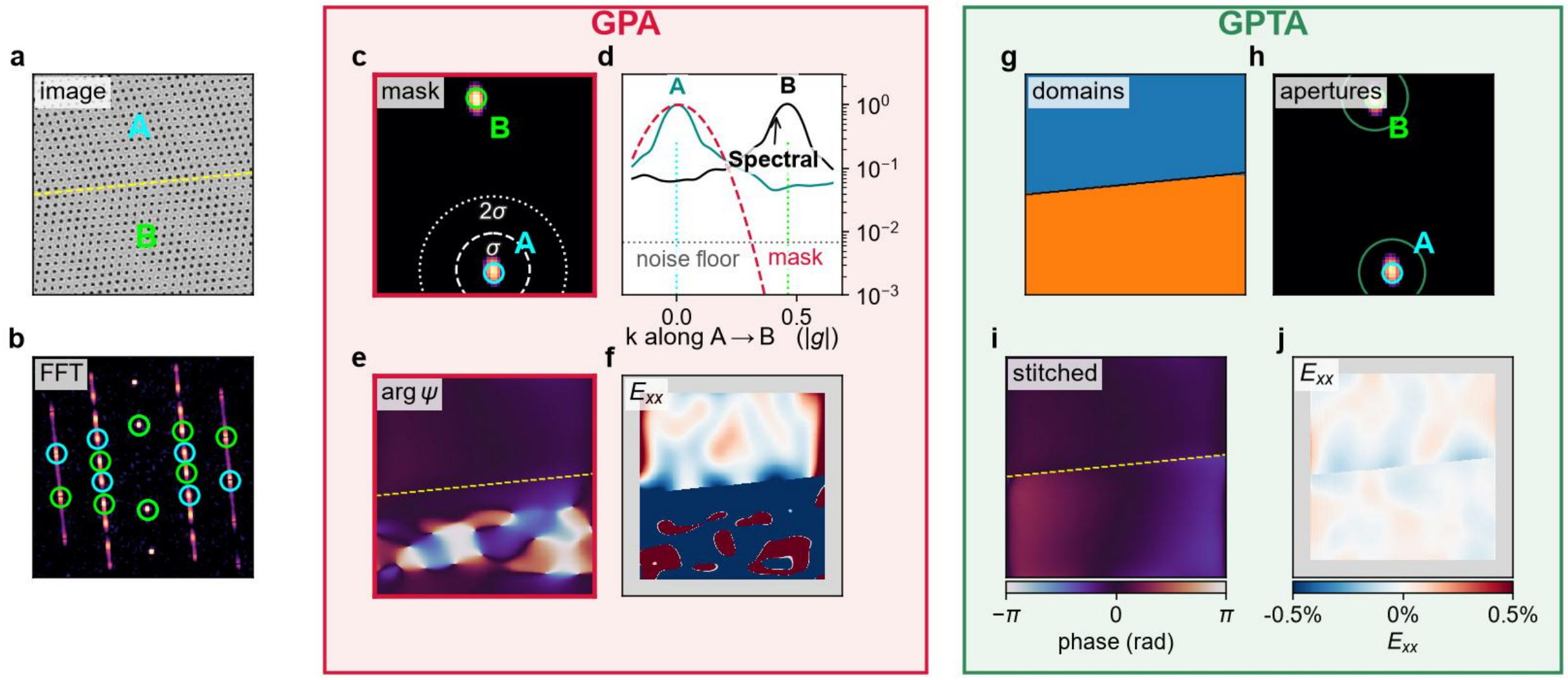

**Supplementary Fig. S6 | Strain mapping across a mirror twin.** A simulated HRTEM image of a mirror twin with variants A and B, whose reflections are split by 0.48|***g***|. **a, b** Image with the variants marked and its diffractogram. **c–f** GPA with a fixed aperture of $\sigma = 0.10|\boldsymbol{g}|$. The aperture over the reflections of the two variants (c), the spectral profile along A → B against the noise level (d), which shows that the aperture passes none of the reflection of B, the phase (e), which is noise over B, and $E_{xx}$ (f, bad area 0.51). **g–j** GPTA. The two domains found by the domain-aware algorithm (g), the aperture placed on the reflections of each variant (h), the stitched phase (i) and $E_{xx}$ (j, bad area 0.00).

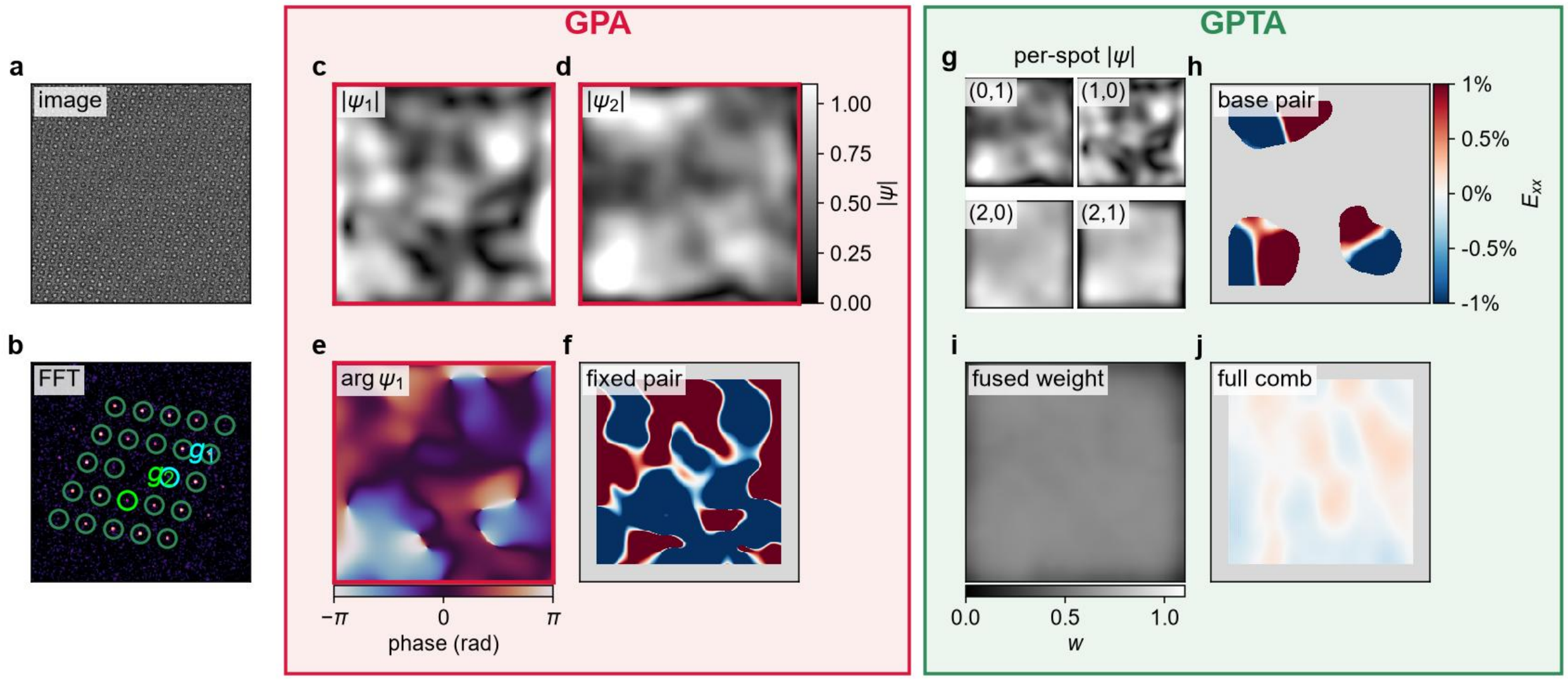

**Supplementary Fig. S7 | All-reflection fusion in a bent foil.** A simulated HRTEM image of a bent foil, in which the loss of reflections is distributed across the field. **a, b** Image and its diffractogram with $\boldsymbol{g}_1$ and $\boldsymbol{g}_2$. **c–f** GPA with a fixed aperture on the two basis reflections. The amplitudes $|\psi(\boldsymbol{g}_1)|$ (c) and $|\psi(\boldsymbol{g}_2)|$ (d), each of which collapses over a different part of the field, the phase of $\boldsymbol{g}_1$ (e), which is noise wherever its amplitude has vanished, and $E_{xx}$ (f, bad area 0.90). **g–j** GPTA. The $|\psi|$ maps of four reflections (g), of which the two higher-order reflections cover the field where the base pair cannot, $E_{xx}$ from the base pair alone (h, bad area 0.96), the fused weight (i) and $E_{xx}$ from the weighted inversion over all reflections (j, bad area 0.00).

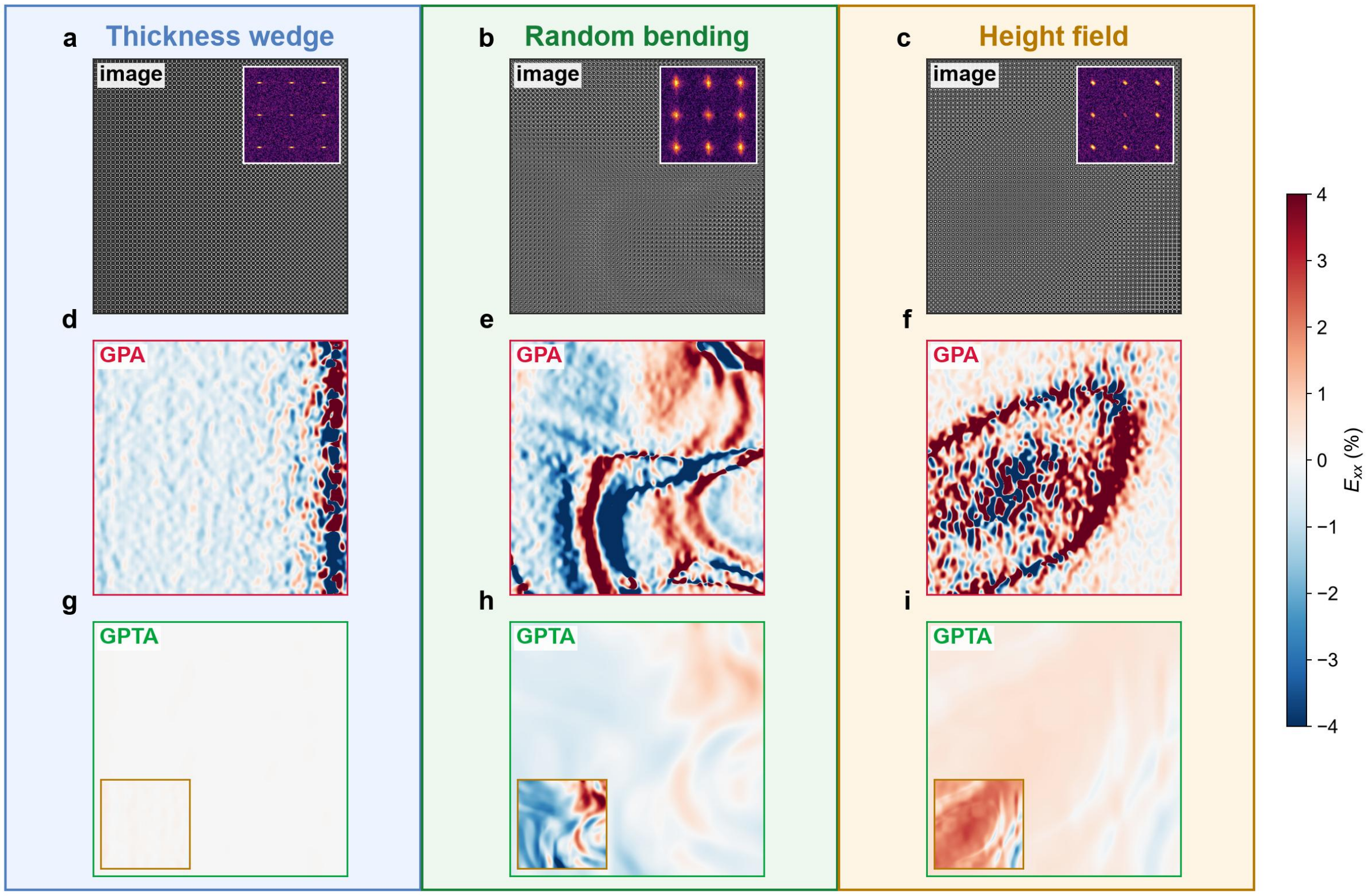

**Supplementary Fig. S8 | Reflection loss from thickness, bending and tilt.** Three multislice simulated images of a strain-free crystal, in each of which a reflection is suppressed by a different physical mechanism. **a–c** Thickness wedge, random bending and height field, each with its diffractogram inset. **d–f** $E_{xx}$ from GPA, with RMS errors of 4.92, 5.54 and 5.27%. **g–i** $E_{xx}$ from GPTA, with RMS errors of 0.021, 0.44 and 0.43%. Because the ground truth is strain-free, every feature in the GPA maps comes from the envelope and not from the lattice.

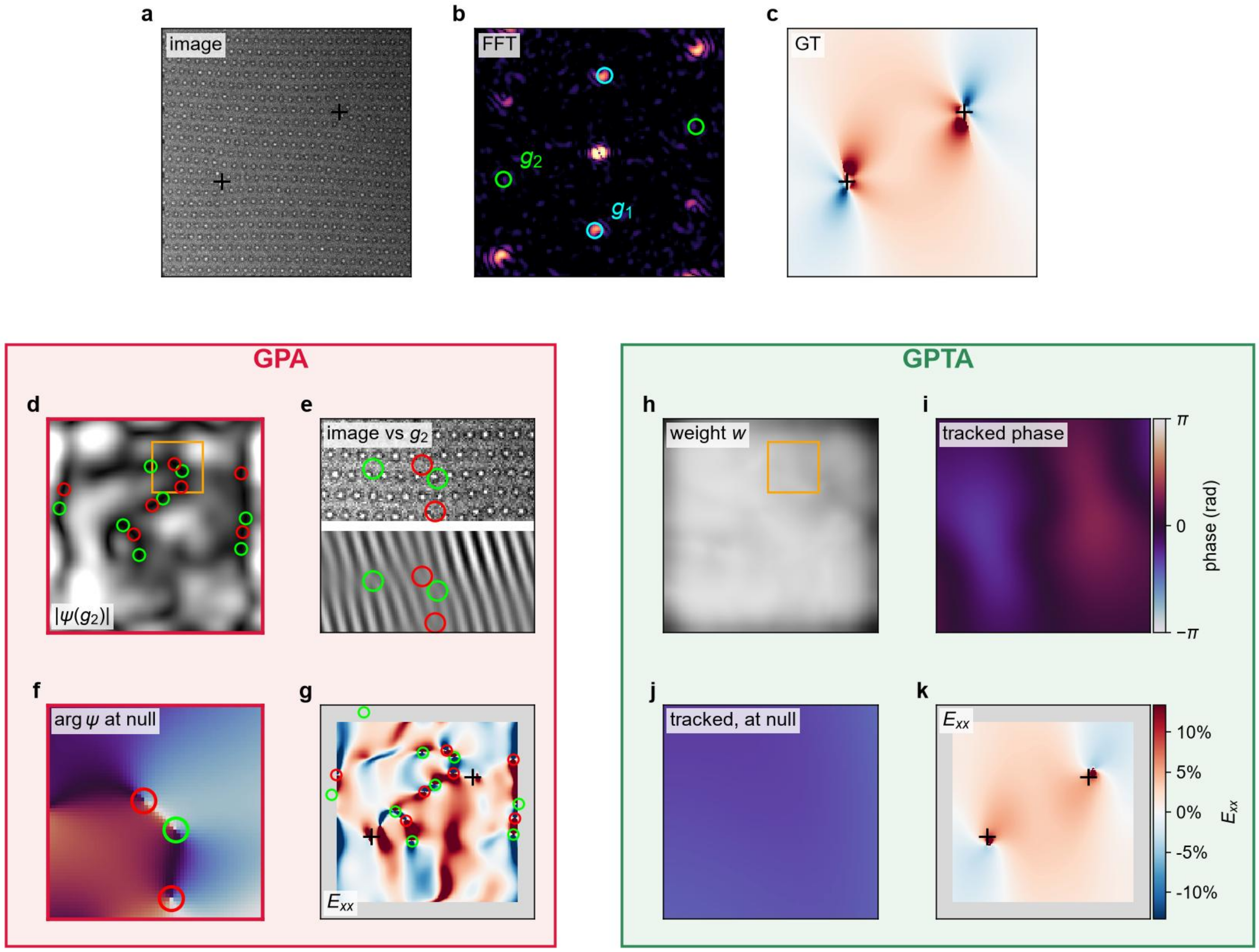

**Supplementary Fig. S9 | Confidence weighting across a reflection null.** A simulated HRTEM image of a dislocation array under a vignette envelope. **a–c** Image, diffractogram with $\boldsymbol{g}_1$ and $\boldsymbol{g}_2$, and ground-truth $E_{xx}$. **d** The amplitude $|\psi(\boldsymbol{g}_2)|$, which the envelope drives below the noise level in the null. **e** The image against the $\boldsymbol{g}_2$ fringes in the same region. **f, g** GPA with a fixed aperture of $\sigma = 0.10|\boldsymbol{g}|$. The phase in the null (f) is noise and nucleates spurious phase vortices, and the resulting $E_{xx}$ (g) has a bad area of 0.76. **h–k** GPTA. The weight map $w$ (h), which follows the amplitude and withdraws confidence from the null, the tracked phase (i) and the same region under it (j), where the reference bridges the null, and $E_{xx}$ (k), with a bad area of 0.01 and no spurious vortices. The tracked reference without the confidence weight leaves a bad area of 0.40.

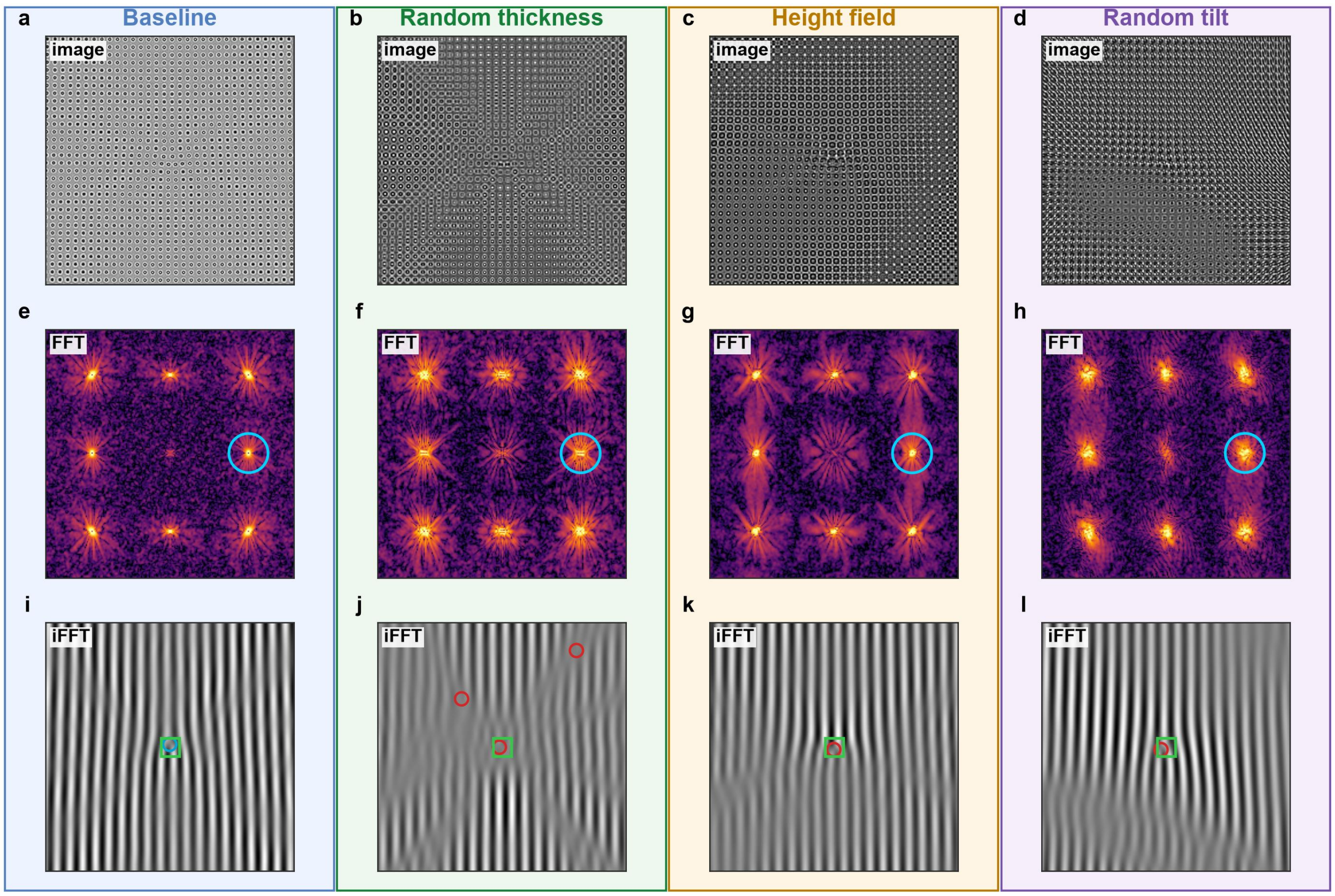

**Supplementary Fig. S10 | False dislocation cores caused by reflection loss.** Four multislice simulated images analyzed by the textbook construction of transform, mask and inverse transform. **a–d** The images, namely a baseline, random thickness, a height field and random tilt. **e–h** Their diffractograms with the mask. **i–l** The reconstructed fringes with the phase singularities marked (green square, the true core; circles, windings found elsewhere). The true core is recovered in all four images. The random-thickness image adds spurious cores where the envelope alone has driven the fringe amplitude through zero, and the other three add none. Spurious cores are therefore a property of the envelope and not of the mask width or of the fitted reflection.

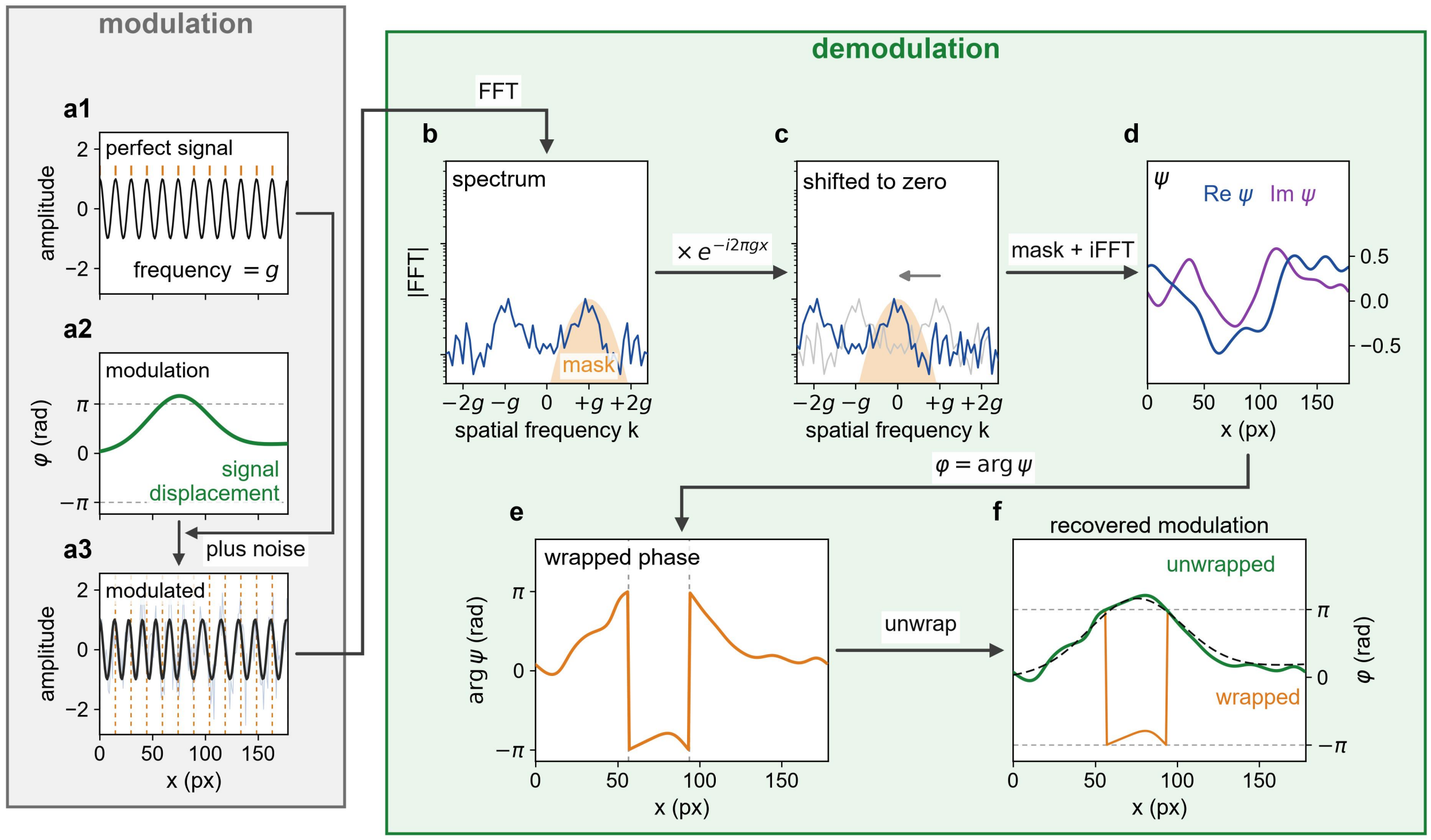

**Supplementary Fig. S11 | Demodulation explained in one dimension.** A one-dimensional fringe pattern whose phase is modulated by a Gaussian bump on a ramp, with added Gaussian noise. **a1–a3** The perfect signal of frequency $g$, the modulation $\varphi(x)$ and the modulated signal with noise. **b** The spectrum with the mask on the $+g$ reflection. **c** The spectrum after multiplication by $\exp(-i2\pi gx)$, which shifts the reflection to zero frequency. **d** The complex field $\psi$ after masking and inverse transformation, shown as Re $\psi$ and Im $\psi$. **e** Its wrapped phase arg $\psi$. **f** The unwrapped phase against the true modulation (dashed). The modulation is recovered together with the noise that the mask passes.

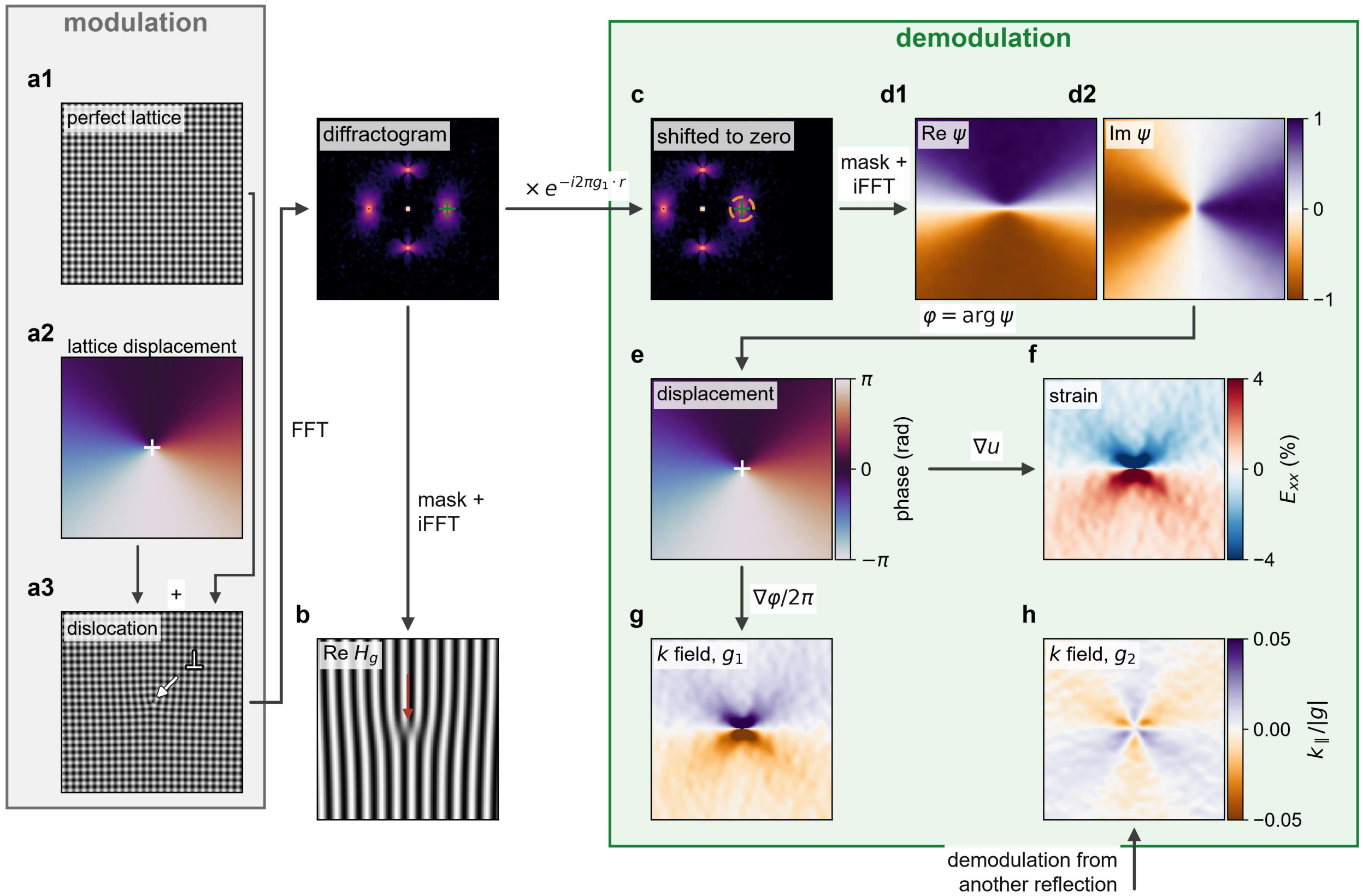

**Supplementary Fig. S12 | Demodulation explained in two dimensions.** A simulated image of an edge dislocation. **a1–a3** The perfect lattice, its displacement as a phase and the dislocation image. **b** The Bragg-filtered image $H_g$, obtained by masking $g_1$ in the diffractogram and transforming back, in which the extra half-plane is visible. **c** The diffractogram of the image multiplied by $\exp(-i2\pi \boldsymbol{g}_1 \cdot \boldsymbol{r})$, in which $\boldsymbol{g}_1$ is shifted to zero, with the mask drawn. **d1, d2** Masking and inverse transformation give the two quadratures Re $\psi$ and Im $\psi$. **e** Their argument $\varphi = \arg \psi$, the displacement as a phase. **f** The strain $E_{xx}$ from its derivative. **g** The local wave-vector field of $\boldsymbol{g}_1$, $\boldsymbol{k}_1 = \boldsymbol{g}_1 + \nabla\varphi/2\pi$. **h** The field of $\boldsymbol{g}_2$ from demodulating the other reflection. Two reflections give the two $\boldsymbol{k}$ fields needed for a strain tensor.

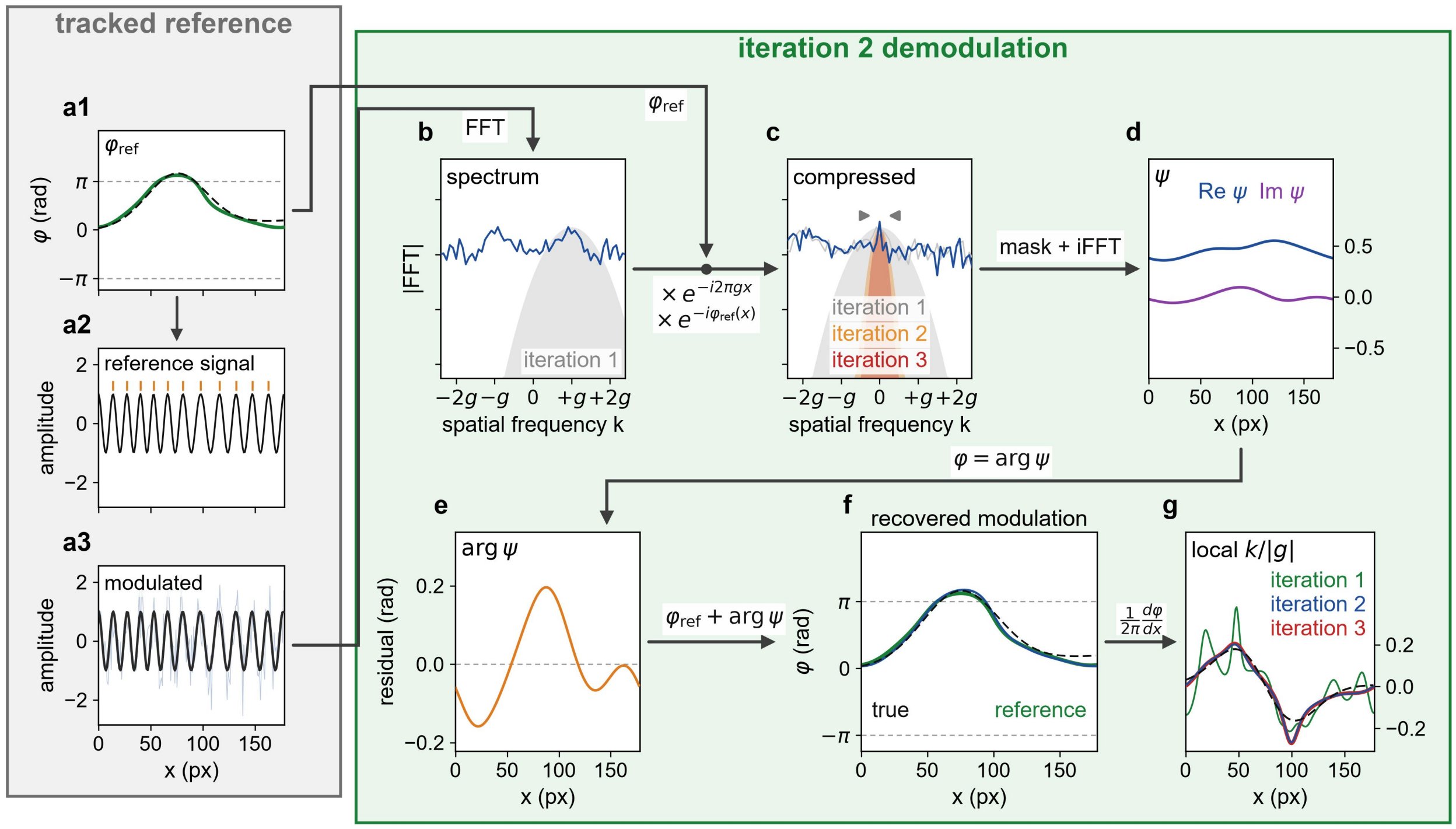

**Supplementary Fig. S13 | The tracked reference in one dimension.** The signal and noise of **Supplementary Fig. S11**. **a1–a3** The reference phase $\varphi_{\text{ref}}$, which is the low-pass-filtered recovery of iteration 1 (dashed, the true modulation), the reference signal $\cos(2\pi gx + \varphi_{\text{ref}})$ and the modulated signal. **b** The spectrum with the mask of iteration 1 (gray) for comparison. **c** The spectrum after multiplication by $\exp(-\mathrm{i}2\pi gx)$ and by $\exp(-\mathrm{i}\varphi_{\text{ref}})$. The frequency swing of the modulation is now in the reference, the residual is compressed to a narrow peak, and the mask can close from $0.35g$ to $0.12g$ and $0.09g$ in iterations 2 and 3. **d** The complex field $\psi$. **e** The residual phase $\arg\psi$. **f** The recovered modulation $\varphi_{\text{ref}} + \arg\psi$ against the truth. **g** The deviation of the local wave vector from $g$ after iterations 1–3, in which the noise falls as the mask closes while the modulation is preserved.

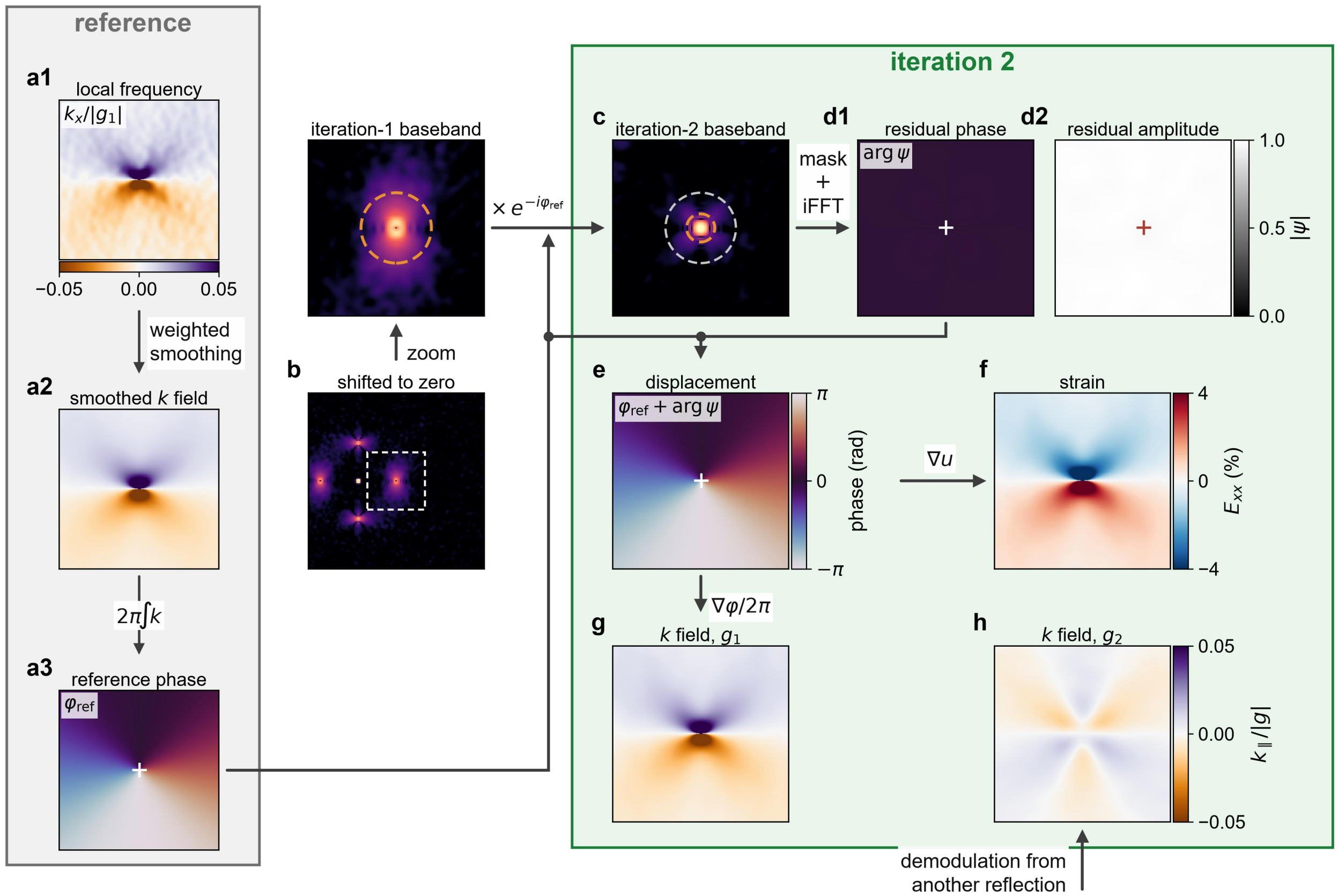

**Supplementary Fig. S14 | The tracked reference in two dimensions.** The dislocation of **Supplementary Fig. S12**. **a1–a3** The local wave vector $k_x/|\boldsymbol{g}_1|$ from iteration 1, the same field after amplitude-weighted smoothing of the complex field, which preserves the winding of the core, and its integral, the reference phase $\varphi_{\text{ref}}$. **b** The diffractogram shifted to zero with the mask of iteration 1, and the baseband enlarged. **c** The baseband after multiplication by $\exp(-i\varphi_{\text{ref}})$, which contracts so that the mask closes from 0.30 to $0.12|\boldsymbol{g}_1|$ (dashed circles). **d1, d2** The residual phase arg $\psi$, now flat, and the residual amplitude $|\psi|$, which is unity except at the core. **e** The displacement $\varphi_{\text{ref}}$ + arg $\psi$. **f** The strain $E_{xx}$. **g, h** The wave-vector fields of $\boldsymbol{g}_1$ and $\boldsymbol{g}_2$.

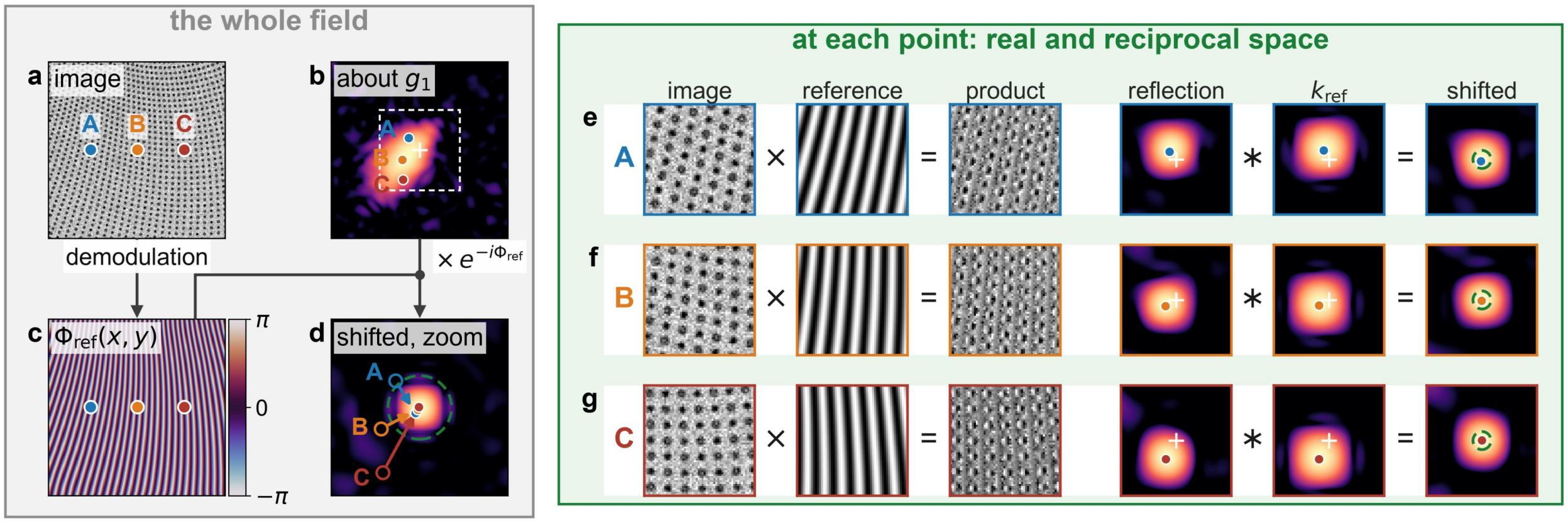

**Supplementary Fig. S15 | The reference phase as a function of position.** The rotating lattice of **Fig. 1a**. **a** Image with three points A, B and C marked. **b** The diffractogram about $\boldsymbol{g}_1$ with the final tracked wave vector of each point. **c** The reference phase $\Phi_{\rm ref}(x, y)$ after two iterations, whose fringes turn with the lattice. **d** The demodulated diffractogram, enlarged. The reflection of each point is brought to the origin by its own reference (arrows), inside the final aperture (dashed). **e–g** The three points. In real space, the image patch multiplied by the local reference wave cos $\Phi_{\rm ref}$ gives a product whose fringes have almost vanished. In reciprocal space, the reflection of the patch convolved with the point $\boldsymbol{k}_{\rm ref}$ of the reference gives the shifted reflection at the origin. The residual $|\boldsymbol{k} - \boldsymbol{k}_{\rm ref}|$ at the three points is at most $0.020|\boldsymbol{g}|$, whereas the three reflections are spread over $0.28|\boldsymbol{g}|$ in b.

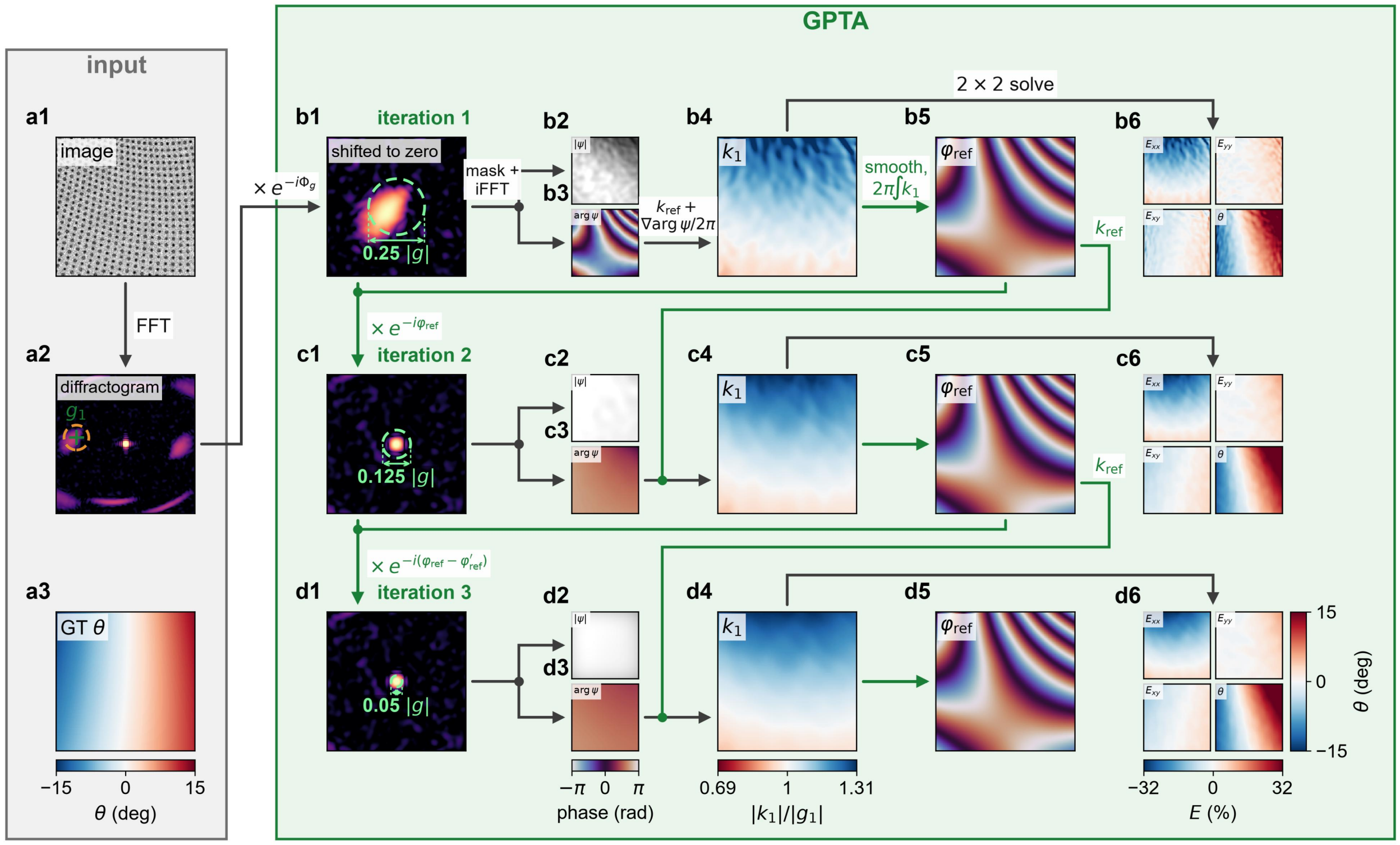

**Supplementary Fig. S16 | Geometric phase tracking analysis iteration by iteration.** The rotating lattice of **Fig. 1a**, analyzed with two basis reflections and apertures of 0.25, 0.125 and 0.05|$\boldsymbol{g}$|. **a1–a3** Image, its diffractogram with $\boldsymbol{g}_1$ and the aperture of iteration 1, and the ground-truth rotation $\theta$. **b–d** Iterations 1–3. The diffractogram of $I$ exp($-\mathrm{i}\Phi_{\mathrm{ref}}$) with the aperture of the iteration (b1, c1, d1), the amplitude $|\psi|$ and the phase arg $\psi$ after masking and inverse transformation (b2, b3), the wave-vector field $\boldsymbol{k}_1 = \boldsymbol{k}_{\mathrm{ref}} + \nabla \arg \psi/2\pi$ (b4), the reference $\varphi_{\mathrm{ref}}$ obtained by smoothing and integrating $\boldsymbol{k}_1$ (b5), which passes to the next iteration, and the output $E_{xx}$, $E_{yy}$, $E_{xy}$ and $\theta$ (b6), with the same layout in rows c and d. The reflection contracts to a point as the reference tracks the lattice, and the $\theta$ error falls from 0.94° to 0.30° and 0.25°.

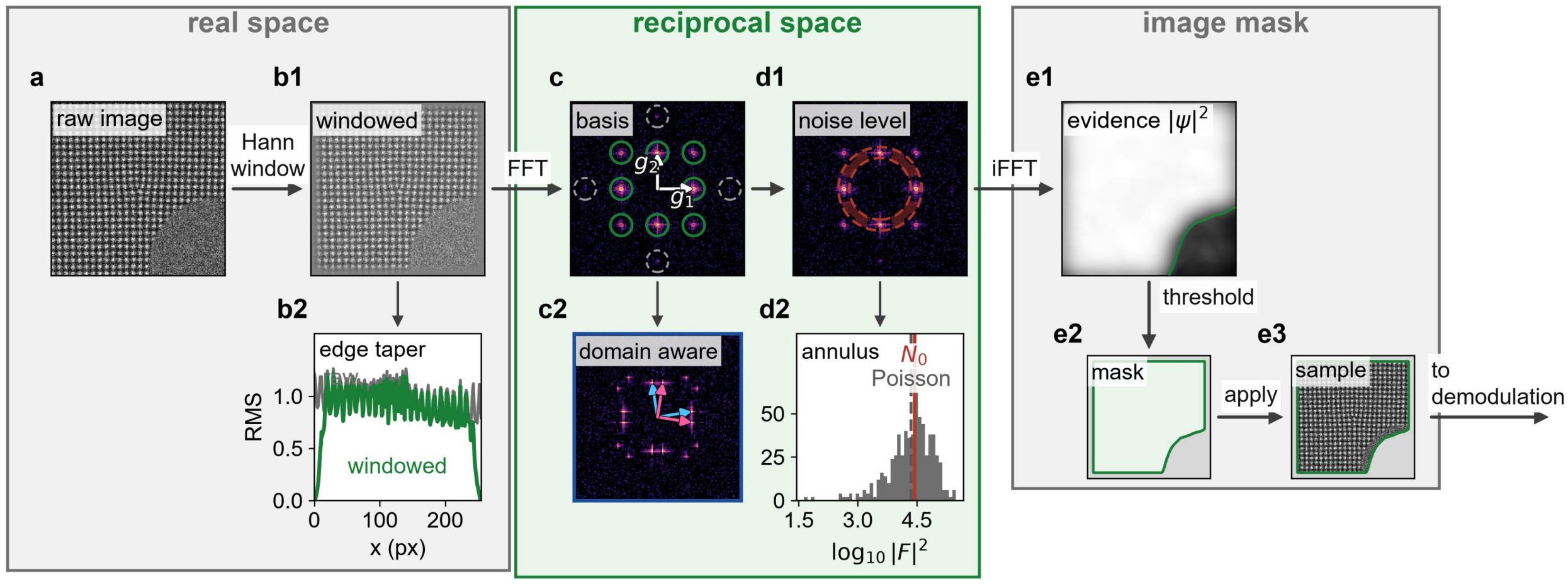

**Supplementary Fig. S17 | Pre-treatment, comprising window, basis, noise level and image mask.** A simulated edge dislocation at low dose with a vacuum corner filled with granular noise. **a** The raw image. **b1, b2** The image after plane detrending and a Hann taper, and the RMS profile at the edge before and after windowing. **c** The basis $\boldsymbol{g}_1$, $\boldsymbol{g}_2$ found from the diffractogram, with the higher orders ringed. **c2** The domain-aware case, two lattices tilted by ±9° whose reflections are found separately. **d1, d2** The noise level. The annulus at $|\boldsymbol{g}|$ from which every reflection is excluded, and the histogram of $|F|^2$ in the annulus, whose median is $N_0$, against the Poisson prediction. **e1–e3** The image mask. The evidence $|\psi|^2$, the mask after thresholding and morphological cleaning, and the mask applied to the image, which excludes the vacuum. With the mask, noise added in the excluded area leaves the measured $\boldsymbol{k}$ unchanged, compared with a change of $0.006|\boldsymbol{g}|$ without it.

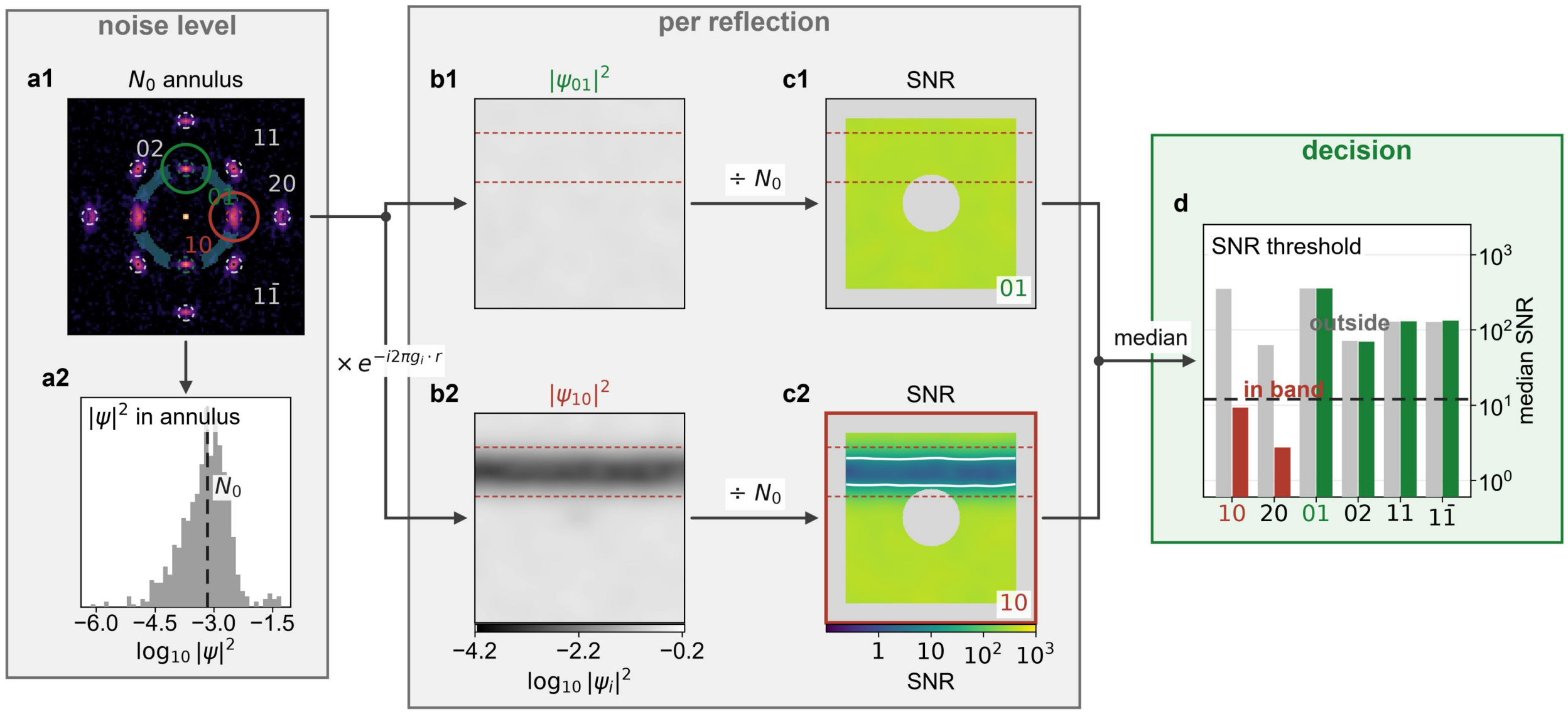

**Supplementary Fig. S18 | Noise level and SNR threshold.** A simulated edge dislocation at low dose with a contrast-reversal band on the (1,0) family, analyzed with six reflections. The band is drawn wider and the threshold higher here than the settings in Methods so that the decision is visible. **a1, a2** The $N_0$ annulus on the diffractogram and the histogram of $|\psi|^2$ in the annulus, whose median is $N_0$. **b1, b2** The demodulated power $|\psi_{01}|^2$ and $|\psi_{10}|^2$ (dashed lines bound the reversal band). **c1, c2** The corresponding SNR $|\psi_i|^2/N_0$ on a logarithmic scale. **d** The median SNR of each reflection outside (gray) and inside (colored) the reversal band against the threshold. Inside the band, the (1,0) and (2,0) reflections fall below the threshold while the other four stay far above it, and outside the band all six pass. At the software default threshold of 3, no reflection in this image is dropped, and the result of **Supplementary Fig. S19** is the same.

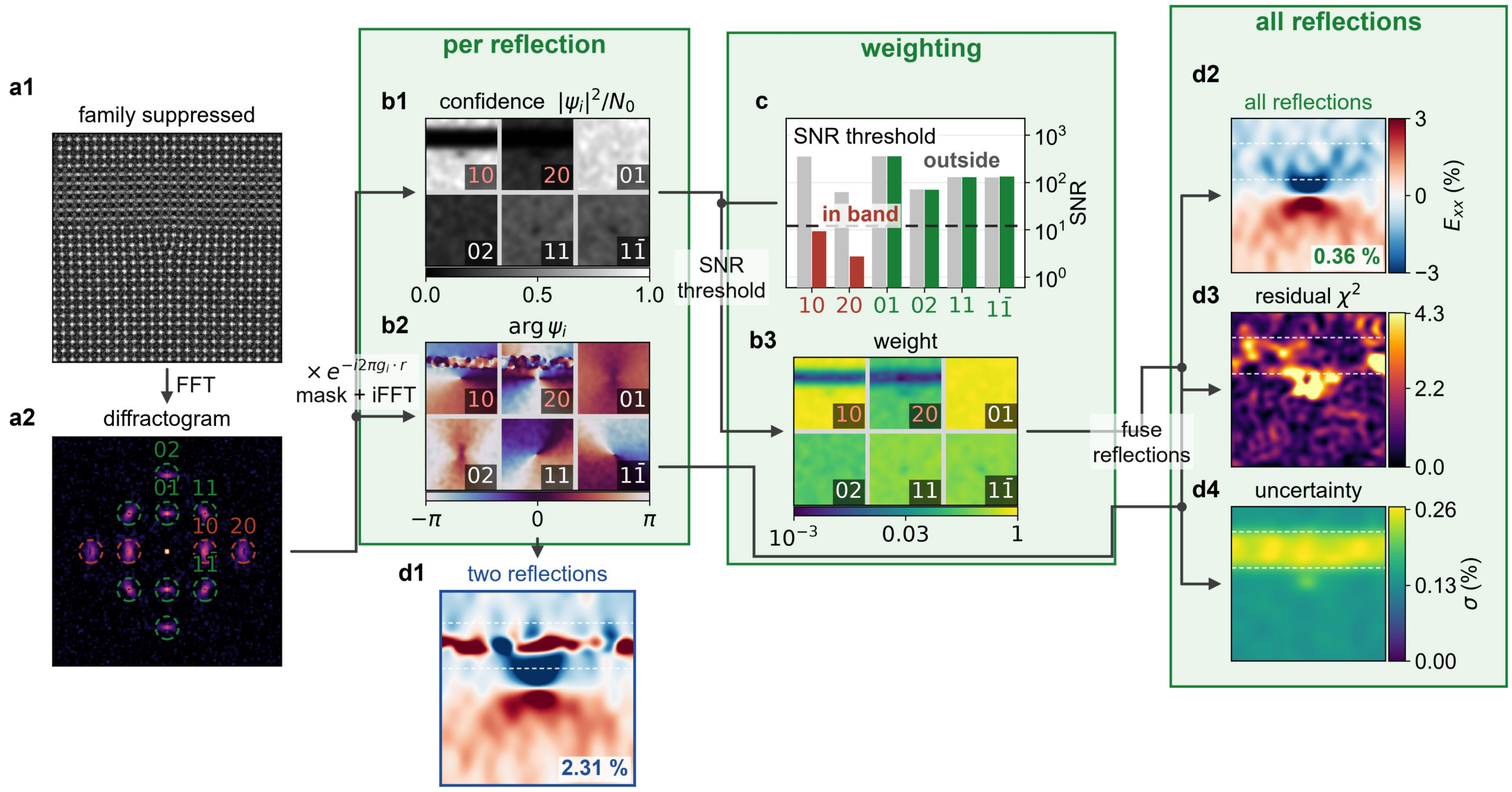

**Supplementary Fig. S19 | Confidence-weighted fusion.** Fusion of six reflections in the image of **Supplementary Fig. S18**, in which one fringe family is suppressed inside a reversal band. **a1, a2** The image and its diffractogram with the six reflections indexed. **b1, b2** The confidence $|\psi_i|^2/N_0$ and the phase arg $\psi_i$ of each reflection. The phase of (1,0) and (2,0) is noise inside the band. **c** The SNR threshold decision. **b3** The resulting weights. **d1, d2** $E_{xx}$ from the two basis reflections alone and from the weighted inversion over all reflections, with RMS errors inside the band of 2.31% and 0.36%. **d3** The inconsistency $\chi^2$ of the inversion, which rises inside the band where one lattice basis does not describe the field. **d4** The uncertainty of $E_{xx}$ from the covariance of the inversion.

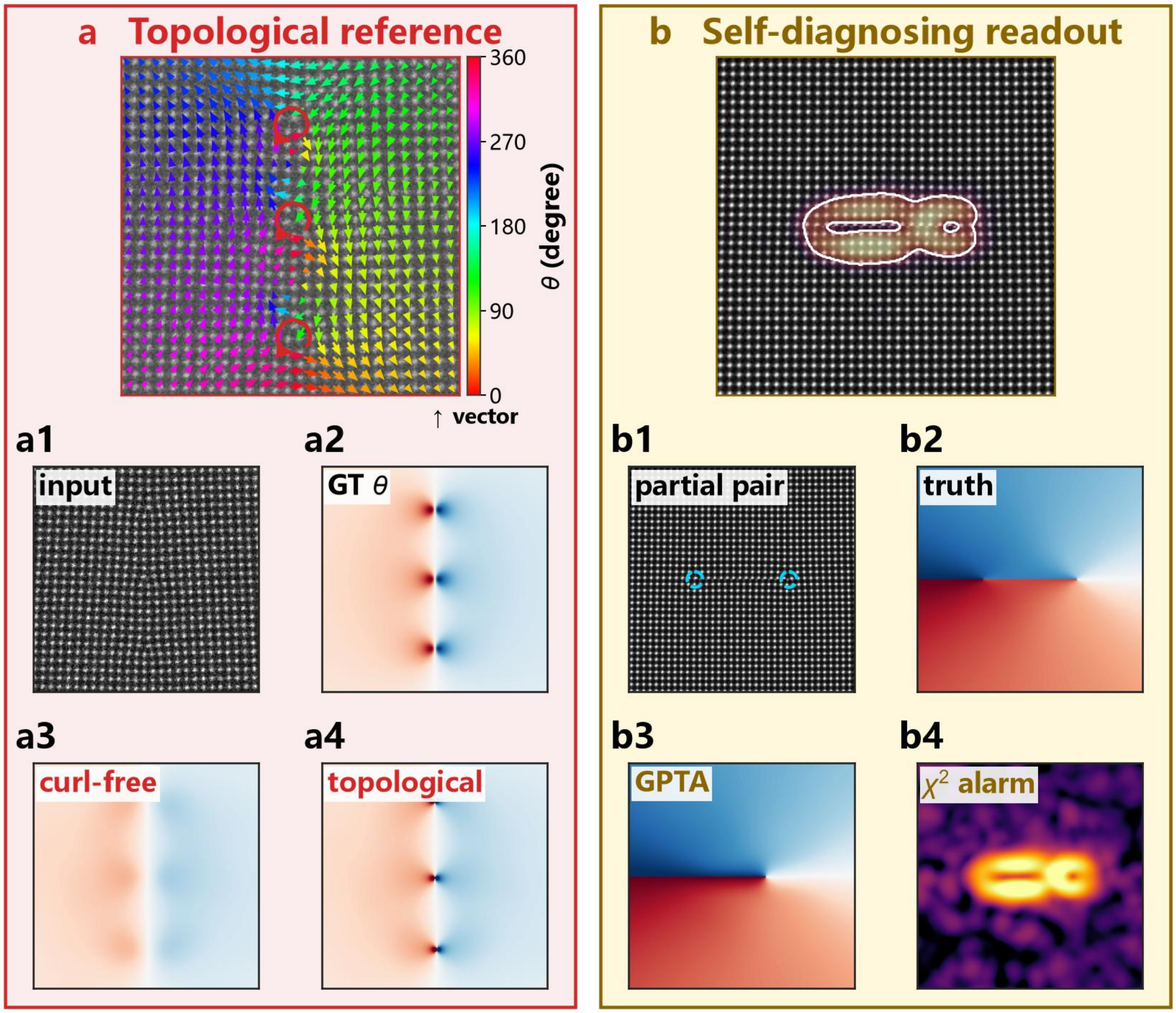

**Supplementary Fig. S20 | Topological reference and self-diagnosing readout.** Two outputs of the iteration that GPA does not provide, a reference that contains dislocation cores as phase vortices and the inconsistency $\chi^2$ of the joint inversion. **a** Topological reference. The measured local $\boldsymbol{k}$ vectors on a wall of three same-sign edge dislocations at 10 e/px, with the vortex cores marked. **a1–a4** Image, ground-truth $\theta$, and $\theta$ with the curl-free and with the topological reference, which places the cores at their true positions. **b** Self-diagnosing readout. The inconsistency $\chi^2$ over a dissociated dislocation. **b1–b4** The pair of partial dislocations, the ground truth, the map from GPTA and the $\chi^2$ map. The inconsistency rises on the ribbon between the partials, where one lattice basis does not describe the field and the displacement error is 0.286 px compared with 0.020 px in the far field. The color wheel indicates the direction of the $\boldsymbol{k}$ vectors in a.

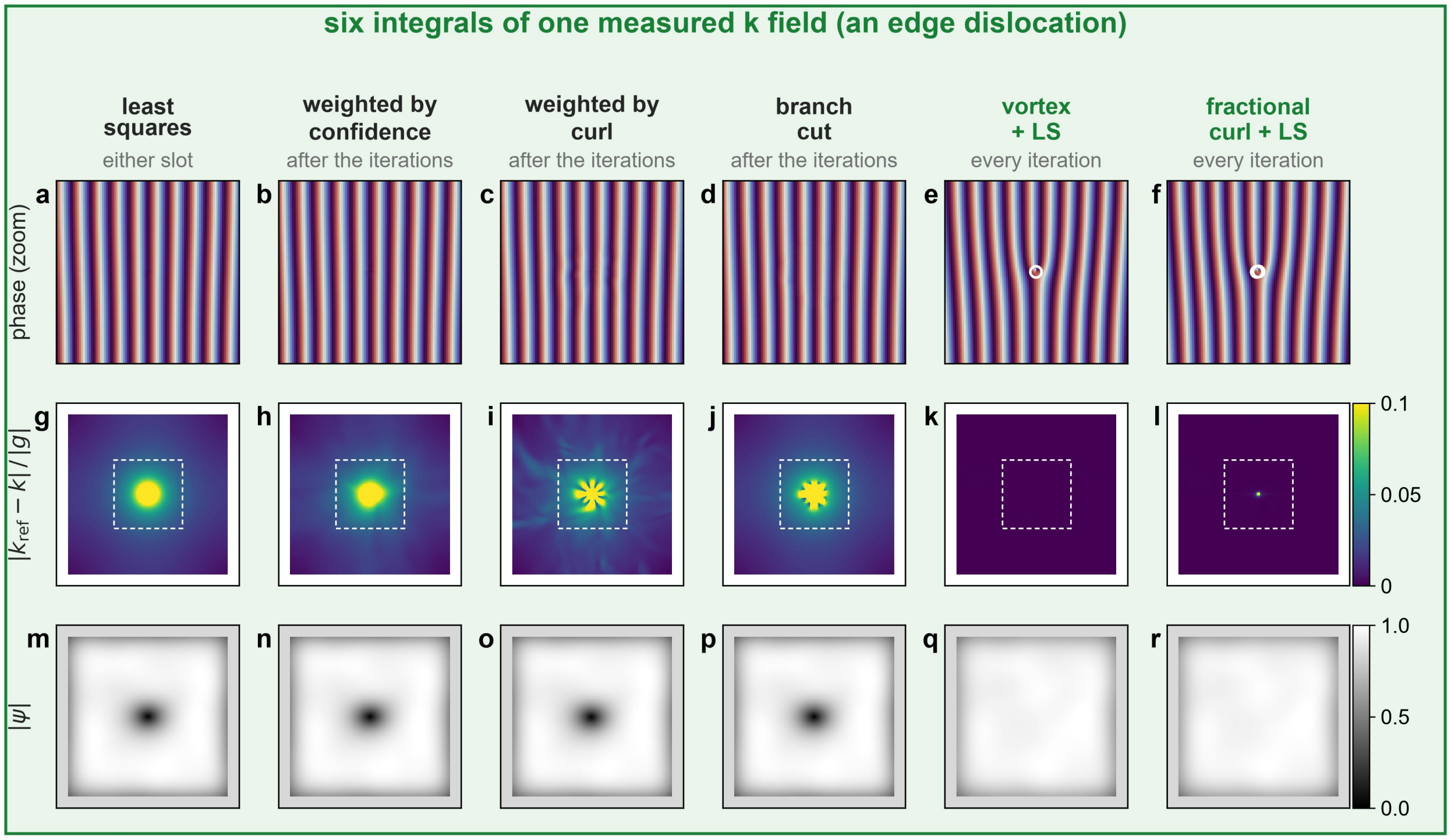

**Supplementary Fig. S21 | Six integrals of one measured wave-vector field.** The fused $\boldsymbol{k}_1$ field of an edge dislocation integrated into a reference phase by the six integrals available in the software. Columns show least squares, the confidence-weighted, curl-weighted and branch-cut integrals, which are applied once after the iteration, and the vortex and fractional-curl integrals, which are applied to the reference at every iteration. **a–f** The phase (zoom on the core). **g–l** The residual $|\boldsymbol{k}_{\mathrm{ref}} - \boldsymbol{k}|/|\boldsymbol{g}|$ (the dashed square indicates the zoom). **m–r** The amplitude $|\psi|$ after demodulation against that reference. The four single-valued integrals leave the winding of the core in the residual, and $|\psi|$ at the core stays below 0.02 of its far-field value. The two integrals that include the circulation reduce the residual to $0.0003|\boldsymbol{g}|$ and restore $|\psi|$ at the core to 0.96. The topological reference in the main text is the vortex integral.

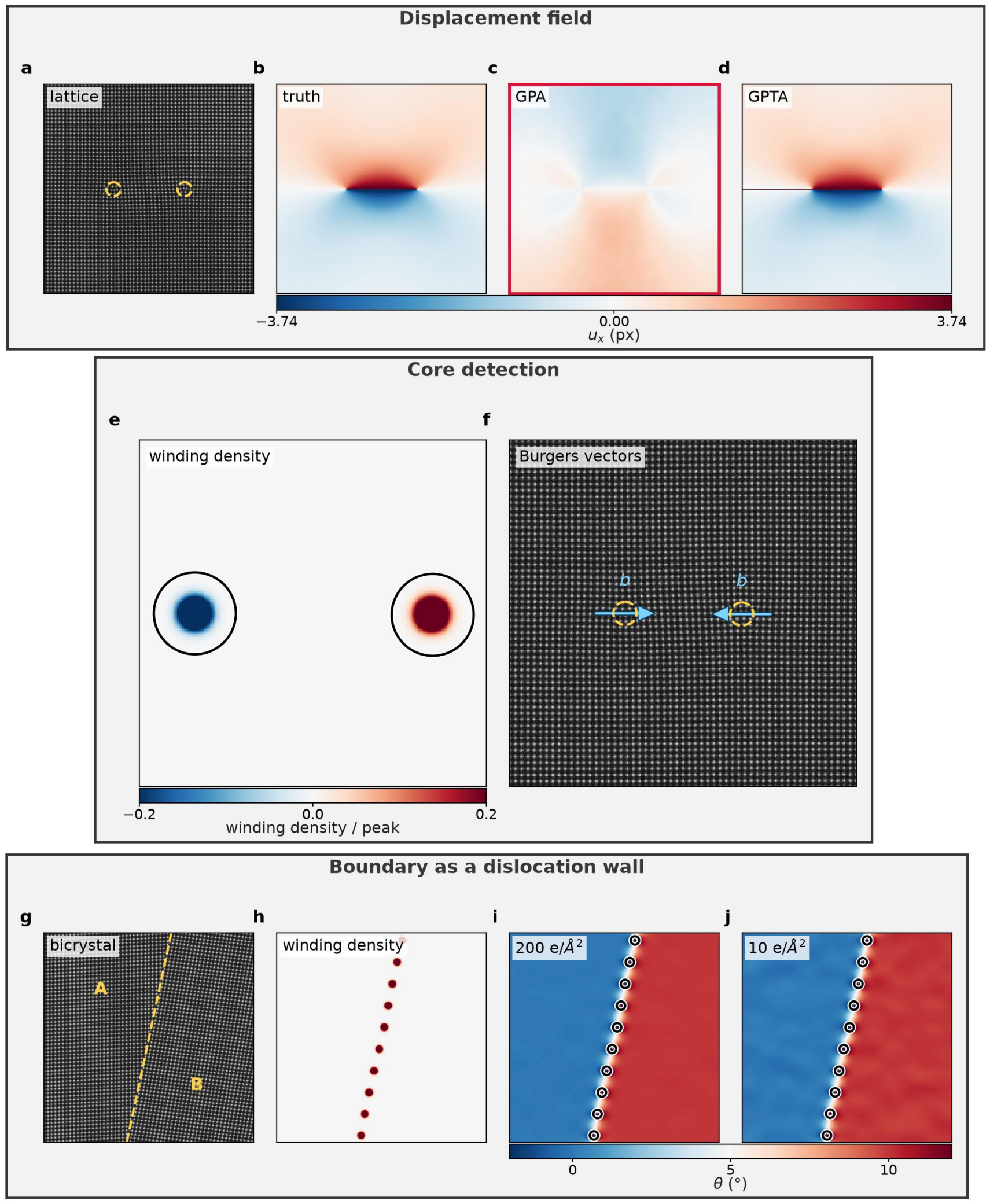

**Supplementary Fig. S22 | Displacement and dislocation measurements from the tracked reference. a–d** An analytic dislocation dipole at 50 e/Å$^2$. The lattice (a), the true displacement $u_x$ (b), $u_x$ from GPA, which unwraps the single-valued phase of each reflection (c, RMS error 1.43 px), and $u_x$ from GPTA (d, RMS error 0.031 px). **e, f** The winding density, which gives −1.00 and +1.00 at the two cores, and the recorded Burgers vectors, which have a magnitude of one lattice vector with the core positions off by at most 0.85 px. **g–j** An analytic bicrystal with a misorientation of 10° (g), its winding density (h) and $\theta$ with the recorded cores at 200 e/Å$^2$ (i) and 10 e/Å$^2$ (j). Ten dislocations are recorded at both doses, consistent with the 10.34 predicted by Frank's rule, and the measured $\theta$ steps are 10.001° and 9.990°.

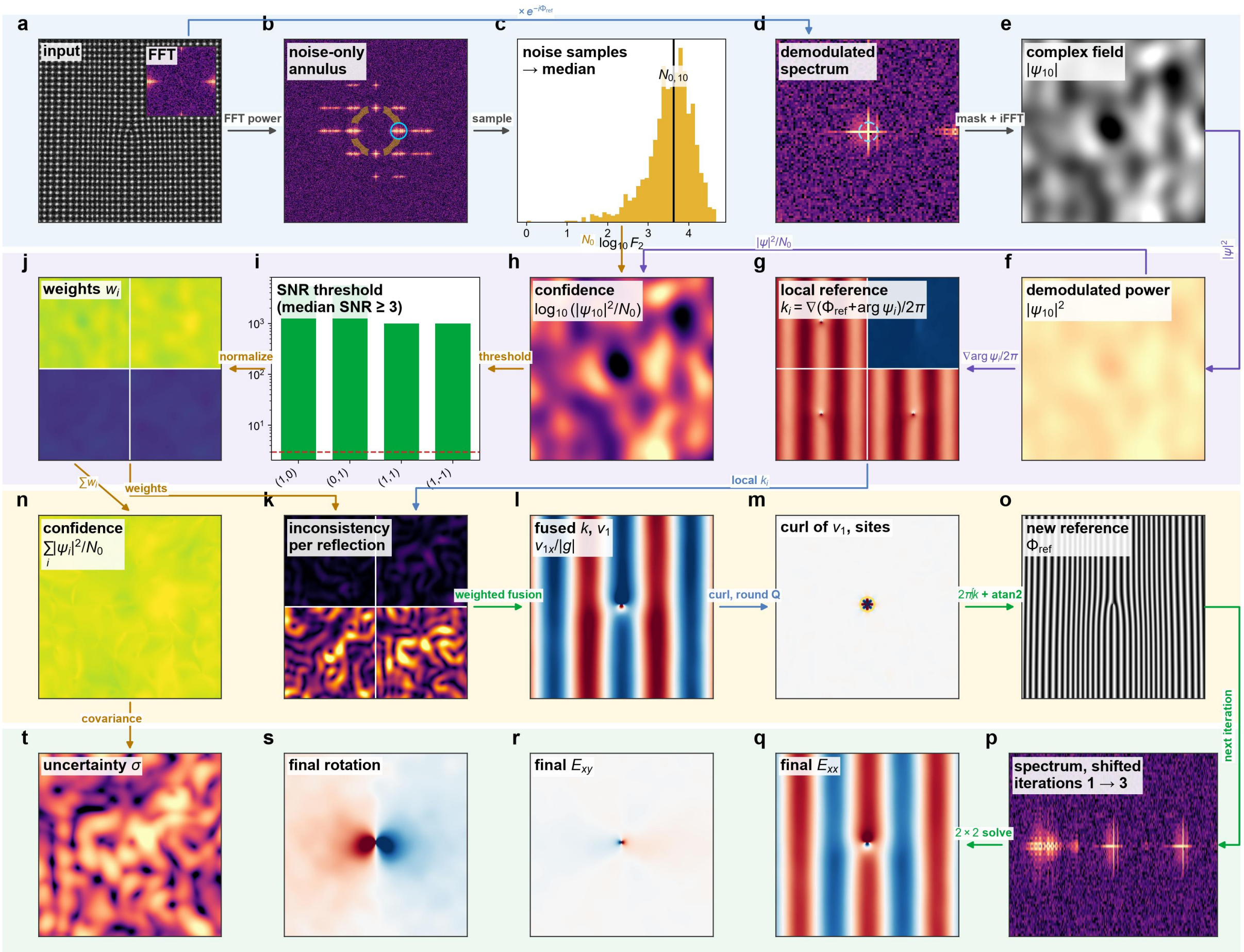

**Supplementary Fig. S23 | Complete geometric phase tracking analysis workflow.** A simulated edge dislocation with smooth strain at 20 e/px, analyzed with four reflections. **a** Input image with its transform inset. **b, c** The noise-only annulus, and the noise samples with their median $N_0$. **d** The demodulated spectrum. **e, f** The modulus $|\psi_{10}|$ of the demodulated field and its power $|\psi_{10}|^2$. **g** The local wave vector $\boldsymbol{k}_i = \nabla(\Phi_{\mathrm{ref}} + \arg\psi_i)/2\pi$ of the four reflections. **h** The confidence $\log_{10}(|\psi_{10}|^2/N_0)$. **i** The SNR threshold. **j** The weights $w_i$. **k** The inconsistency of each reflection with the fused basis. **l** The fused $\boldsymbol{k}$ of $\boldsymbol{v}_1$. **m** The curl of $\boldsymbol{v}_1$, which marks the site of the dislocation. **n** The total confidence $\Sigma|\psi_i|^2/N_0$. **o** The new reference $\Phi_{\mathrm{ref}}$, rebuilt with the dislocation as a phase vortex. **p** The demodulated spectrum over iterations 1–3. **q–s** The final $E_{xx}$, $E_{xy}$ and rotation. **t** The uncertainty from the covariance of the weighted solve. All four reflections pass the SNR threshold, and one vortex of charge −1 is detected at the core.

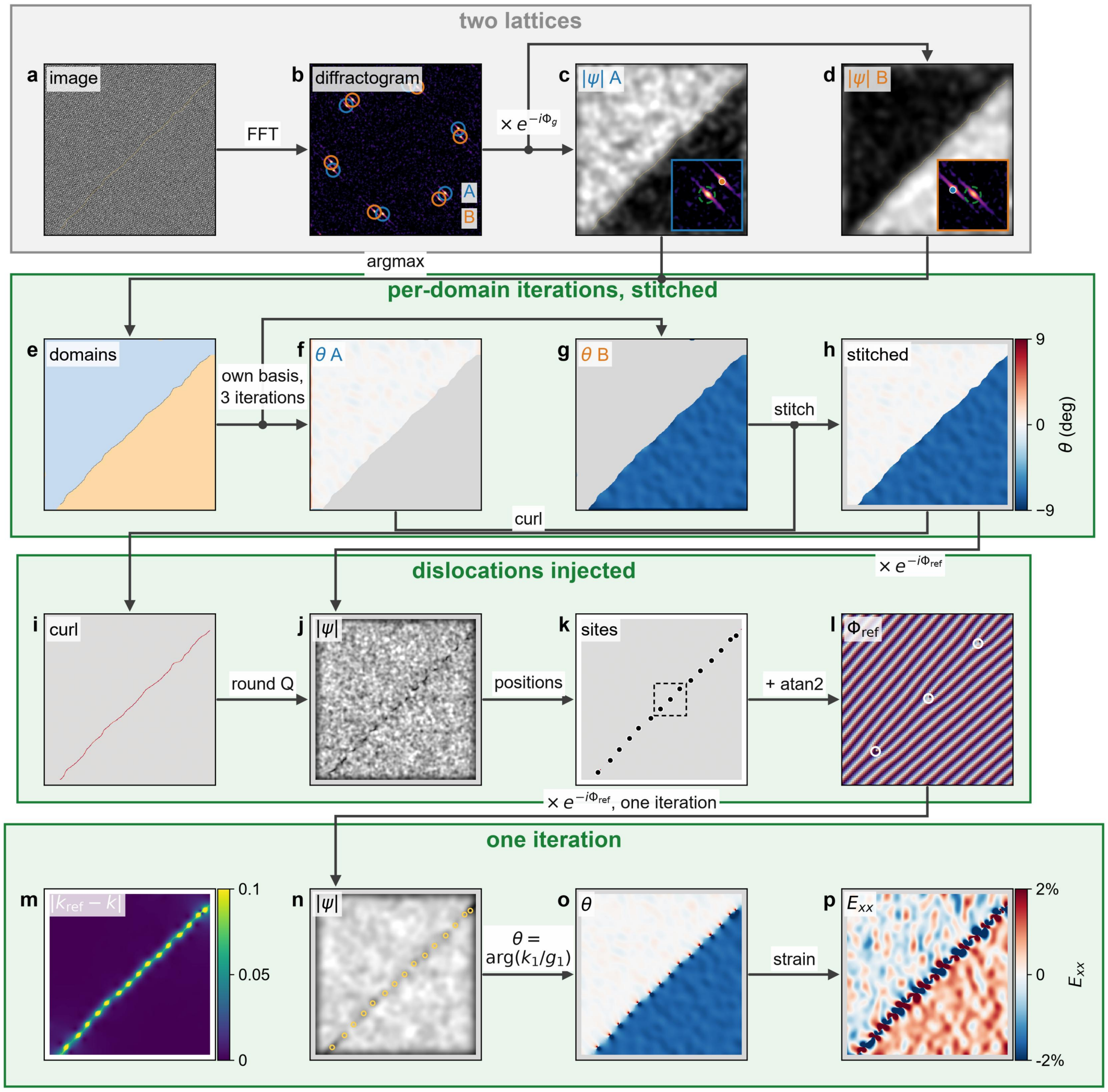

**Supplementary Fig. S24 | Domain awareness, with per-domain iterations, stitching and injected dislocations.** A benchmark bicrystal at 50 e/px with a misorientation of 7.2°. **a, b** Image and its diffractogram with the reflections of lattice A (blue) and lattice B (orange). **c, d** The amplitude $|\psi|$ of the reflections of each lattice, which is high on its own side of the boundary. **e** The domains, with each pixel assigned to the lattice with the highest confidence. **f, g** $\theta$ from three iterations in each domain with its own basis. **h** The stitched map. **i** The curl of the stitched wave-vector field, which localizes the boundary. **j** $|\psi|$ from one demodulation against the stitched reference, whose holes mark the cores. **k** The dislocation sites, an even comb along the boundary whose spacing follows Frank's rule (41.4 px compared with 42.5 px expected). **l** The reference $\Phi_{\rm ref}$ with the sites added as phase vortices. **m–p** One further iteration. The residual $|\boldsymbol{k}_{\rm ref} - \boldsymbol{k}|$, which is small everywhere except at the cores (m), $|\psi|$, whose holes on the boundary shrink (n), $\theta$ (o) and $E_{xx}$ (p), in which the cores appear as the singular fields of edge dislocations.

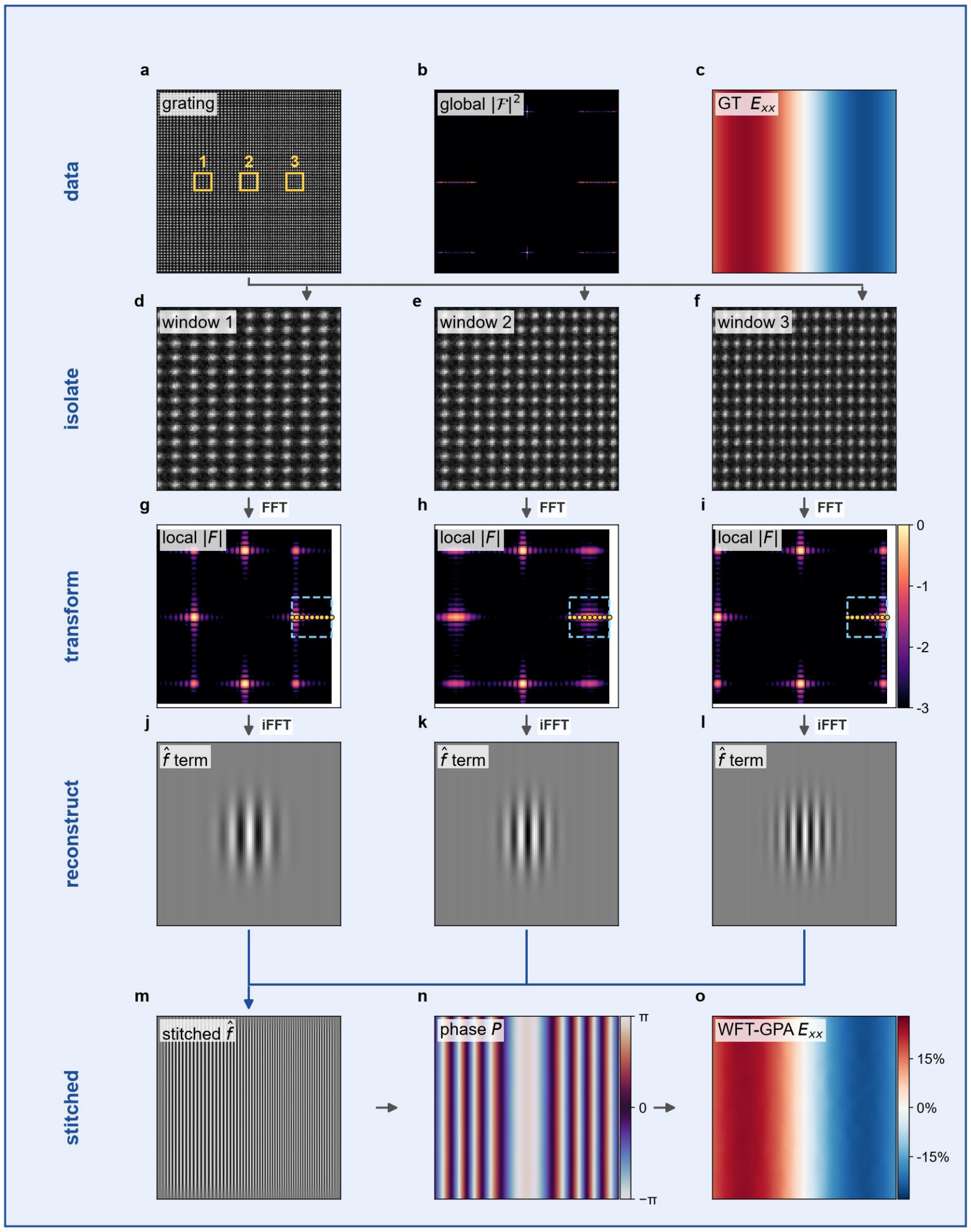

**Supplementary Fig. S25 | Windowed geometric phase analysis.** Windowed geometric phase analysis (S-GPA, Zhang et al., Ultramicroscopy 171 (2016) 34) replaces the single global transform of GPA with a bank of local transforms. The simulated HRTEM image contains a sinusoidal strain grating with a peak strain of 25%. **a** The grating with three window positions marked. **b** The global transform. **c** Ground-truth $E_{xx}$. **d–f** The three windowed blocks, where the local strain is +25, 0 and −25%. **g–i** Their transforms, each of which follows its own local frequency. **j–l** The corresponding filtered terms. **m** The inverse windowed transform, which stitches every windowed contribution into one field. **n** The resulting phase. **o** $E_{xx}$ from S-GPA (labeled WFT-GPA in the figure), with an RMS error of 0.233%.

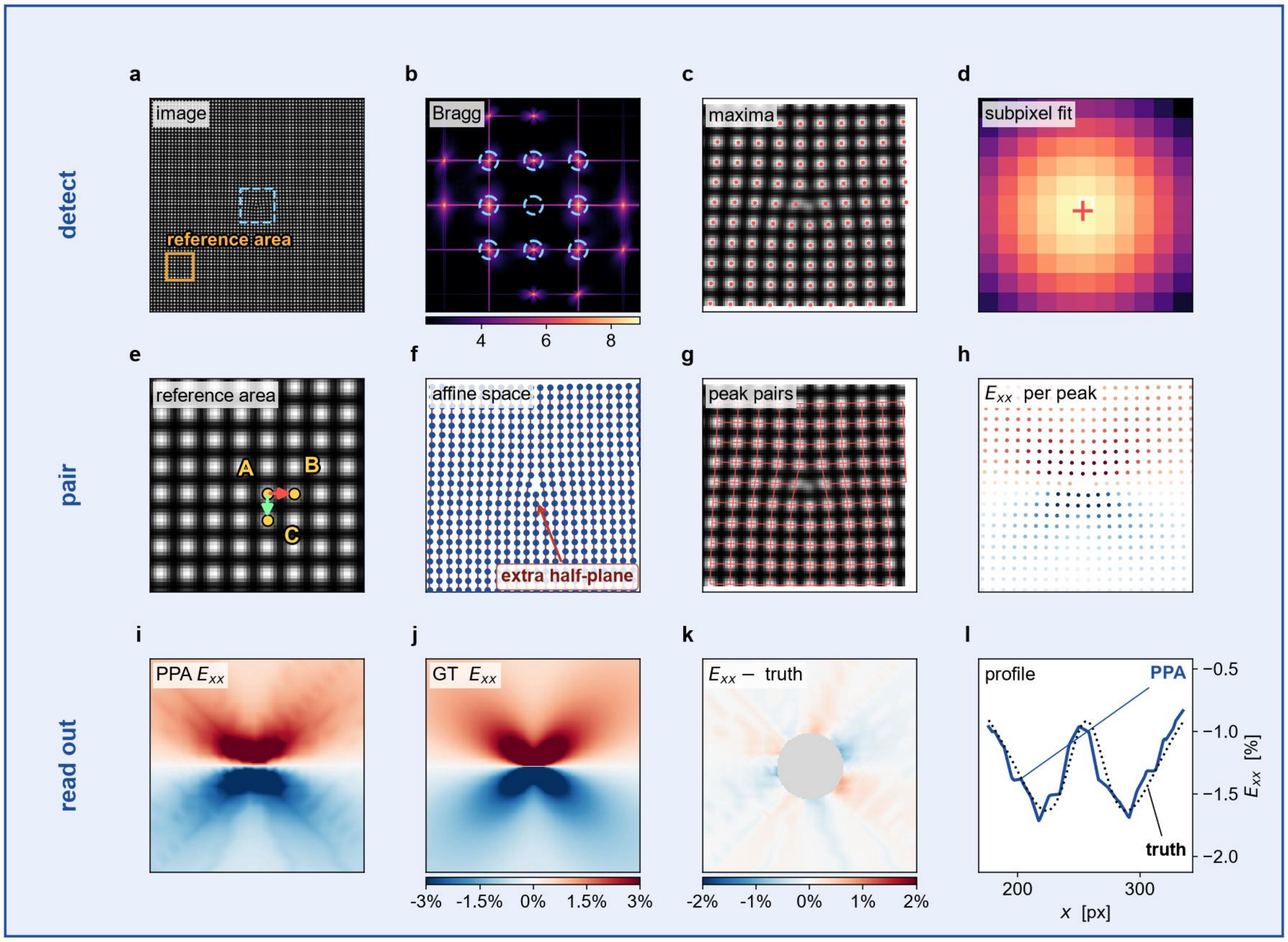

**Supplementary Fig. S26 | Peak-pair analysis.** Peak-pair analysis (PPA, Galindo et al., Ultramicroscopy 107 (2007) 1186) is a real-space method, shown here on a noiseless simulated edge dislocation. **a** The image. **b** The Bragg filter with the reflections it uses circled. **c** The local maxima of the filtered image. **d** The quadratic sub-pixel fit. **e** The reference area far from the core, whose three points A, B and C give the basis vectors ***a*** and ***b***. **f** The affine space in which a perfect lattice is an integer square grid and the dislocation appears as an extra half-plane. **g** The pairs. **h** The strain per peak, obtained from each pair vector. **i** The interpolated $E_{xx}$. **j** Ground truth. **k** Their difference, with an RMS of 0.130% away from the core. **l** A profile through the core.

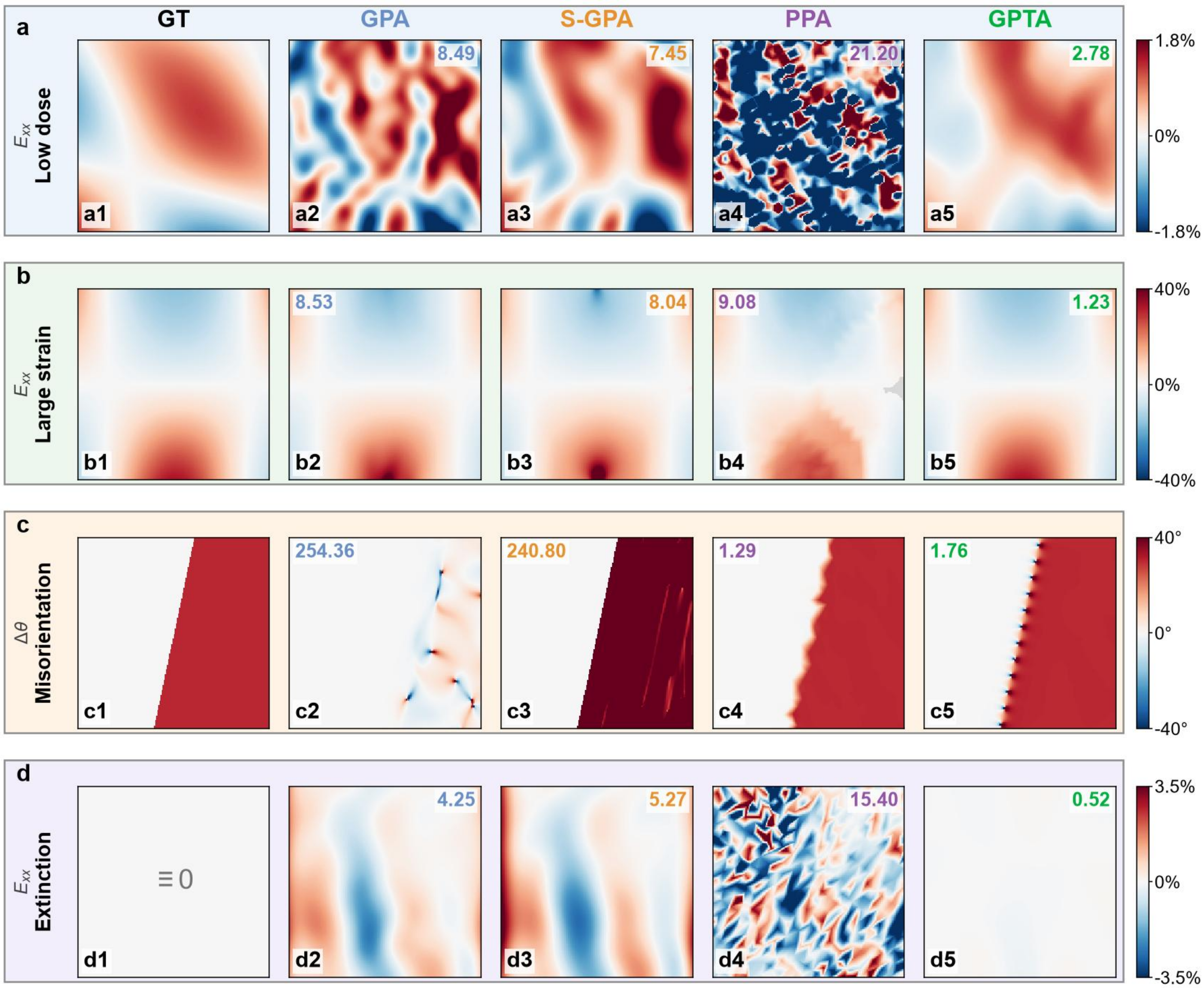

**Supplementary Fig. S27 | Ground truth and four methods on one image of each regime of Fig. 5.** One additional image from each regime of **Fig. 5**, analyzed with the same reference basis by GPA, S-GPA, PPA and GPTA. Each row shows the ground truth followed by the four methods, and each map is labeled with its $k_{50}$ in units of $10^{-3}|\boldsymbol{g}|$. **a1–a5** Low dose (2 e/px), $E_{xx}$. **b1–b5** Large smooth strain with a peak of 25%, $E_{xx}$. **c1–c5** A 30° bicrystal, $\theta$. PPA is the most accurate method in this case. **d1–d5** Reflection extinction in a bent foil with zero true strain, $E_{xx}$. GPTA uses its single-reference analysis in a, b and d and its domain-aware algorithm in c.

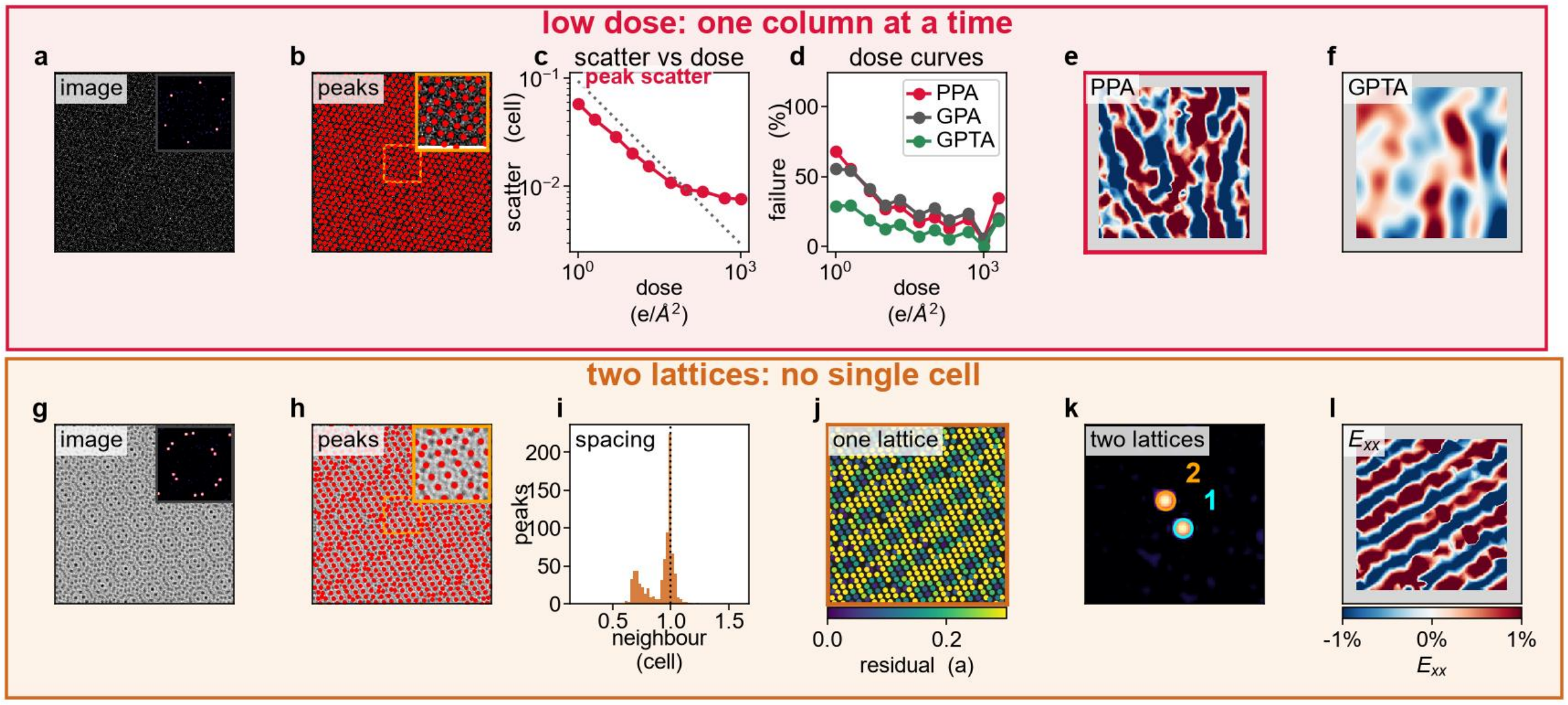

**Supplementary Fig. S28 | Two limits of peak-pair analysis, low dose and two lattices. a–f** Low dose. **a** Image at 1 e/Å$^2$ with its diffractogram inset. **b** The peaks detected by PPA, which finds almost every column. **c** The scatter of the peaks about the best single lattice against dose, which grows sixfold from 100 to 1 e/Å$^2$. **d** The failure rate over the image set against dose for PPA, GPA and GPTA. **e, f** $E_{xx}$ from PPA and GPTA, with bad areas of 0.24 and 0.00. **g–l** Two superimposed lattices. **g** Image, whose diffractogram (inset) contains two sets of reflections. **h** The detected peaks, which belong to both lattices. **i** The histogram of the nearest-neighbor spacing of the peaks, which splits into two populations. **j** The residual of a fit of all peaks to one lattice. No single cell fits the peaks. **k** The reflections of the two lattices in the diffractogram. **l** $E_{xx}$ from PPA with one reference cell, which turns the beating of the two lattices into stripes of false strain.

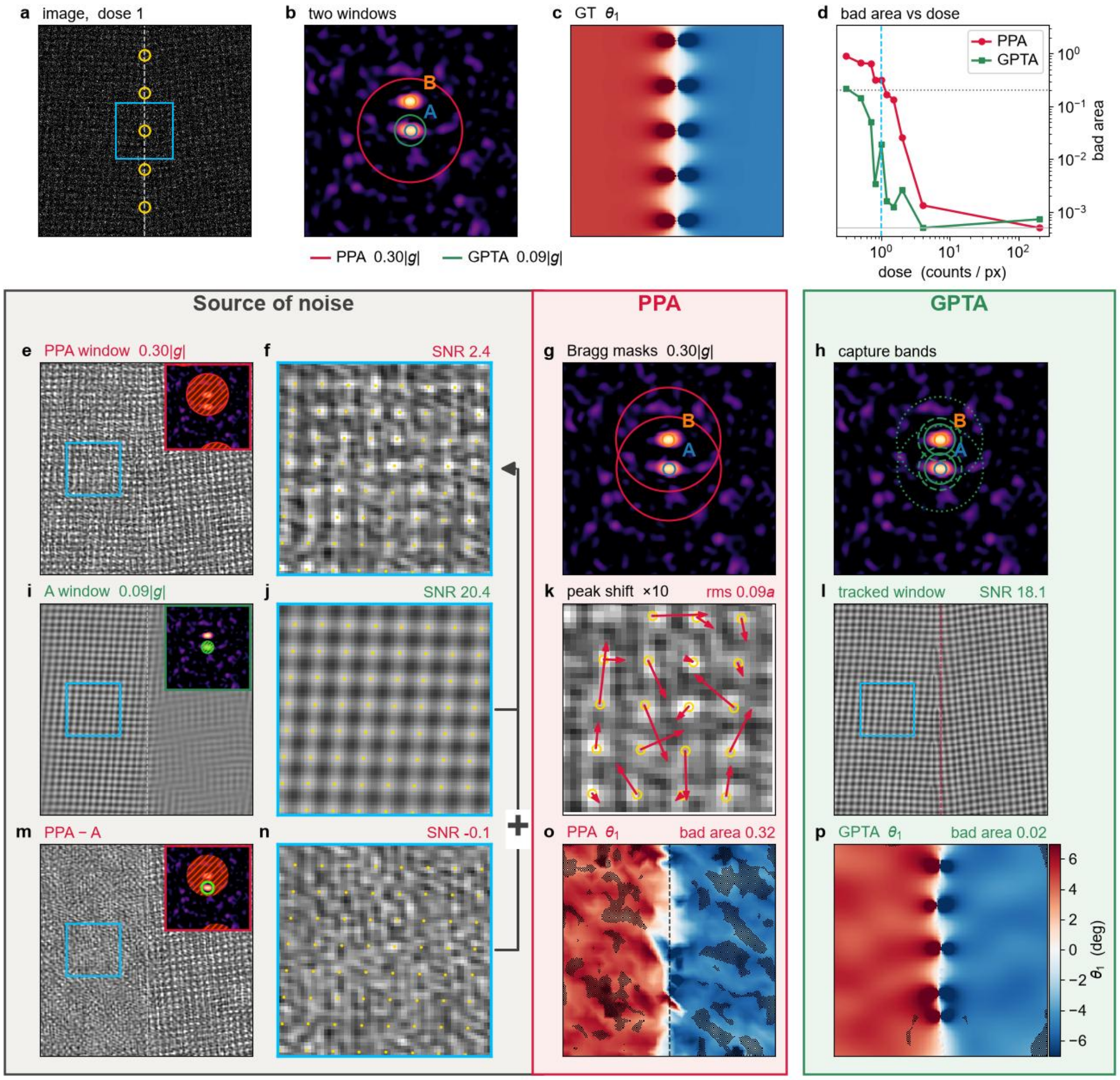

**Supplementary Fig. S29 | Peak-pair and tracking analyses on a low-dose tilt wall.** A simulated HRTEM image of a bicrystal with a misorientation of 9.68° and a wall of five edge dislocations, at a dose of 1 count per pixel. **a** The image. **b** The reciprocal-space windows of PPA (0.30|***g***|) and GPTA (0.09|***g***|) on the split reflections A and B. **c** Ground-truth $\theta_1$. **d** Bad area against dose. **e–n** The same image seen through three reciprocal-space windows, in full-field and zoomed views, namely the union of masks used by PPA (e, f, SNR 2.4), the final aperture of GPTA (i, j, SNR 20.4) and their difference (m, n). **g, h** The masks of PPA and the apertures of GPTA on both grains. **k** The peak displacement, magnified tenfold. **l** The window following the tracked reference at every pixel. **o** $\theta_1$ from PPA, bad area 0.32. **p** $\theta_1$ from GPTA with the domain-aware algorithm, bad area 0.02. The advantage is confined to low dose. At 200 counts per pixel, PPA reaches a bad area of 0.000 and is the more precise of the two.

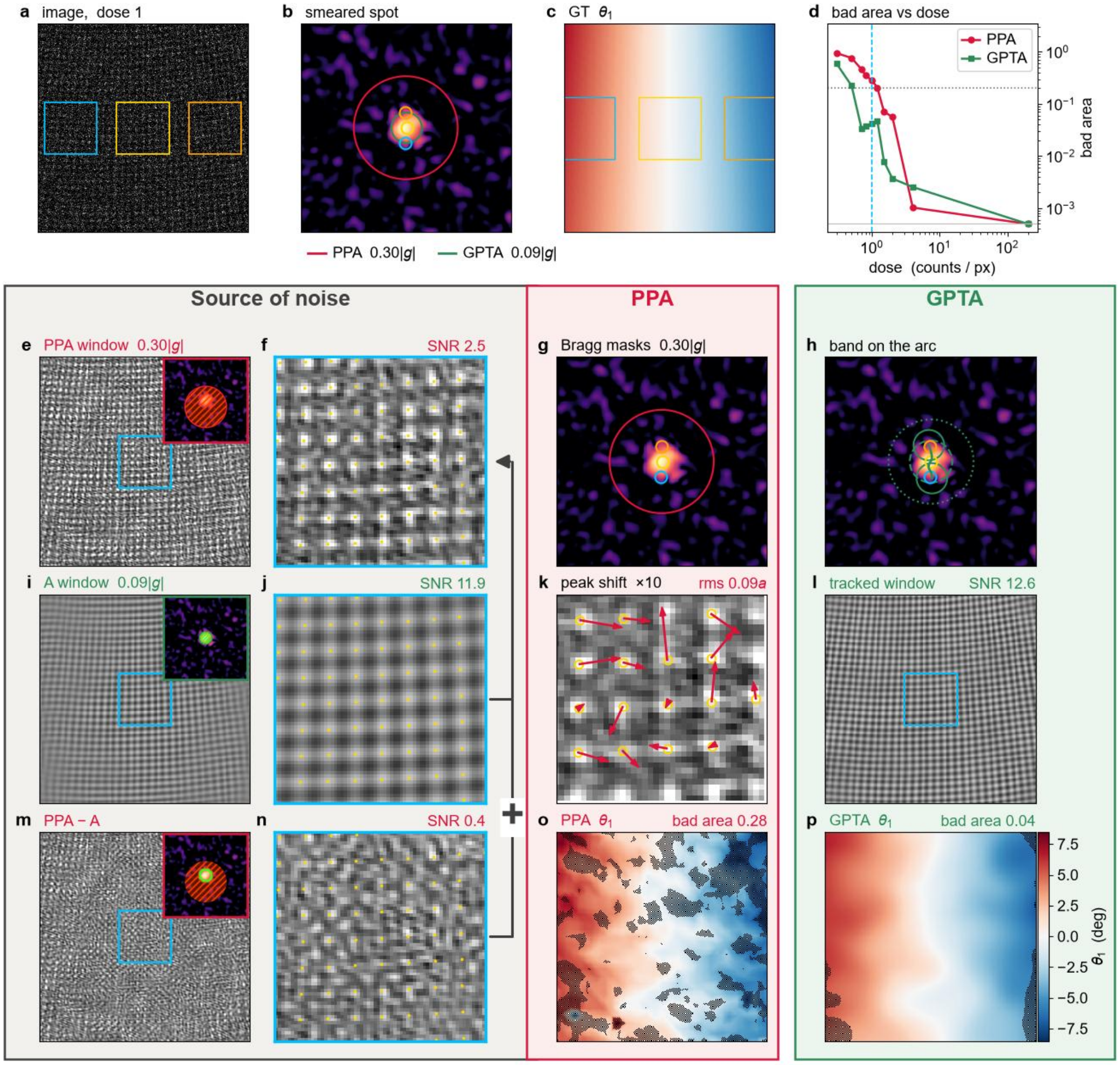

**Supplementary Fig. S30 | Peak-pair and tracking analyses on a low-dose lattice gradient.** A simulated HRTEM image of a circular bend, in which the lattice turns by 15° across the field without a boundary or a dislocation, at a dose of 1 count per pixel. The reflection is smeared over a patch 4.8 aperture widths wide. **a** The image. **b** The smeared reflection. **c** Ground-truth $\theta_1$. **d** Bad area against dose. **e–n** The window decomposition of **Supplementary Fig. S29**, namely the fixed masks of PPA (e, f, SNR 2.5), the tracked aperture of GPTA (i, j, SNR 11.9) and their difference (m, n). **g, h** The masks of PPA and the aperture of GPTA on the arc. **k** The peak displacement, magnified tenfold. **l** The tracked window. **o** $\theta_1$ from PPA, bad area 0.28. **p** $\theta_1$ from GPTA, bad area 0.04. At 200 counts per pixel, both methods reach a bad area of 0.000, so the two differ only at low dose.

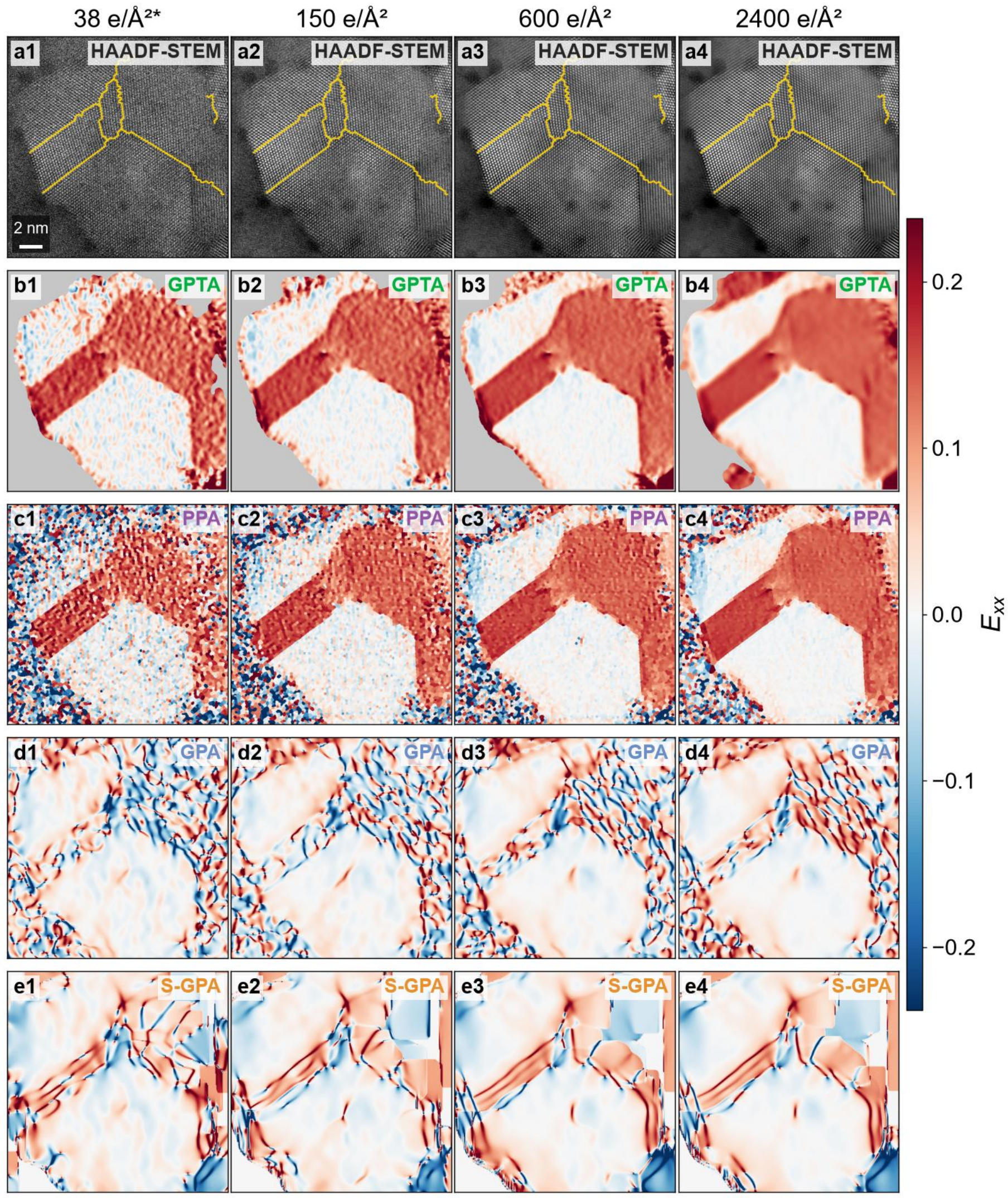

**Supplementary Fig. S31 | Pt polycrystal over a dose series.** HAADF-STEM images of a Pt [110] polycrystal with six grains. Columns show frame averages of 0.25, 1, 4 and 16 frames, headed by their dose. The 0.25-frame column (asterisk) is a calibrated Poisson injection and not measured data. Rows show the images with the grain boundaries from the 40-frame average overlaid (a1–a4), and the $E_{xx}$ maps from GPTA (b1–b4), classic PPA (c1–c4), GPA at $\sigma = 0.10|\boldsymbol{g}|$ (d1–d4) and S-GPA (e1–e4). The map resolution of GPTA is set for each dose as described in Supplementary Note 12.3. The color scale is wide enough to include the misoriented grains, which appear as large uniform strain under the single global re-indexing used for all methods. That color represents orientation and not strain. Scale bar, 2 nm.

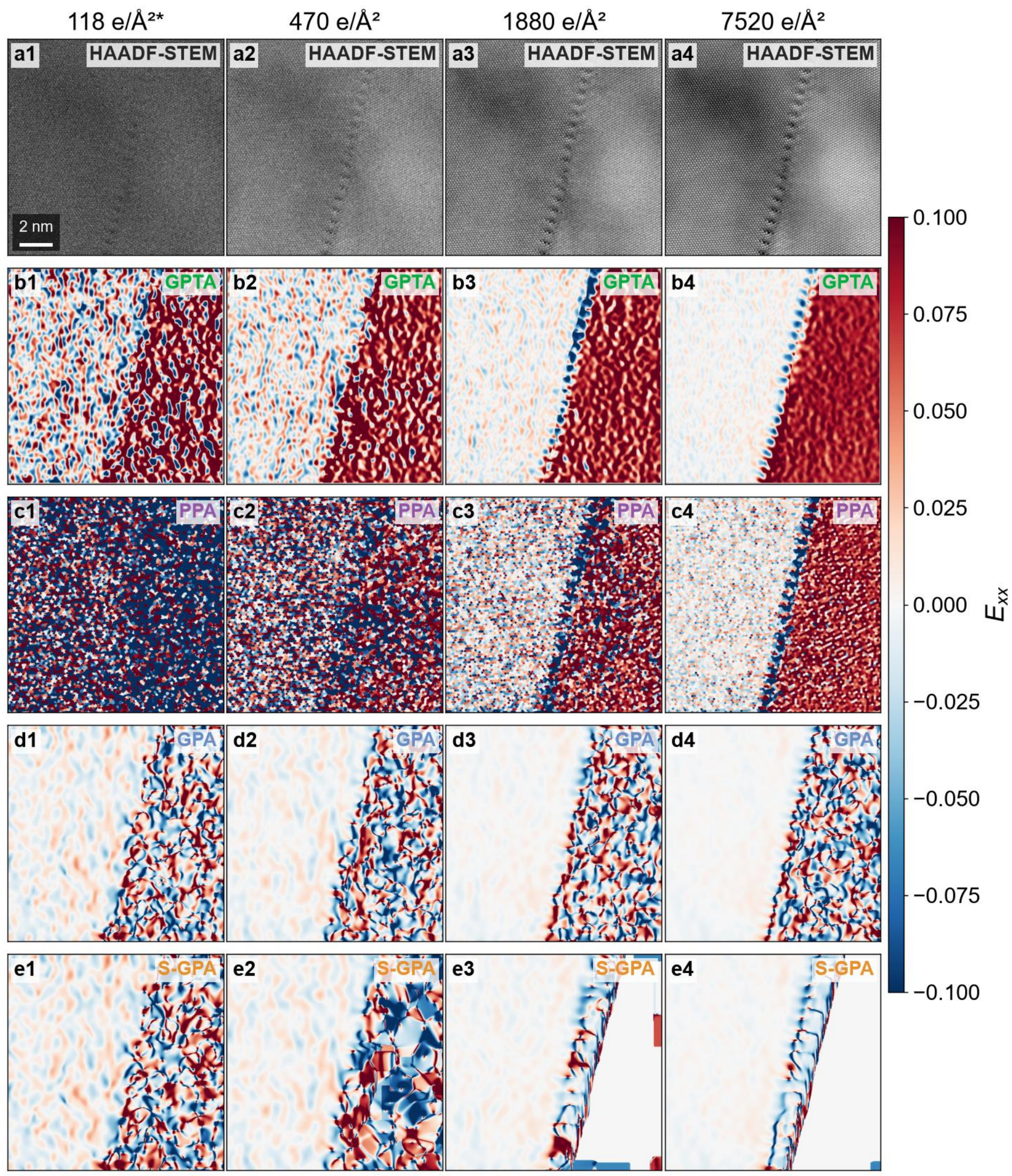

**Supplementary Fig. S32 | Pt bicrystal over a dose series.** HAADF-STEM images of a Pt [111] bicrystal with a 23.5° tilt boundary. Columns show frame averages of 0.25, 1, 4 and 16 frames, headed by their dose, and the 0.25-frame column (asterisk) is a calibrated Poisson injection. **a1–a4** Images. **b1–b4** $E_{xx}$ from GPTA. **c1–c4** PPA with the classic construction, used here because the updated peak-finding step assigns the two grains to one domain (Supplementary Note 12.5). **d1–d4** GPA at $\sigma = 0.10|\boldsymbol{g}|$. **e1–e4** S-GPA. At high dose, PPA resolves the boundary as sharply as GPTA, so the comparison concerns the low-dose frame averages. The misoriented grain saturates the color scale because the display uses one global re-indexing and a linearized strain. Scale bar, 2 nm.

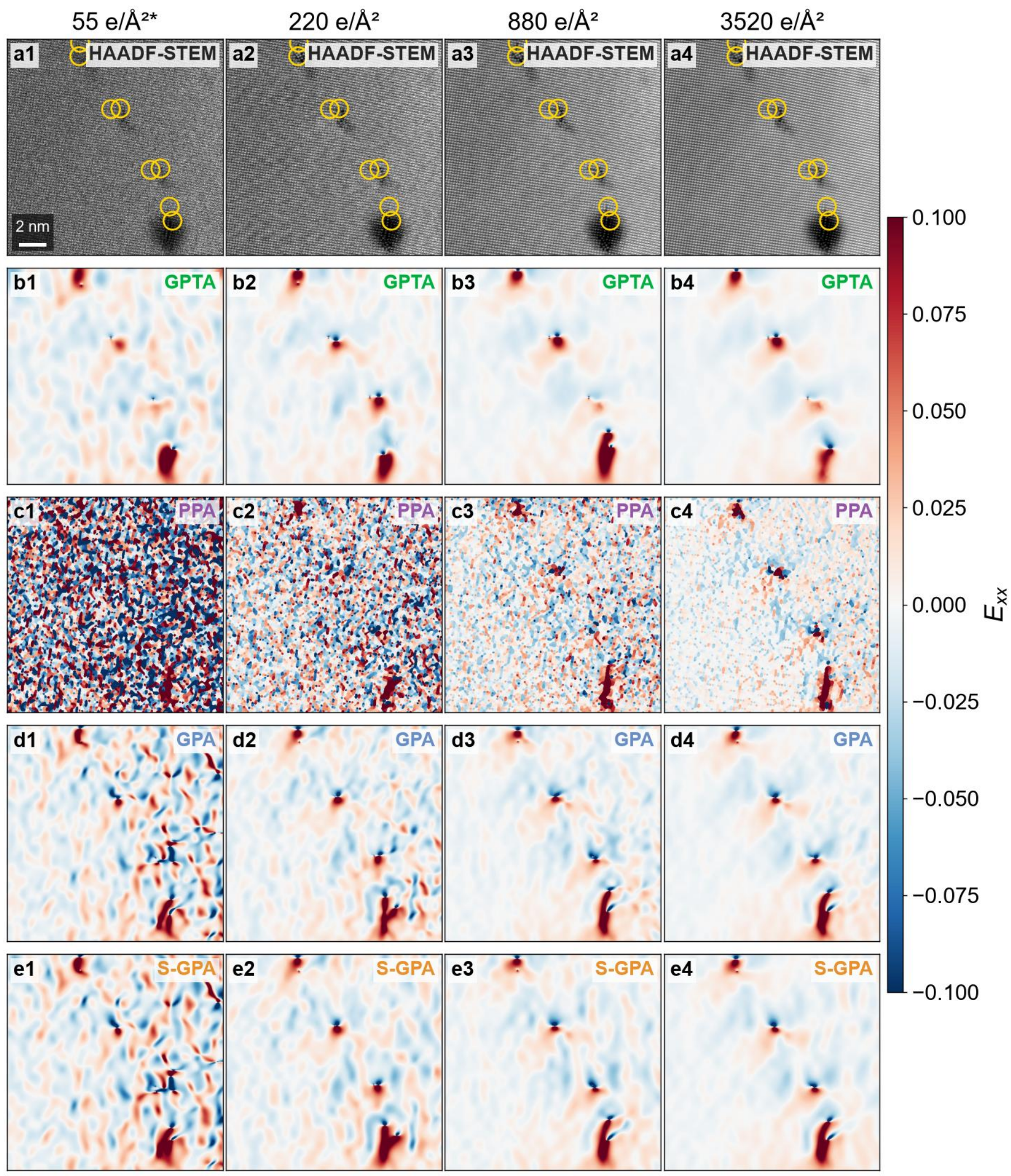

**Supplementary Fig. S33 | Pt dislocation region over a dose series.** HAADF-STEM images of a Pt [001] region in which a low-angle wall of 3.0° crosses the field with eight dislocation cores along it. The cores (circles) are located as phase singularities of the Bragg-filtered 24-frame average, so that no method contributed to their positions. Columns show frame averages of 0.25, 1, 4 and 16 frames, headed by their dose, and the 0.25-frame column (asterisk) is a calibrated Poisson injection. Rows show the images (a1–a4) and the $E_{xx}$ maps from GPTA with the topological reference (b1–b4), classic PPA (c1–c4), GPA at $\sigma = 0.10|\boldsymbol{g}|$ (d1–d4) and S-GPA (e1–e4). With one estimator for every method, $\alpha = \nabla \times \boldsymbol{k}$, the median core offset of GPTA at one frame is 2.5 px, within the lattice period of 5.81 px, compared with 6.5–12.6 px for the other three methods. Scale bar, 2 nm.

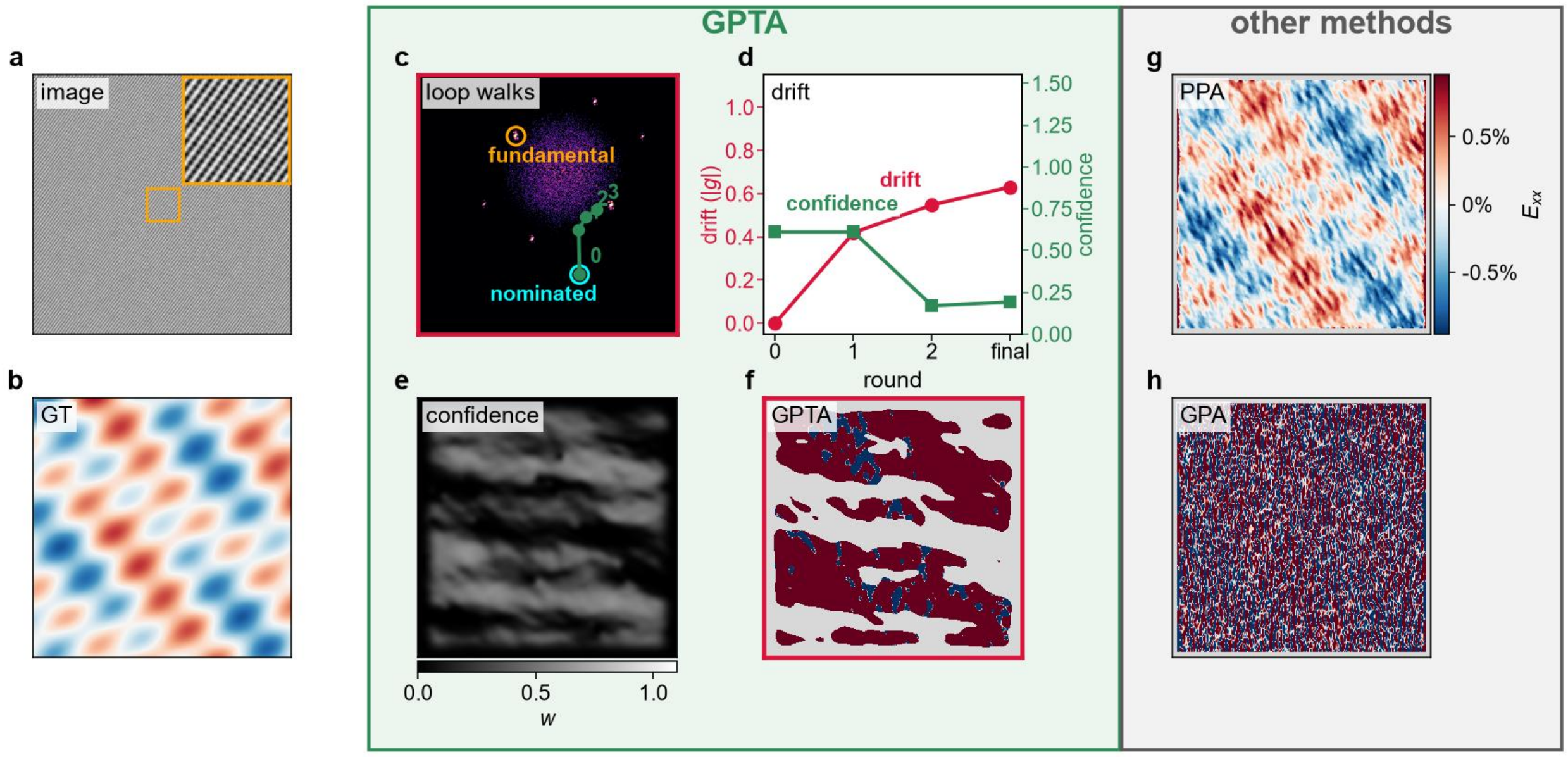

**Supplementary Fig. S34 | Drift of the tracked reference to a neighboring reflection.** A denoised simulated HRTEM image of a dislocation-free single crystal, in which the denoising network leaves the reflection selected by the basis finder, (1,1), far weaker than the neighboring (1,0) fundamental. **a** The image with a magnified inset. **b** Ground truth. **c** The position of the tracked reference in reciprocal space over the iterations, which moves from the selected reflection toward the fundamental. **d** The distance from the selected reflection and the confidence after each iteration. The confidence falls from 0.61 to 0.19 as the reference drifts. **e** The final confidence map. **f** $E_{xx}$ from GPTA, bad area 1.00. **g** $E_{xx}$ from PPA, bad area 0.00. **h** $E_{xx}$ from GPA, which holds its aperture and passes the bad-area criterion (0.16) without resolving the field (correlation 0.04 with the ground truth). The freedom that recovers twins, boundaries and rotation is the same freedom that lets the reference drift to a neighboring reflection, and the confidence reports the drift.

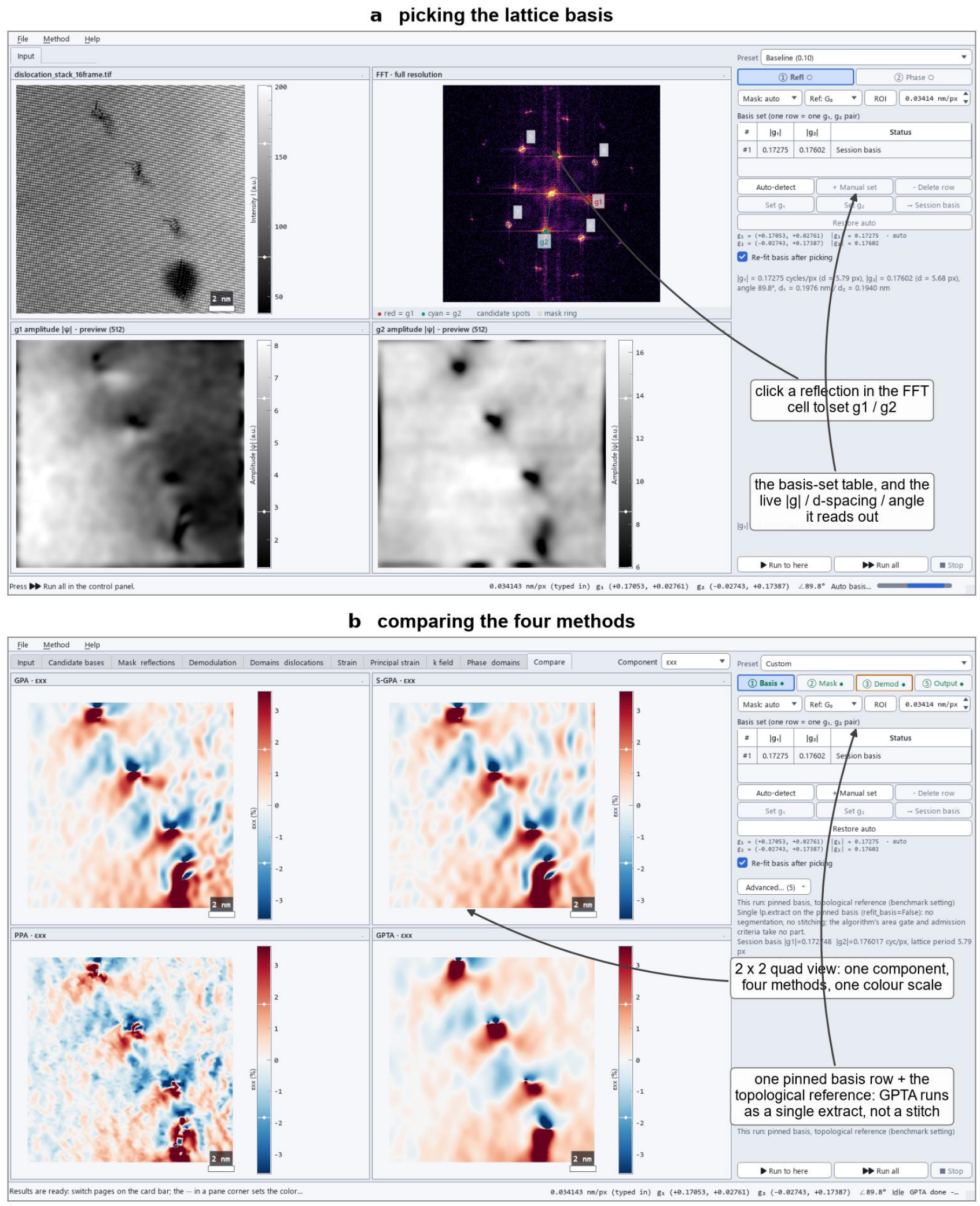

**Supplementary Fig. S35 | Lattice Studio desktop application.** Two screenshots from one session on the 16-frame average of the dislocation stack in **Supplementary Fig. S33**. **a** The Input card after the lattice basis has been picked, showing the loaded frame, the clickable diffractogram with the basis reflections $\boldsymbol{g}_1$ (red) and $\boldsymbol{g}_2$ (cyan), the basis table and the amplitude previews $|\psi|$ of the two reflections, in which the cores are already visible. **b** The Compare card, which displays $E_{xx}$ from GPA, S-GPA, PPA and GPTA in a 2 × 2 view on one color scale. GPTA uses the topological reference at the setting of **Supplementary Fig. S33**.